\documentclass[reprint,prd,longbibliography,eqsecnum,amsfonts,floatfix,nofootinbib,superscriptaddress]{revtex4-1}
\usepackage[pdftex]{graphicx}
\usepackage{amsmath}
\usepackage{multirow}
\usepackage{url}
\usepackage{color}
\usepackage{hyperref}
\usepackage{fnbreak}
\newcommand{\sref}[1]{Sec.~\ref{#1}}
\newcommand{\aref}[1]{Appendix~\ref{#1}}
\newcommand{\fref}[1]{Fig.~\ref{#1}}
\newcommand{\tref}[1]{Table~\ref{#1}}
\newcommand{\Sref}[1]{Section~\ref{#1}}

\newcommand{\adj}[1]{{#1}^{\dag}}
\newcommand{\trans}[1]{{#1}^{\text{tr}}}
\newcommand{\mc}[1]{\mathcal{#1}}
\newcommand{\A}{\mc{A}}
\newcommand{\Ap}{\A_{+}}
\newcommand{\Ac}{\A_{\times}}
\newcommand{\cosi}{\cos\iota}
\newcommand{\abs}[1]{\left\lvert#1\right\rvert}
\newcommand{\smallabs}[1]{\lvert#1\rvert}
\newcommand{\nint}[1]{\left\lfloor#1\right\rceil}
\newcommand{\lint}[1]{\left\lfloor#1\right\rfloor}
\newcommand{\hint}[1]{\left\lceil#1\right\rceil}
\newcommand{\khat}{\hat{k}}

\newcommand{\deltaf}{\delta\!f}
\newcommand{\deltat}{\delta t}
\newcommand{\Tsft}{T_{\text{sft}}}
\newcommand{\Tmax}{T_{\text{max}}}
\newcommand{\Tobs}{T_{\text{obs}}}
\newcommand{\Tmid}{T_{\text{mid}}}
\newcommand{\Tasc}{t_{\text{asc}}}
\newcommand{\Ndet}{N_{\text{det}}}
\newcommand{\Npair}{N_{\text{pairs}}}
\newcommand{\Porb}{P_{\text{orb}}}
\newcommand{\un}[1]{\text{\,#1}}
\newcommand{\cft}[1]{\widetilde{#1}}

\newcommand{\tbin}{\mathfrak{t}}

\newcommand{\tdet}{t}
\newcommand{\tmid}{t}
\newcommand{\tnorm}{\theta}
\newcommand{\rorb}{\vec{r}_{\text{orb}}}
\newcommand{\rdet}{\vec{r}_{\text{det}}}

\newcommand{\kbest}{\tilde{k}}
\newcommand{\kappabest}{\tilde{\kappa}}
\newcommand{\II}{K}
\newcommand{\JJ}{L}
\newcommand{\kI}{k}
\newcommand{\kJ}{\ell}
\newcommand{\jt}{{j}}
\newcommand{\cmplxi}{\mathrm{i}}
\newcommand{\mism}{\mu}
\newcommand{\Gmat}{\mathbf{G}}
\newcommand{\Wmat}{\mathbf{W}}
\newcommand{\vvec}{\mathbf{v}}
\newcommand{\zvec}{\mathbf{z}}
\newcommand{\onemat}{\boldsymbol{1}}
\newcommand{\zeromat}{\boldsymbol{0}}
\newcommand{\muvec}{\boldsymbol{\mu}}

\newcommand{\rhoth}{\rho^{\text{th}}}
\newcommand{\fap}{\alpha}
\newcommand{\rhoave}{\varrho^{\text{ave}}}

\newcommand{\hsens}{h_0^{\text{sens}}}

\newcommand{\ev}[1]{E\left[#1\right]}

\newcommand{\wsqbar}{\overline{w^2}}
\newcommand{\wsqbarbeta}{\overline{w^2_\beta}}
\newcommand{\lamvec}{\boldsymbol{\lambda}}
\newcommand{\Tbar}{\overline{\tdet}}
\newcommand{\tbar}{\overline{t}}
\newcommand{\Tdiff}{\Delta\tdet}
\newcommand{\tbinbar}{\overline{\tbin}}
\newcommand{\tbindiff}{\Delta\tbin}
\newcommand{\Trun}{\Tobs}
\newcommand{\ddiff}{\Delta d}
\newcommand{\Tdrift}{T_{\text{drift}}}
\newcommand{\fdotdrift}{\smallabs{\dot{f}}_{\text{drift}}}
\DeclareMathOperator{\Var}{Var}

\DeclareMathOperator{\sinc}{sinc}
\DeclareMathOperator{\erfc}{erfc}
\DeclareMathOperator{\Real}{Re}
\DeclareMathOperator{\Imag}{Im}
\DeclareMathOperator{\Tr}{Tr}

\newcommand{\coeff}{coefficient}

\newcommand{\Naive}{Naive}

\newcommand{\dcc}{LIGO-P1200142-v7}

\def\commitDATE{ Wed May 20 17:21:53 2015 +0200}

 \newcommand{\onebinpct}{77.4}
\newcommand{\onebinXisq}{0.774}
\newcommand{\twobinXisq}{0.903}
\newcommand{\twobinXisqratio}{1.17}
\newcommand{\LSCRaHr}{16}
\newcommand{\LSCRaMin}{19}
\newcommand{\LSCRaSec}{55.0850}
\newcommand{\LSCNegDecDeg}{15}
\newcommand{\LSCNegDecMin}{38}
\newcommand{\LSCNegDecSec}{24.9}
\newcommand{\LSCApSec}{1.44}
\newcommand{\LSCdApSec}{0.18}

\newcommand{\GallPorbSec}{68023.70}
\newcommand{\GalldPorbSec}{0.04}

\newcommand{\GallTascGPS}{897753994}
\newcommand{\GalldTascGPS}{100}
\newcommand{\orbitAcc}{1.23\times 10^{-8}}

\newcommand{\sensHLVmin}{5\times 10^{-26}}
\newcommand{\htorque}{3.4\times 10^{-26}}

 \newcommand{\Xisqaverect}{0.774}
\newcommand{\XisqaveTukey}{0.699}
\newcommand{\XisqaveHann}{0.601}
 \begin{document}
\title{Model-based cross-correlation search for
  gravitational waves from Scorpius X-1}
\author{John T Whelan}
\email{john.whelan@ligo.org}
\affiliation{  School of Mathematical Sciences
  and
  Center for Computational Relativity and Gravitation,
  Rochester Institute of Technology,
  85 Lomb Memorial Drive, Rochester, New York 14623, USA}
\affiliation{Max-Planck-Institut f\"{u}r Gravitationsphysik
  (Albert-Einstein-Institut), D-30167 Hannover, Germany}
\author{Santosh Sundaresan}
\affiliation{  Indian Institute of Science Education and Research,
  Kolkata, Mohanpur Campus, Nadia District, WB 741252, India}
\author{Yuanhao Zhang}
\email{yuanhao.zhang@ligo.org}
\affiliation{  School of Physics and Astronomy
  and
  Center for Computational Relativity and Gravitation,
  Rochester Institute of Technology,
  84 Lomb Memorial Drive, Rochester, New York 14623, USA}
\author{Prabath Peiris}
\affiliation{  School of Physics and Astronomy
  and
  Center for Computational Relativity and Gravitation,
  Rochester Institute of Technology,
  84 Lomb Memorial Drive, Rochester, New York 14623, USA}
\date{\commitDATE
}
\begin{abstract}
  We consider the cross-correlation search for periodic gravitational
  waves and its potential application to the low-mass x-ray binary
  Sco X-1.  This method coherently combines data not only from
  different detectors at the same time, but also data taken at
  different times from the same or different detectors.  By adjusting
  the maximum allowed time offset between a pair of data segments to
  be coherently combined, one can tune the method to trade off
  sensitivity and computing costs.  In particular, the detectable
  signal amplitude scales as the inverse fourth root of this coherence
  time.  The improvement in amplitude sensitivity for a search with a
  maximum time offset of one hour, compared with a directed stochastic
  background search with 0.25-Hz-wide bins is about a factor of 5.4.
  We show that a search of one year of data from the Advanced LIGO
  and Advanced Virgo detectors with a coherence time of one hour would be able
  to detect gravitational waves from Sco X-1 at the level predicted by
  torque balance over a range of signal frequencies from 30 to 300\,Hz;
  if the coherence time could be increased to ten hours, the range would
  be 20 to 500\,Hz.  In
  addition, we consider several technical aspects of the
  cross-correlation method: We quantify the effects of spectral
  leakage and show that nearly rectangular windows still lead to the
  most sensitive search.  We produce an explicit parameter-space
  metric for the cross-correlation search, in general, and as applied to
  a neutron star in a circular binary system.  We consider the effects
  of using a signal template averaged over unknown amplitude
  parameters: The quantity to which the search is sensitive is a given
  function of the intrinsic signal amplitude and the inclination of
  the neutron-star rotation axis to the line of sight, and the peak of
  the expected detection statistic is systematically offset from the
  true signal parameters.  Finally, we describe the potential loss
  of signal-to-noise ratio due to unmodeled effects such as signal
  phase acceleration within the Fourier transform time scale and
  gradual evolution of the spin frequency.
\end{abstract}
\preprint{\dcc}
\maketitle

\section{Introduction}

The low-mass x-ray binary (LMXB) Scorpius X-1
(Sco X-1)\cite{Steeghs:2001rx} is one of the most promising
potential sources of gravitational waves (GWs) which may be
observed by the generation of GW
detectors---such as Advanced LIGO\cite{aLIGO}, Advanced Virgo\cite{adV}
and KAGRA\cite{KAGRA}---which will begin operation in 2015 with the first
Advanced LIGO observing run, and Advanced Virgo and KAGRA observations
expected to follow in the coming years.  Sco X-1 is
presumed to be a binary consisting of a neutron star which is
accreting matter from a low-mass companion; its parameters are
summarized in \tref{tab:ScoX1}.
\begin{table}[tbp]
  \caption{Parameters of the low-mass x-ray binary Scorpius X-1.  Since
    the sky position is determined to microarcsecond or better
    accuracy, the relevant astrophysical parameters with residual
    uncertainty are those describing the orbit.  Those are the
    projected semimajor axis $a_p=a\sin i$ of the neutron star's
    orbit, the orbital period $\Porb$, and the time $\Tasc$ at
    which the neutron star crosses the ascending node (moving
    away from the observer), measured in the solar-system
    barycenter.  The orbital eccentricity of Sco X-1 is believed
    to be small\cite{Steeghs:2001rx}, and the present work
    presumes the orbit to be circular for simplicity; consideration of
    eccentric orbits would add two search parameters which are
    determined by the eccentricity and the argument of
    periapse\cite{Messenger:2011rg,Leaci:2015bka}. Note that the
    observational constraint in \cite{Steeghs:2001rx} is not on $a_p$
    itself, but on the radial velocity
    amplitude $K_1=\frac{2\pi a_p}{\Porb}$ of the primary.
    We could have formulated
    the parameter space in terms of $K_1$ and $\Porb$ rather than
    $a_p$ and $\Porb$, but this has no significant impact on the
    accuracy of the method, since the uncertainty in $a_p$ is
    dominated by that associated with $K_1$.
                            Finally, note that the
    orbital reference time $\Tasc$ (which we quote as the time of
    ascension of the compact object, $1/4$ cycle before the time
    of inferior conjunction of the companion quoted in
    \cite{Galloway2014}) can be propagated
    to a later epoch by adding an integer number of periods, at
    the cost of increasing the uncertainty due to the uncertainty
    in the period itself.}
  \label{tab:ScoX1}
  \begin{center}
    \begin{tabular*}{\columnwidth}{@{\extracolsep{\fill}}lcc}
      \hline
      \hline
      Parameter & Value & Reference(s) \\
      \hline
      Right ascension
      & $\LSCRaHr^{\mathrm{h}}\LSCRaMin^{\mathrm{m}}\LSCRaSec^{\mathrm{s}}$
      & \cite{Abbott:2006vg} from \cite{Bradshaw1999} \\
      Declination
      & $-\LSCNegDecDeg^{\circ}\LSCNegDecMin'\LSCNegDecSec''$
      & \cite{Abbott:2006vg} from \cite{Bradshaw1999} \\
      Distance (kpc) & $2.8\pm0.3$
      & \cite{Bradshaw1999} \\
      $a_p$ (sec) & $\LSCApSec\pm\LSCdApSec$
      & \cite{Abbott:2006vg} from \cite{Steeghs:2001rx} \\
      $\Tasc$ (GPS sec) & $\GallTascGPS\pm\GalldTascGPS$
      & \cite{Galloway2014} \\
      $\Porb$ (sec) & $\GallPorbSec\pm\GalldPorbSec$
      & \cite{Galloway2014} \\
      \hline
      \hline
    \end{tabular*}
  \end{center}
\end{table}
Nonaxisymmetric deformations in the neutron star can give rise to
gravitational radiation, most of which is emitted at twice the
rotation frequency of the neutron
star\cite{Jaranowski:1998qm}.\footnote{Additionally, unstable
  rotational modes of the neutron star, or $r$ modes
  \cite{Andersson:1998qs} can lead to GW at 4/3 of the neutron star's
  rotational frequency.}  Such deformations can be maintained by the
accretion of matter onto the neutron star.  It has been conjectured
\cite{Bildsten:1998ey} that the neutron star's rotation may be in an
approximate equilibrium state, where the spin-up torque due to
accretion is balanced by the spin-down due to gravitational waves.
Scorpius X-1's high x-ray flux implies a high accretion rate, which
makes it the most promising potential source of observable GWs among
known LMXBs\cite{Watts:2008qw}.

Since Sco X-1 is not seen as a pulsar, its
rotation frequency is unknown.  There is also residual uncertainty
in the orbital parameters which determine the Doppler modulation of
the signal, monochromatic in the neutron star's rest frame, which
reaches the solar-system barycenter (SSB).
This parameter uncertainty limits the
effectiveness of the usual coherent search for periodic gravitational
waves\cite{Jaranowski:1998qm}.  The first search for GW from Sco X-1
with the first generation of interferometric GW detectors, using data
from the second LIGO science run\cite{Abbott:2006vg}, was limited to
six hours of data for this reason.  A subsequent search with data
from the fourth LIGO science run \cite{Abbott:2007tw} used a variant
of the cross-correlation method developed to search for stochastic GW
backgrounds, treating Sco X-1 as a random unpolarized monochromatic
source with a known sky location\cite{Ballmer:2005uw}.\footnote{Other
  methods have been developed, specialized to search for LMXBs.  These
  include summing over contributions from sidebands created by Doppler
  modulation\cite{Messenger:2007zi,Aasi:2014qak}, searching for such
  modulation patterns in doubly-Fourier-transformed
  data\cite{Goetz:2011bd,Aasi:2014sda}, and fitting a polynomial
  expansion in the Doppler-modulated GW
  phase\cite{vanderPutten:2010zz}.}

The stochastic analysis formed the inspiration for a new method to
search for periodic gravitational waves with a model-based
cross-correlation statistic which takes into account the signal model
for continuous GW emission from a rotating neutron
star\cite{Dhurandhar:2007vb}. (This method has also been adapted
\cite{Chung:2011da} to search for young neutron stars in supernova
remnants.)  The present work further develops some of the details of
this method and the specifics of applying it to search for
gravitational waves from Sco X-1 and, by extension, other LMXBs.

The paper is organized as follows: \Sref{s:crosscorr} reviews the
basics of the method and the construction of the combined
cross-correlation statistic using a new, streamlined formalism.
\Sref{s:stats} works out the statistical properties of the
cross-correlation statistic, including the first careful determination
of the effects of signal leakage and the unknown value of the
inclination angle of the neutron star's axis to the line of sight.  It
also considers in detail how the sensitivity of the model-based
cross-correlation search should compare to the directed unmodeled
cross-correlation search for a monochromatic stochastic background.
\Sref{e:paramspace} considers two effects related to the dependence of
the statistic on phase-evolution parameters such as frequency and
binary orbital parameters: a systematic offset of the maximum in
parameter space from the true signal parameters (which depends on the
unknown inclination angle), and the quadratic falloff of the signal
away from its maximum.  The latter is encoded in a parameter space
metric, which we construct in general as well as for the LMXB search
both in its exact form and in limiting form relevant if the
observation time is long compared to the orbital period.  In
\sref{s:deviations} we consider limitations to the method from
inaccuracies in the signal model, either due to slight variations in
frequency (``spin wandering'') arising from an inexact torque-balance
equilibrium, or due to phase acceleration during a stretch of data to
be Fourier transformed.  Finally, in \sref{s:ScoX1} we summarize our
results and consider the expected sensitivity of this search to
Sco X-1.

\section{Cross-Correlation Method}

\label{s:crosscorr}

The cross-correlation method is derived and described in detail in
\cite{Dhurandhar:2007vb}.  In this section, we review the
fundamentals, using a more streamlined formalism and including a more
careful treatment of signal-leakage issues and nuisance parameters.

\subsection{Short-time Fourier transforms}

\label{s:sft}

Because the signal of interest is nearly monochromatic, with
slowly varying signal parameters, it is convenient to describe the
analysis in the frequency domain by dividing the available data into
segments of length $\Tsft$ and calculating a short-time Fourier
transform (SFT) from each.  Since the sampling time $\deltat$ is
typically much less than the SFT duration $\Tsft$, we can
approximate the discrete Fourier transform of the data by a
finite-time continuous Fourier transform.  If we use the index $\II$ to
label both the choice of detector and the selected time interval, which
has midpoint $\tmid_\II$, the SFT will be\footnote{Note that the factor
  $e^{-\cmplxi\pi f_\kI\Tsft}$ appears in Eq.~(2.25) of
  \cite{Dhurandhar:2007vb} with the wrong sign in the exponent.
  However, given \eqref{e:fkdef} for integer $\kI$, this phase
  correction is simply the sign $(-1)^\kI$ so the complex conjugate
  does not change it.}
\begin{equation}
  \label{e:sft0}
  \begin{split}
    \cft{x}_{\II\kI}
    &= \sum_{\jt=0}^{N-1} x_\II(\tmid_\II - \Tsft/2 + \jt\,\deltat)
    \,e^{-\cmplxi2\pi \jt\,\deltat\,\kI/\Tsft}\,\deltat
    \\
    &\approx  e^{-i\pi f_\kI\Tsft}\int_{\tmid_\II-\Tsft/2}^{\tmid_\II+\Tsft/2}
    x_\II(\tdet)\,e^{-\cmplxi2\pi f_\kI(\tdet-\tmid_\II)}\,d\tdet
    \\
    &= (-1)^\kI\int_{\tmid_\II-\Tsft/2}^{\tmid_\II+\Tsft/2}
    x_\II(\tdet)\,e^{-\cmplxi2\pi f_\kI(\tdet-\tmid_\II)}\,d\tdet
    \ ,
  \end{split}
\end{equation}
where the frequency corresponding to the $\kI$th bin of the SFT is
\begin{equation}
  \label{e:fkdef}
  f_\kI = \kI\,\deltaf = \frac{\kI}{\Tsft}
  \ .
\end{equation}

In practice, the data are often multiplied by a window function
$w_\jt=w\left(\frac{\jt\,\deltat-\tmid_\II}{\Tsft}\right)$ before
being Fourier transformed, so that \eqref{e:sft0} becomes
\begin{multline}
  \label{e:sftwin}
  \cft{x}^w_{\II\kI}
  = \sum_{\jt=0}^{N-1} w_\jt\,x_{\II\jt}\,e^{-\cmplxi2\pi \jt\kI/N}\,\deltat
  \\
  \approx (-1)^\kI\int_{\tmid_\II-\Tsft/2}^{\tmid_\II+\Tsft/2}
  w\left(\frac{\tdet-\tmid_\II}{\Tsft}\right)\,x_\II(\tdet)
  \,e^{-\cmplxi2\pi f_\kI(\tdet-\tmid_\II)}\,d\tdet
  \ .
\end{multline}
In this work we assume that the windowing function is nearly
rectangular with some small transition at the beginning and end, so
that leakage of undesirable spectral features is suppressed, but the
effects of windowing on the signal and noise can otherwise be ignored.
The implications of other window choices are considered in
\aref{app:windowing}.

\subsection{Mean and variance of Fourier components}

Let the data
\begin{equation}
x_\II(\tdet) = h_\II(\tdet) + n_\II(\tdet)
\end{equation}
in SFT $\II$ consist of the signal $h_\II(\tdet)$ plus random instrumental
noise $n_\II(\tdet)$ with one-sided power spectral density (PSD)
$S_\II(\abs{f})$ so that its expectation value is
\begin{equation}
  \label{e:ntmean}
  \ev{n_\II(\tdet)}=0
\end{equation}
and\footnote{Strictly speaking, we should allow for data from adjacent
  SFT intervals in the same detector to be correlated, but we assume
  that the autocorrelation function
  $K_n(\tdet-\tdet')=\int_{-\infty}^{\infty}\frac{S_n(\abs{f})}{2}
  e^{-\cmplxi2\pi f(\tdet-\tdet')}\,df$ falls off quickly compared to
  $\Tsft$, so that we can neglect the correlation between noise in
  different time intervals.}
\begin{equation}
  \label{e:ntvar}
  \ev{n_\II(\tdet)n_\JJ(t')}
  = \delta_{\II\JJ}\int_{-\infty}^{\infty}\frac{S_\II(\abs{f})}{2}
  e^{-\cmplxi2\pi f(t-t')}\,df
  \ .
\end{equation}
If we write the noise contribution to the SFT labeled by $\II$ as
\begin{equation}
  \begin{split}
  \cft{n}_{\II\kI}
  &= \sum_{\jt=0}^{N-1} n_{\II\jt}\,e^{-\cmplxi2\pi \jt\kI/N}\,\deltat
  \\
  &\approx \int_{\tmid_\II-\Tsft/2}^{\tmid_\II+\Tsft/2}
  n_\II(\tdet)\,e^{-\cmplxi2\pi (t-[\tmid_\II-\Tsft/2])f_\kI}\,d\tdet
  \end{split}
\end{equation}
then \eqref{e:ntmean} implies $\ev{\cft{n}_{\II\kI}} = 0$ and we can use
\eqref{e:ntvar} to show that
\begin{equation}
  \label{e:expectation_rectangular}
  \ev{\cft{n}_{\II\kI}\cft{n}_{\JJ\kJ}^*}
  \approx \delta_{\II\JJ}\,\delta_{\kI\kJ}\,\Tsft\,\frac{S_\II(f_\kI)}{2}
  \ .
\end{equation}
(As detailed in \aref{app:windowing}, this is not the case for
nontrivial windowing, where noise contributions from different
frequency bins are correlated.)  If we can estimate the noise PSD
$S_\II(f_\kI)$, we can ``normalize'' the data to define (as in
\cite{T0900149-v5})
\begin{equation}
  \label{e:zKkdef}
  z_{\II{\kI}} = \cft{x}_{\II{\kI}} \sqrt{\frac{2}{\Tsft S_\II}}
\end{equation}
which has mean
\begin{equation}
  \label{e:meanztaut}
  \ev{z_{\II{\kI}}} = \mu_{\II{\kI}}
  = \cft{h}_{\II{\kI}} \sqrt{\frac{2}{\Tsft S_\II}}
\end{equation}
unit covariance
\begin{equation}
  \ev{(z_{\II{\kI}}-\mu_{\II{\kI}})
    (z_{\JJ{\kJ}}-\mu_{\JJ{\kJ}})^*}
  = \delta_{\II\JJ}\,\delta_{{\kI}{\kJ}}
\end{equation}
and zero ``pseudocovariance''
\begin{equation}
  \ev{(z_{\II{\kI}}-\mu_{\II{\kI}})
    (z_{\JJ{\kJ}}-\mu_{\JJ{\kJ}})}
  = 0
\end{equation}
(This is because the real and imaginary parts of each $z_{\II{\kI}}$ are
independent and identically distributed.)

\subsection{Signal contribution to SFT}

The signal from a rotating deformed neutron star is determined by
various parameters of the system, which can be divided
into the following categories \cite{Jaranowski:1998qm}.
\begin{enumerate}
\item Amplitude parameters: intrinsic signal amplitude $h_0$,
  the angles $\iota$ and $\psi$ which define the orientation of the
  neutron star's rotation axis ($\iota$ is the inclination to the line
  of sight and $\psi$ is a polarization angle from celestial west to
  the projection of the rotation axis onto the plane of the sky), and the
  signal phase $\Phi_0$ at some reference time.
\item Phase-evolution parameters: intrinsic phase evolution
  (frequency and frequency derivatives) of the signal, as well as
  parameters such as sky location and binary orbital parameters which
  govern the Doppler modulation of the signal.
\end{enumerate}
Those parameters determine the signal received by a gravitational-wave
detector at time $t$ as
\begin{equation}
  h(\tdet) = h_0
  \left(
    F_+ \Ap \cos \Phi(\tdet) + F_\times \Ac \sin \Phi(\tdet)
  \right)
\end{equation}
where $F_+$ and $F_\times$ are the antenna pattern
functions\cite{Jaranowski:1998qm,Whelan:2009jr} which change slowly
with time as the Earth rotates.
The signal contribution to a SFT can be estimated by
\begin{equation}
  \label{e:signal}
  \begin{split}
    h_\II(\tdet)
    \approx h_0
    \bigl\{
    &
    F^{\II}_+ \Ap \cos (\Phi_\II + 2\pi f_\II [\tdet-\tmid_\II])
    \\
    & + F^{\II}_\times \Ac \sin (\Phi_\II + 2\pi f_\II [\tdet-\tmid_\II])
    \bigr\}
  \end{split}
\end{equation}
where we have Taylor expanded the phase about the time $\tmid_\II$:
\begin{equation}
  \label{e:phaselinear}
  \Phi(\tbin(\tdet)) \approx \Phi_\II + 2\pi f_\II (\tdet-\tmid_\II)
  \ .
\end{equation}
The validity of this approximation will be one of the limiting factors
which determines the choice of SFT duration $\Tsft$, as detailed
in \sref{s:sftlength}.

\begin{figure}[tbp]
  \begin{center}
    \includegraphics[width=\columnwidth]{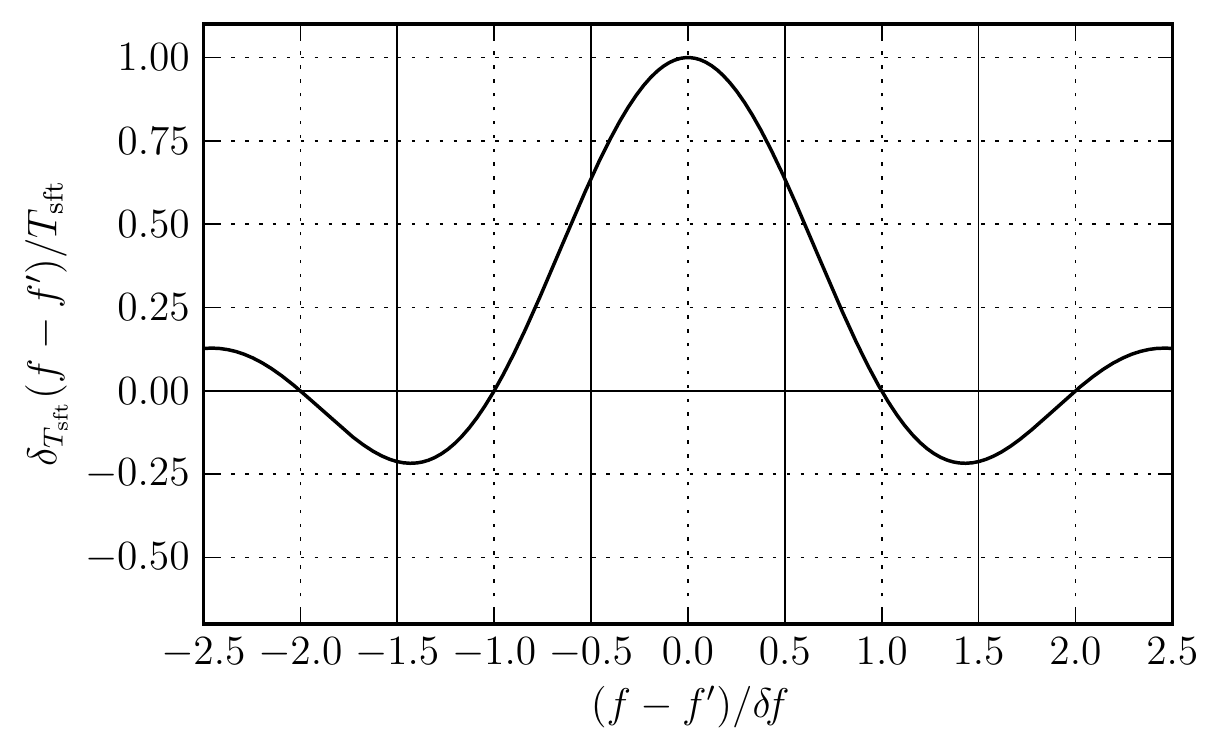}
  \end{center}
  \caption{Plot of $\delta_{\Tsft}(f-f')$ defined in
    \eqref{e:deltarect} which determines the signal contribution to a
    given frequency bin of a short Fourier transform (SFT) of duration
    $\Tsft$ according to \eqref{e:signalsft}.  Since the spacing
    between frequency bins is $\deltaf=1/\Tsft$, there will be, for a
    given signal frequency $f_\II$, one bin whose value of
    $\kappa_{\II\kI}=(f_\kI-f_\II)/\deltaf$ lies between each pair of
    vertical solid lines.}
  \label{f:deltaT}
\end{figure}

The form of \eqref{e:signal} includes the following parameters and
definitions:
\begin{enumerate}
\item $\Ap = \frac{1+\cos^2\iota}{2}$ and $\Ac = \cos\iota$
  depend on the inclination $\iota$ of the rotation axis to the line
  of sight.
\item The antenna patterns $F_+^{\II}$ and $F_\times^{\II}$ depend on the
  detector in question, the sidereal time at $\tmid_\II$, the sky position
  $\alpha,\delta$, and the polarization angle $\psi$.
\item The relationship $\tbin(\tdet)$ between the SSB time and the time at
  the detector depends on the sky position and
  time.\footnote{Specifically, if $\rdet$ is the position of the
    detector and $\khat$ is the unit vector pointing from the source
    to the SSB, $\tbin(\tdet)\approx t - \rdet\cdot\khat/c$.}  Thus the
  phase $\Phi(\tbin(\tdet))$ depends on time, detector, $\Phi_0$,
  $f_0,f_1,\ldots$, sky position and---in the case of a binary---the
  binary orbital parameters.
\end{enumerate}
The signal contribution to bin $\kI$ of SFT $\II$ is
\begin{equation}
  \label{e:signalsft}
  \cft{h}_{\II\kI} \approx h_0 (-1)^\kI
  e^{\cmplxi\Phi_\II} \frac{F^{\II}_+ \Ap - \cmplxi F^{\II}_\times \Ac}{2}
  \delta_{\Tsft}(f_\kI-f_\II)
\end{equation}
where we have defined
\begin{equation}
  \begin{split}
    \label{e:deltarect}
    \delta_{\Tsft}(f_\kI-f_\II)
    &= \int_{\tmid_\II-\Tsft/2}^{\tmid_\II+\Tsft/2}
    e^{-\cmplxi2\pi(f_\kI-f_\II)(\tdet-\tmid_\II)}\,d\tdet
    \\
    &= \Tsft\,\sinc([f_\kI-f_\II]\Tsft)
  \end{split}
\end{equation}
in terms of the normalized sinc function $\sinc\alpha =
\frac{\sin\pi\alpha}{\pi\alpha}$.  This is plotted in
\fref{f:deltaT}.\footnote{Previous sensitivity estimates
  \cite{Dhurandhar:2007vb,Chung:2011da} noted that
  $\delta_{\Tsft}(0)=\Tsft$ and therefore replaced each of the
  finite-time delta functions with the SFT length $\Tsft$, but a
  more careful treatment requires that we keep track of spectral
  leakage caused by the signal frequency not being centered in a SFT
  bin.}
The signal contribution will be largest in the $\kbest_\II$th Fourier
bin, defined by
\begin{equation}
  \label{e:bestbin}
  \kbest_\II := \nint{\frac{f_\II}{\deltaf}} = \nint{f_\II\Tsft}
\end{equation}
whose frequency $f_{\kbest_\II}$ is closest to $f_\II$.  (We have
introduced the notation that $\nint{\alpha}$ is the closest integer to
$\alpha$.)  It will prove useful to define, similarly to
\cite{T0900149-v5},\footnote{Note that our definition of
  $\kappa_{\II\kI}$ differs by a sign from the one used in
  \cite{T0900149-v5}.}
\begin{equation}
  \label{e:kappadef}
  \kappa_{\II\kI} = \kI - f_\II\Tsft = \frac{f_{\kI}-f_\II}{\deltaf}
  \equiv \kappabest_\II + (\kI-\kbest_\II)
\end{equation}
where
\begin{equation}
  \label{e:kappabest}
  \kappabest_\II
  = \frac{f_{\kbest_\II}-f_\II}{\deltaf}=\kbest_\II-f_\II\Tsft
  \ ,
\end{equation}
so that $-\frac{1}{2}\le\kappabest_\II\le\frac{1}{2}$.
A simple search would consider, from each SFT
$\II$, only the Fourier component $\cft{x}_{\II\kbest_\II}$ closest in
frequency to the signal frequency $f_{\II}$ at the search parameters.
However, as we will see, the sensitivity of the search can be improved
by including contributions from additional adjacent bins, so we
indicate by $\mc{K}_{\II}$ the set of bins to be considered from SFT
$\II$, and we will construct a detection statistic using
$\cft{x}_{\II\kI}$ for all $\kI\in\mc{K}_{\II}$.

We can then write
\begin{equation}
  \label{e:signalsft2}
  \cft{h}_{\II\kI} \approx h_0 (-1)^\kI\sinc(\kappa_{\II\kI})
  e^{\cmplxi\Phi_\II} \frac{F^{\II}_+ \Ap - \cmplxi F^{\II}_\times \Ac}{2}
  \Tsft
\end{equation}
which means that, from \eqref{e:meanztaut}
\begin{multline}
  \ev{z_{\II{\kI}}} = \mu_{\II{\kI}}
  \\
  \approx h_0 (-1)^\kI\sinc(\kappa_{\II\kI})
  e^{\cmplxi\Phi_\II} \frac{F^{\II}_+ \Ap - \cmplxi F^{\II}_\times \Ac}{2}
  \sqrt{\frac{2\Tsft}{S_\II}}
\end{multline}

\subsection{Construction of the cross-correlation statistic}

For a given choice of signal parameters, which determine
$\kappabest_\II$ for each SFT, and therefore $\kappa_{\II\kI}$ for
each Fourier component, it is useful to define\footnote{Note that
  computations can be made more efficient by use of the identity
  $\sinc(\kappa_{\II\kI})
  =(-1)^{\kbest_\II-k}\frac{\sin(\pi\kappabest_\II)}{\pi\kappa_{\II\kI}}$
  so $(-1)^\kI\sinc(\kappa_{\II\kI}) =
  (-1)^{\kbest_\II}\sin(\pi\kappabest_\II)\frac{1}{\kappa_{\II\kI}}$
  where only the final factor depends on the bin index
  $k\in\mc{K}_{\II}$.}
\begin{equation}
  \label{e:zbinssum}
  \begin{split}
    z_\II
    &= \frac{\sum_{\kI\in\mc{K}_{\II}}(-1)^\kI\sinc(\kappa_{\II\kI})z_{\II\kI}}
    {\sqrt{\sum_{\kI'\in\mc{K}_{\II}}\sinc^2(\kappa_{\II\kI'})}}
    \\
    &\equiv \frac{1}{\Xi_{\II}}
    \sum_{\kI\in\mc{K}_{\II}}(-1)^\kI\sinc(\kappa_{\II\kI})z_{\II\kI}
  \end{split}
\end{equation}
This is still normalized so that
\begin{subequations}
  \label{e:covcompts}
  \begin{gather}
    \ev{(z_{\II}-\mu_{\II}) (z_{\JJ}-\mu_{\JJ})^*}
    = \delta_{\II\JJ}
    \\
    \ev{(z_{\II}-\mu_{\II}) (z_{\JJ}-\mu_{\JJ})}
    = 0
  \end{gather}
\end{subequations}
where now
\begin{equation}
  \label{e:meancompts}
  \mu_\II
  \approx h_0
  e^{\cmplxi\Phi_\II} \frac{F^{\II}_+ \Ap - \cmplxi F^{\II}_\times \Ac}{2}
  \Xi_{\II}\sqrt{\frac{2\Tsft}{S_\II}}
\end{equation}
If we define vectors indexed by SFT number, we can write
\eqref{e:covcompts} and \eqref{e:meancompts} in matrix form as
\begin{subequations}
  \label{e:zvecstats}
  \begin{gather}
    \ev{\zvec} = \muvec
    \\
    \ev{(\zvec-\muvec)\adj{(\zvec-\muvec)}} = \onemat
    \\
    \ev{(\zvec-\muvec)\trans{(\zvec-\muvec)}} = \zeromat
  \end{gather}
\end{subequations}
where $\onemat$ is the identity matrix, $\zeromat$ is a matrix of
zeros, $\trans{(\cdot)}$ indicates the matrix transpose and
$\adj{(\cdot)}$ the matrix adjoint (complex conjugate of the
transpose).

A real cross-correlation statistic $\rho$ can be constructed by
defining a Hermitian matrix $\Wmat$ and constructing
$\rho=\adj{\zvec}\Wmat\zvec=\Tr(\Wmat\zvec\adj{\zvec})$.
[Our chosen form of $\Wmat$ will be defined in \eqref{e:Wdef}.]
Equation \eqref{e:zvecstats} tells us that
\begin{equation}
  \ev{\zvec\adj{\zvec}} = \onemat + \muvec\adj{\muvec}
\end{equation}
where the second term is a matrix with elements
\begin{equation}
  \mu_{\II}\mu_{\JJ}^*
  = h_0^2 \Xi_{\II}\Xi_{\JJ}
  e^{{\cmplxi}\Delta\Phi_{\II\JJ}}
  \Gamma_{\II\JJ}
  \frac{2\Tsft}{\sqrt{S_\II S_\JJ}}
\end{equation}
where $\Delta\Phi_{\II\JJ}=\Phi_\II-\Phi_\JJ$ is the difference
between the modeled signal phases in the two SFTs
and $\Gamma_{\II\JJ}$ is a geometrical factor which depends on $\iota$ and
$\psi$ as follows [compare Eq.~(3.10) of \cite{Dhurandhar:2007vb}]:
\begin{equation}
  \label{e:Gammadef}
  \begin{split}
    \Gamma_{\II\JJ} =&\,
    \frac{1}{4}
    \bigl(
      F^\II_+ F^\JJ_+\Ap^2 + F^\II_\times F^\JJ_\times\Ac^2
      \\
      &\,
      \phantom{
        \frac{1}{4}
        \bigl(
      }
      +{\cmplxi}[F^\II_+ F^\JJ_\times-F^\II_\times F^\JJ_+]\Ap\Ac
    \bigr)
    \\
    =&\,
    \frac{1}{4}
    \biggl(
    \frac{\Ap^2+\Ac^2}{2}\,(a^\II a^\JJ + b^\II b^\JJ)
    \\
    &\,
    + {\cmplxi}\, \Ap\Ac\, (a^\II b^\JJ - b^\II a^\JJ)
    \\
    &\,
    \phantom{\frac{1}{4}\biggl(}
    +\, \frac{\Ap^2-\Ac^2}{2}\,
    \bigl[
    (a^\II a^\JJ - b^\II b^\JJ)\, \cos 4\psi
    \\
    &\,
    \phantom{
      \frac{1}{4}\biggl(
      +\,
      \frac{\Ap^2-\Ac^2}{2}\,
      \bigl[
    }
    + (a^\II b^\JJ + b^\II a^\JJ)\, \sin 4\psi
    \bigr]
    \biggr)
  \end{split}
\end{equation}
where we have used the fact that the $\psi$ dependence of the antenna
patterns $F^{\II}_{+,\times}$ can be written in terms of the amplitude
modulation (AM) {\coeff}s $a^{\II}$ and $b^{\II}$ as
\begin{alignat}{3}
  \label{e:antennarot}
  F^{\II}_+\, &= &a^{\II}&\,\cos2\psi\, &+&\,b^{\II}\,\sin2\psi
  \\
  F^{\II}_\times\, &= -\,&a^{\II}&\,\sin2\psi\, &+&\,b^{\II}\,\cos2\psi
  \ .
\end{alignat}
The AM {\coeff}s\cite{Jaranowski:1998qm} are determined by the
relevant sky position, detector and sidereal time.  They can be
defined\cite{Prix:2007zh} as
$a^{\II}={\varepsilon}_+^{ab}{d}^{\II}_{ab}$ and
$b^{\II}={\varepsilon}_\times^{ab}{d}^{\II}_{ab}$ where
${\varepsilon}_+^{ab}$ and ${\varepsilon}^{ab}_\times$ are a
polarization basis defined using one basis vector pointing west along
a line of constant declination and one pointing north along a line of
constant right ascension.  Note that $\iota$ and $\psi$ are properties
of the source which do not change for different SFT pairs, while
$a^{\II}$ and $b^{\II}$ depend only on the SFT (detector and sidereal
time) and sky position.  It is also useful to note that the
combinations
\begin{subequations}
  \begin{align}
    F^{\II}_+ F^{\JJ}_+ + F^{\II}_\times F^{\JJ}_\times
    &= a^{\II} a^{\JJ} + b^{\II} b^{\JJ} \equiv 10\Gamma^{\text{ave}}_{\II\JJ}\\
    \label{e:abba}
    F^{\II}_+ F^{\JJ}_\times - F^{\II}_\times F^{\JJ}_+
    &= a^{\II} b^{\JJ} - b^{\II} a^{\JJ} \equiv 10\Gamma^{\text{circ}}_{\II\JJ}
  \end{align}
\end{subequations}
are independent of $\psi$.

Since terms in $\Gamma_{\II\JJ}$ change signs if we vary $\cosi$ and
$\psi$, which are unknown, it is convenient, as proposed in
\cite{Dhurandhar:2007vb}, to work with the average over those
quantities, which picks out the ``robust'' part:
\begin{equation}
  \Gamma^{\text{ave}}_{\II\JJ} = \left\langle\Gamma_{\II\JJ}\right\rangle_{\cosi,\psi}
  = \frac{1}{10} (a^\II a^\JJ + b^\II b^\JJ)
\end{equation}
Note that $\Gamma^{\text{ave}}_{\II\JJ}$ is real and non-negative,
while $\Gamma_{\II\JJ}$ is complex.  On the other hand,
$\Gamma_{\II\JJ}$ can be factored into $\gamma_{\II}\gamma_{\JJ}^*$,
while $\Gamma^{\text{ave}}_{\II\JJ}$ cannot.  If we define (again as
in \cite{T0900149-v5}, but with a different overall normalization)
``noise-weighted AM {\coeff}s'' $\widehat{a}^\II$ and
$\widehat{b}^\II$ by dividing by $\sqrt{\frac{S_\II}{2\Tsft}}$ and
construct $\widehat{\Gamma}_{\II\JJ}$ from those, we can write
\begin{equation}
  \mu_{\II}\mu_{\JJ}^*
  = h_0^2 \Xi_{\II}\Xi_{\JJ} e^{{\cmplxi}\Delta\Phi_{\II\JJ}}
  \widehat{\Gamma}_{\II\JJ}
  = h_0^2 \widehat{G}_{\II\JJ}
\end{equation}
or, as a matrix equation, $\muvec\adj{\muvec} = h_0^2
\widehat{\Gmat}$.  Note that \cite{Dhurandhar:2007vb} did not consider
issues of spectral leakage responsible for $\Xi_{\II}$, and used a
different convention for the placement of complex conjugates in atomic
cross-correlation term, so their $\tilde{\mc{G}}_{\II\JJ}$ would be
equal to $\frac{G_{\II\JJ}^*}{\Xi_{\II}\Xi_{\JJ}}$ in the present
notation.  Similarly, our
$\frac{\widehat{G}_{\II\JJ}^*}{\Xi_{\II}\Xi_{\JJ}}$ corresponds to the
combination $\frac{\tilde{\mc{G}}_{\II\JJ}}{\sqrt{\sigma^2_{\II\JJ}}}$
from \cite{Dhurandhar:2007vb}.\footnote{Note that eq~(3.10) of
  \cite{Dhurandhar:2007vb} is also missing a factor of
  $(-1)^{\kbest_\II-\kbest_\JJ}$ which should appear in
  $\cft{h}^*_{\II\kbest_\II}\cft{h}_{\JJ\kbest_\JJ}$.  This omission
  was pointed out in \cite{Chung:2011da}, but eq~(5) of
  \cite{Chung:2011da} included the wrong sign in the phase correction
  and failed to stress that the relevant frequency is $f_{\kbest_\II}$
  rather than $f_\II$.}

As noted in \cite{Dhurandhar:2007vb}, an ``optimal'' combination of
cross-correlation terms would use a weight $\Wmat$ proportional to
$\widehat{\Gmat}$.  However, as described above, we work with
$\widehat{G}^{\text{ave}}_{\II\JJ}=\Xi_{\II}\Xi_{\JJ}
e^{{\cmplxi}\Delta\Phi_{\II\JJ}}
\widehat{\Gamma}^{\text{ave}}_{\II\JJ}$ in order to avoid specifying
the parameters $\cosi$ and $\psi$.  For reasons of computational cost
to be detailed later, we limit the possible set of SFT pairs $\II\JJ$
included in the cross-correlation to some set $\mc{P}$, in particular
by requiring that $\II<\JJ$ and $\abs{\tdet_{\II}-\tdet_{\JJ}}<\Tmax$.  Then
we define the Hermitian weighting matrix $\Wmat$ by
\begin{equation}
  \label{e:Wdef}
  W_{\II\JJ} =
  \begin{cases}
    N \widehat{G}^{\text{ave}}_{\II\JJ}
    & \II\JJ\in\mc{P}
    \\
    N (\widehat{G}^{\text{ave}}_{\II\JJ})^*
    & \JJ\II\in\mc{P}
    \\
    0 & \text{otherwise}
  \end{cases}
\end{equation}
so that the cross-correlation statistic is
\begin{equation}
  \label{e:CCbins}
  \begin{split}
    \rho
    &=
    \adj{\zvec}\Wmat\zvec=\Tr(\Wmat\zvec\adj{\zvec})
    \\
    &= N\sum_{\II\JJ\in\mc{P}}
    \left(
      \widehat{G}^{\text{ave}}_{\II\JJ} z_{\II}^*z_{\JJ}
      + \widehat{G}^{\text{ave}\,*}_{\II\JJ} z_{\II}z_{\JJ}^*
    \right)
    \\
    &=
    N\sum_{\II\JJ\in\mc{P}}
    \widehat{\Gamma}^{\text{ave}}_{\II\JJ}
    \sum_{\kI\in\mc{K}_{\II}}
    \sum_{\kJ\in\mc{K}_{\JJ}}
    (-1)^{\kI-\kJ}\sinc(\kappa_{\II\kI})\sinc(\kappa_{\JJ\kJ})
    \\
    &
    \phantom{
      N\sum_{\II\JJ\in\mc{P}}
      \widehat{\Gamma}^{\text{ave}}_{\II\JJ}
      \sum_{\kI\in\mc{K}_{\II}}
    }
    \times
    \left(
      e^{{\cmplxi}\Delta\Phi_{\II\JJ}}z_{\II\kI}^*z_{\JJ\kJ}
      + e^{-{\cmplxi}\Delta\Phi_{\II\JJ}}z_{\II\kI}z_{\JJ\kJ}^*
    \right)
  \end{split}
\end{equation}

Since we assume that the list of pairs $\mc{P}$ includes no
autocorrelations, the matrix $\Wmat$ contains no diagonal
elements,\footnote{Note that if we analogously
constructed the matrix to include \emph{only} diagonal terms, i.e.,
constructed a statistic only out of auto-correlations, the statistic
would be equivalent to that used in the PowerFlux method\cite{:2007tda}.}
which implies $\Tr(\Wmat)=0$.
We will later introduce, and use when convenient, the notation that
$\alpha$ labels a (nonordered) pair of SFTs $\II\JJ\in\mc{P}$.

\section{Statistics and Sensitivity}

\label{s:stats}

In this section we consider in detail the statistical properties of
the cross-correlation statistic $\rho$ which were sketched in a basic
form in \cite{Dhurandhar:2007vb}.  In particular, we consider the
impact on the expected sensitivity of spectral leakage and unknown
amplitude parameters, and compare the sensitivity of a
cross-correlation search to the directed stochastic search by analogy
to which it was defined.

\subsection{Mean and variance of cross-correlation statistic}

\label{s:meanvar}

The expectation value of the cross-correlation statistic is
\begin{equation}
  \begin{split}
    \ev{\rho} &= \ev{\Tr(\Wmat\zvec\adj{\zvec})}
    = \Tr(\Wmat) + h_0^2 \Tr(\Wmat\widehat{\Gmat})
    \\
    &= h_0^2 \Tr(\Wmat\widehat{\Gmat})
    = \adj{\muvec}\Wmat\muvec
  \end{split}
\end{equation}
where we have used the fact that $\Wmat$ is traceless.
The variance is
\begin{equation}
  \Var(\rho) = \ev{\rho^2} - \ev{\rho}^2
  = \ev{\adj{\zvec}\Wmat\zvec\adj{\zvec}\Wmat\zvec}
  - (\adj{\muvec}\Wmat\muvec)^2
\end{equation}
The first term can be evaluated by writing $\zvec = (\zvec-\muvec) + \muvec$;
after some simplification we have
\begin{equation}
  \begin{split}
    \Var(\rho)
    =&\, \ev{\adj{(\zvec-\muvec)}\Wmat(\zvec-\muvec)
      \adj{(\zvec-\muvec)}\Wmat(\zvec-\muvec)}
    \\
    &+ 2 \adj{\muvec}\Wmat^2\muvec
  \end{split}
\end{equation}
Ordinarily we would need to know something about the fourth moment of the
noise distribution to evaluate the expectation value, but since
$\Wmat$ contains no diagonal elements, and the different elements of
$\zvec-\muvec$ are independent of each other, the expectation value
can be evaluated using only the variance-covariance matrix of $\zvec$
to give
\begin{equation}
  \label{e:Varrho}
  \Var(\rho)
  = \Tr{\Wmat^2} + 2 \adj{\muvec}\Wmat^2\muvec
  = \Tr{\Wmat^2} + 2 h_0^2 \Tr{\Wmat^2\widehat{\Gmat}}
\end{equation}
We choose the normalization constant $N$ so that $\rho$ has unit
variance in the limit $h_0^2\rightarrow 0$, i.e.,
\begin{equation}
  1 = \Tr(\Wmat^2)
  = \sum_{\II}\sum_{\JJ} W_{\II\JJ}W_{\JJ\II}
  = 2 N^2 \sum_{\II\JJ\in\mc{P}} \abs{\widehat{G}^{\text{ave}}_{\II\JJ}}^2
\end{equation}
i.e.,
\begin{equation}
  N^{-2}
  = 2 \sum_{\II\JJ\in\mc{P}} \abs{\widehat{G}^{\text{ave}}_{\II\JJ}}^2
  = 2 \sum_{\II\JJ\in\mc{P}} \Xi_{\II}^2\Xi_{\JJ}^2
  \left(\widehat{\Gamma}^{\text{ave}}_{\II\JJ}\right)^2
\end{equation}

\begin{figure}[tbp]
  \begin{center}
    \includegraphics[width=\columnwidth]{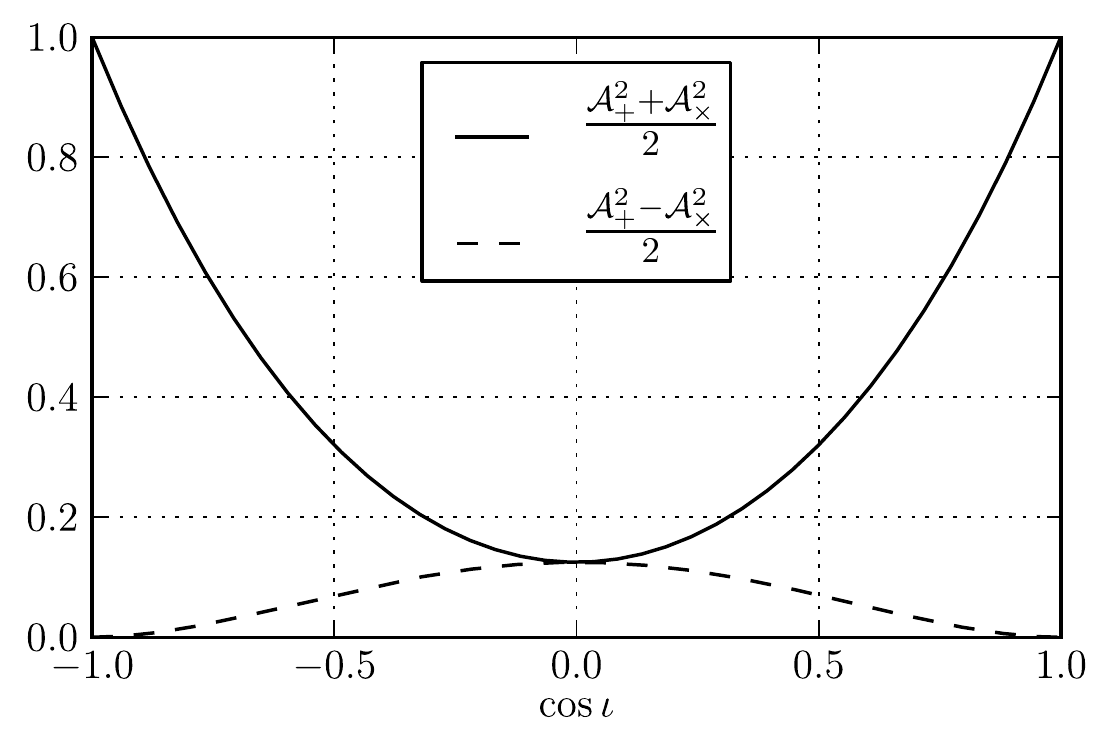}
  \end{center}
  \caption{Plot of $\frac{\Ap^2+\Ac^2}{2}$ and
    $\frac{\Ap^2-\Ac^2}{2}$, the {\coeff}s of the two contributions to
    $\Real\Gamma_{\II\JJ}$ in \eqref{e:ReGamma}.  The factor
    $\frac{\Ap^2+\Ac^2}{2}$ is also equal to
    $\frac{2}{5}\frac{(h^{\text{eff}}_0)^2}{h_0^2}$ where
    $(h^{\text{eff}}_0)^2$ is the combination of $h_0$ and $\cos\iota$
    approximately measured by the cross-correlation statistic, as
    shown in, e.g., Eq.~\eqref{e:evrhosig}}
  \label{f:Ap2Ac2}
\end{figure}

Written in terms of SFT pairs, the expectation value of the statistic is
\begin{equation}
  \label{e:evrho}
  \begin{split}
    \ev{\rho} &= h_0^2 \Tr(\Wmat\widehat\Gmat)
    \\
    &= N h_0^2 \sum_{\II\JJ\in\mc{P}}
    \left(
      \widehat{G}^{\text{ave}}_{\II\JJ} \widehat{G}_{\II\JJ}^*
      + \widehat{G}^{\text{ave}\,*}_{\II\JJ} \widehat{G}_{\II\JJ}
    \right)
    \\
    &= Nh_0^2\,2\sum_{\II\JJ\in\mc{P}}
    \Xi_{\II}^2\Xi_{\JJ}^2
    \widehat{\Gamma}^{\text{ave}}_{\II\JJ}
    \Real\widehat{\Gamma}_{\II\JJ}
  \end{split}
\end{equation}
Looking at \eqref{e:Gammadef} we see that the real part of
$\Gamma_{\II\JJ}$ has a piece proportional to $\Gamma^{\text{ave}}_{\II\JJ}$
and a piece that depends on $\psi$:
\begin{equation}
  \label{e:ReGamma}
  \Real\Gamma_{\II\JJ}
  = \frac{5}{2}\frac{\Ap^2+\Ac^2}{2}\Gamma^{\text{ave}}_{\II\JJ}
  + \frac{\Ap^2-\Ac^2}{2}
  (F^\II_+ F^\JJ_+ - F^\II_\times F^\JJ_\times)
\end{equation}
The sum over SFT pairs $\II\JJ$ can be broken down as a sum over
detector pairs, over time offsets $\tdet_{\II}-\tdet_{\JJ}$, and over the
time stamp $\frac{1}{2}(\tdet_{\II}+\tdet_{\JJ})$ halfway between the
time stamps of the SFTs in the pair.  In an idealized long observing
run, if the detector noise is uncorrelated with sidereal time, the sum
over $\frac{1}{2}(\tdet_{\II}+\tdet_{\JJ})$ means we are averaging the
two expressions $(a^{\II}
a^{\JJ} + b^{\II} b^{\JJ})^2$ and $(a^{\II}
a^{\JJ} + b^{\II} b^{\JJ})(F^\II_+ F^\JJ_+ - F^\II_\times
F^\JJ_\times)$ (the latter of which
depends on the polarization angle $\psi$) over
sidereal time.  Because the former is positive definite and the latter
is not, this average tends to suppress the $\psi$-dependent term.
This is in addition to the fact that
$\frac{\Ap^2+\Ac^2}{2}\ge\frac{\Ap^2-\Ac^2}{2}$,
possibly substantially, depending on the value of $\iota$, as
illustrated in \fref{f:Ap2Ac2}.
If we neglect the second term in \eqref{e:ReGamma}, \eqref{e:evrho}
becomes
\begin{equation}
  \label{e:evrhosig}
  \begin{split}
    \ev{\rho} &\approx Nh_0^2
    \frac{5}{2}\frac{\Ap^2+\Ac^2}{2}
    2 \sum_{\II\JJ\in\mc{P}}
    \Xi_\II^2\Xi_\JJ^2
    \left(\widehat{\Gamma}^{\text{ave}}_{\II\JJ}\right)^2
    \\
    &= (h^{\text{eff}}_0)^2
    \sqrt{
      2\sum_{\II\JJ\in\mc{P}}
      \Xi_\II^2\Xi_\JJ^2
      \left(\widehat{\Gamma}^{\text{ave}}_{\II\JJ}\right)^2
    }
  \end{split}
\end{equation}
where
\begin{equation}
  \label{e:h0eff}
  h_0^{\text{eff}}
  = h_0 \sqrt{\frac{5}{2}\frac{\Ap^2+\Ac^2}{2}}
\end{equation}
is the combination of $h_0$ and $\cosi$ that we can estimate by
filtering with the averaged template.

Since we have normalized the statistic so that $\Var(\rho)=1$ for weak
signals, the expectation value \eqref{e:evrhosig} is an expected
signal-to-noise ratio for a signal with a given $h_0^{\text{eff}}$.
This means that if we define a SNR threshold $\rhoth$ such that
$\rho>\rhoth$ corresponds to a detection, the signal will be
detectable if
\begin{equation}
  h_0^{\text{eff}} \gtrsim
  \sqrt{\rhoth}
  \left(
    2\sum_{\II\JJ\in\mc{P}}
    \Xi_\II^2\Xi_\JJ^2
    \left(\widehat{\Gamma}^{\text{ave}}_{\II\JJ}\right)^2
  \right)^{-1/4}
\end{equation}

\subsection{Impact of spectral leakage on estimated sensitivity}
\label{s:kappa}

Finally, we consider the impact of the leakage factors of the form
$\Xi_\II^2= \sum_{\kI\in\mc{K}_{\II}}\sinc^2(\kappa_{\II\kI})$ on the
expectation value.  Expanding these expressions, we have
\begin{multline}
  \ev{\rho} \approx
  (h^{\text{eff}}_0)^2
  \biggl(
  2\sum_{\II\JJ\in\mc{P}}
  \left(\widehat{\Gamma}^{\text{ave}}_{\II\JJ}\right)^2
  \\
  \times
  \sum_{\kI\in\mc{K}_{\II}}\sinc^2(\kappa_{\II\kI})
  \sum_{\kJ\in\mc{K}_{\JJ}}\sinc^2(\kappa_{\JJ\kJ})
  \biggr)^{1/2}
\end{multline}
If we choose only the ``best bin'' $k_\II=\kbest_\II$ from each SFT,
defined by \eqref{e:bestbin}, we have
\begin{equation}
  \Xi_\II^2 = \sinc^2(\kappabest_{\II})
\end{equation}
If, instead of the best bin whose frequency $f_{\kbest_\II}$ is
closest to $f_\II$, we take the $m$ closest bins to define $\mc{K}_\II$,
the sum becomes
\begin{equation}
  \begin{split}
    \Xi_\II^2
    &= \sum_{\kI\in\mc{K}_{\II}}\sinc^2(\kappa_{\II\kI})
    = \sum_{\kI =\kbest_\II-\hint{(m-1)/2}}^{\kbest_\II+\lint{(m-1)/2}}
    \sinc^2(\kappa_{\II\kI})
    \\
    &= \sum_{s=-\hint{(m-1)/2}}^{\lint{(m-1)/2}}
    \sinc^2(\kappabest_\II + s)
  \end{split}
\end{equation}
where $\lint{\alpha}\le\alpha$ and $\hint{\alpha}\ge\alpha$ are the
integers below and above $\alpha$, respectively.  Note that, because
of the identity\footnote{This is most easily proved by writing
  $\sinc(\kappa+s)=\int_{-1/2}^{1/2}e^{i 2\pi (\kappa+s)t}\,dt$ and
  using $\sum_{s=-\infty}^{\infty}e^{\cmplxi2\pi s(t-t')} =
  \sum_{s=-\infty}^{\infty}\delta(t-t'+s)$.}
$\sum_{s=-\infty}^{\infty} \sinc^2(\kappa+s) = 1$, valid for any
$\kappa$, the best we can do by including more bins is $\Xi_{\II}^2
\le 1$ and therefore\footnote{Previous sensitivity estimates
  \cite{Dhurandhar:2007vb,Chung:2011da} were missing the factor of
  $\Xi_{\II}^2\Xi_{\JJ}^2$ and therefore slightly overestimated the
  sensitivity.}
\begin{equation}
 \label{e:sensbound}
 \ev{\rho} \le
  (h^{\text{eff}}_0)^2
  \sqrt{
    2\sum_{\II\JJ\in\mc{P}}
    \left(\widehat{\Gamma}^{\text{ave}}_{\II\JJ}\right)^2
  }
\end{equation}
The sensitivity associated with the inclusion of a finite number of
bins from each SFT will depend on the value of
$-\frac{1}{2}\le\kappabest_\II\le\frac{1}{2}$ corresponding to the
signal frequency $f_\II$ in each SFT.  We can get an estimate of this by
assuming that, over the course of the analysis, the Doppler shift
evenly samples the range of $\kappabest$ values, and writing
\begin{equation}
  \label{e:evrhoXi}
  \begin{split}
    \ev{\rho}
    &\approx
    (h^{\text{eff}}_0)^2
    \sqrt{
      2 \langle\Xi^2\rangle^2\sum_{\II\JJ\in\mc{P}}
      \left(\widehat{\Gamma}^{\text{ave}}_{\II\JJ}\right)^2
    }
    \\
    &=
    (h^{\text{eff}}_0)^2
    \langle\Xi^2\rangle
    \sqrt{
      2\sum_{\II\JJ\in\mc{P}}
      \left(\widehat{\Gamma}^{\text{ave}}_{\II\JJ}\right)^2
    }
  \end{split}
\end{equation}
with
\begin{equation}
  \langle\Xi^2\rangle
  =
  \left\langle \sum_s \sinc^2(\kappa+s) \right\rangle_\kappa
\end{equation}
where $\langle\cdot\rangle_\kappa$ indicates an average over the
possible offsets within the bin.  We can numerically evaluate
\begin{equation}
  \label{e:Xiave}
  \begin{split}
    \langle\Xi^2\rangle
    &= \sum_s \left\langle \sinc^2(\kappa+s) \right\rangle_\kappa
    \\
    &= \sum_{s=-\hint{(m-1)/2}}^{\lint{(m-1)/2}}
    2\int_{0}^{1/2}\sinc^2(\kappa+s)\,d\kappa
    \\
    &= 2\int_{0}^{m/2}\sinc^2\!\kappa\,d\kappa
  \end{split}
\end{equation}
as shown in \tref{tab:bincontribs}.
\begin{table}[tbp]
  \caption{Contributions to $\langle\Xi^2\rangle$, defined in
    \eqref{e:Xiave}, from inclusion of multiple SFT bins.  We see that
    using a single bin from each SFT leads to only around \onebinpct\%
    of the maximum sensitivity given by \eqref{e:sensbound}, but that we
    can recover over 90\% of this sensitivity by using two bins and over
    95\% by using four bins from each SFT.  This table applies for
    rectangularly windowed data; using other window options further
    reduces the expected SNR, as described in \aref{app:windowing}.  The
    table also assumes that the various Doppler modulations move the
    signal frequency around to accomplish an average over the fractional
    offset of the signal frequency from the center of the bin.  The
    validity of this approximation is explored in \cite{T1200431-v1}.}
  \begin{center}
    \begin{tabular*}{\columnwidth}{@{\extracolsep{\fill}}lcccccc}
\hline
\hline
$m$ & 1 & 2 & 3 & 4 & 5 & 6 \\
\hline
Contribution & 0.774 & 0.129 & 0.028 & 0.019 & 0.009 & 0.007 \\
Cumulative & 0.774 & 0.903 & 0.931 & 0.950 & 0.959 & 0.966 \\
\hline
\hline
\end{tabular*}
   \end{center}
  \label{tab:bincontribs}
\end{table}

Since most cross-correlation searches will be computationally limited,
the question of how many bins to include from each SFT is one of
optimization of resources.  The value of $\ev{\rho}$ for a given
$h^{\text{eff}}_0$, and therefore the sensitivity of the search, can
be increased by including more frequency bins from each SFT, but this
will involve more computations and therefore more computational
resources.  If instead those resources were put into a search with a
larger $\Tmax$, the value of $\sum_{\II\JJ\in\mc{P}}
\left(\widehat{\Gamma}^{\text{ave}}_{\II\JJ}\right)^2$ would be
higher.  {\Naive}ly, one might expect the computing cost to scale with
the number of terms to be combined, and therefore with the square of
the number of bins taken from each SFT.  So increasing from $m=1$ to
$m=2$ could take up to $4$ times the computing cost.  On the other
hand, for a fixed number of bins, we suppose that the cost will scale
with the number of SFT pairs to be included times the number of
parameter space points to be searched.  Typical behavior will be for
the density of points in parameter space to scale with $\Tmax^d$ for
some integer value of $d$; as described in \sref{s:metric}, for a
search over frequency and two orbital parameters of an LMXB, as long
as $\Tmax$ is small compared to the binary orbital period, $d=3$.
Since the number of SFT pairs at fixed observation time will also
scale like $\Tmax$, the overall computing cost will scale like
$\Tmax^{d+1}$, and quadrupling the computing time would mean
multiplying the possible $\Tmax$, and thus the number of terms in the
sum \eqref{e:evrhoXi} by $4^{\frac{1}{d+1}}$.  This would increase
$\ev{\rho}$ for a given $h^{\text{eff}}_0$ by a factor of
$4^{\frac{1}{2(d+1)}}=2^{\frac{1}{d+1}}$.  For $d=3$, this is
$2^{1/4}\approx 1.19$, which is very slightly more than the benefit
$\frac{\twobinXisq}{\onebinXisq}\approx\twobinXisqratio$ from
including a second bin from each SFT.  However, the assumption that
computing cost scales like $m^2$ is likely an overestimate (since most
of the operations can be done once per SFT rather than once per pair),
so it is generally advisable to use at least two bins from each SFT.

\subsection{Sensitivity estimate for unknown amplitude parameters}

\label{s:hsens}

The cross-correlation statistic is normalized so that
$\Var(\rho)\approx 1$ and, according to \eqref{e:evrhoXi}, and now
adopting the notation that $\alpha$ refers to an unordered allowed
pair of SFTs,
\begin{equation}
  \label{e:evrhoalpha}
  \ev{\rho}
  = (h^{\text{eff}}_0)^2
  \langle\Xi^2\rangle
  \sqrt{
    2\sum_{\alpha}
    \left(\widehat{\Gamma}^{\text{ave}}_{\alpha}\right)^2
  }
  = (h^{\text{eff}}_0)^2 \rhoave
\end{equation}
where $h^{\text{eff}}_0$ is the combination of $h_0$ and $\cos\iota$
given in \eqref{e:h0eff}, and $\rhoave$ is a property of the search
which can be determined from noise spectra, AM {\coeff}s, and choices
of SFT pairs, without knowledge of signal parameters other than the
approximate frequency and orbital parameters.  Even if the noise in
each data stream is Gaussian distributed, the statistic, which
combines the data quadratically, will not be.  It was observed in
\cite{Dhurandhar:2007vb} that each individual cross correlation
between SFTs is Bessel distributed; the of the optimal sum is
considered in \aref{app:stats} both in its exact form and a numerical
approximation.  For simplicity, in what follows we assume that the
central limit theorem allows us to treat the statistic as
approximately Gaussian, with mean $(h^{\text{eff}}_0)^2 \rhoave$ and
unit variance.\footnote{Note that this approximation is less accurate
  in the tails of the distribution.  Unfortunately, for a search over many
  independent templates, the most interesting statistic will
  necessarily be in the tails.  For example, with $10^8$ templates,
  even a $1\%$ false alarm probability for the loudest statistic value
  would correspond to a single-template false alarm probability of
  $10^{-10}$.  See\cite{CrossCorrMDC} for specific examples of this.}

We consider the sensitivity estimates in \cite{Dhurandhar:2007vb},
which implicitly assume the values of $\iota$ and $\psi$ are known and
used to construct the expected cross correlation used in weighting the
terms in the statistic.  [In our notation this would mean using
$\widehat{G}_{\II\JJ}$ rather than $\widehat{G}^{\text{ave}}_{\II\JJ}$
in the definition \eqref{e:Wdef} of $\Wmat$.]  Here we perform the
analogous calculation, assuming we're using
$\widehat{G}^{\text{ave}}_{\alpha}$ in the construction of the
statistic.  Thus the probability of exceeding a threshold $\rhoth$
will be
\begin{multline}
  P(\rho>\rhoth|h_0,\iota,\psi)
  = \int_{\rhoth}^{\infty} f(\rho|h_0,\iota,\psi)\,d\rho
  \\
  \approx
  \frac{1}{\sqrt{2\pi}}
  \int_{\rhoth}^{\infty}
  \exp
  \left(
    -\frac{1}{2}
    \left[
      \rho
      -
      (h^{\text{eff}}_0)^2 \rhoave
    \right]^2
  \right)
  \, d\rho
  \\
  = \frac{1}{2}
  \erfc\left(\frac{\rhoth- (h^{\text{eff}}_0)^2 \rhoave}{\sqrt{2}}\right)
  = \frac{1}{2}
  \erfc\left(\frac{\rhoth-h_0^2\varrho(\iota)}{\sqrt{2}}\right)
\end{multline}
where
\begin{equation}
  \varrho(\iota)
  \approx \frac{5}{2}\frac{\Ap^2+\Ac^2}{2} \varrho^{\text{ave}}
  = \frac{5}{16}(1+6\cos^2\iota+\cos^4\iota) \varrho^{\text{ave}}
\end{equation}
The threshold associated with a false alarm probability $\fap$ is
\begin{equation}
  \rhoth = \sqrt{2} \erfc^{-1}(2\fap)
\end{equation}
but the sensitivity $\hsens$ associated with a false dismissal probability
$\beta$ will now be defined, following a procedure analogous to the one
in \cite{Wette:2011eu}, by marginalizing over the unknown
inclination $\iota$ (since we have neglected the $\psi$ dependence in
$\ev{\rho}$)\footnote{Note that if we had kept the $\psi$-dependent
  term in \eqref{e:ReGamma}, the resulting $\ev{\rho}/h_0^2$ would
  depend not only on both $\iota$ and $\psi$, but also on the detector
  geometry and pairs of SFTs and a numerical solution to the
  equivalent of \eqref{e:margeff} would have to be performed anew for
  basically each sensitivity estimate.}
\begin{equation}
  \label{e:margeff}
  \begin{split}
    1 - \beta &= P(\rho>\rhoth|h_0=\hsens)
    \\
    &= \left\langle
      P(\rho>\rhoth|h_0=\hsens,\iota,\psi)
    \right\rangle_{\cos\iota,\psi}
    \\
    &=
    \frac{1}{2}
    \left\langle
      \erfc\left(\frac{\rhoth-(\hsens)^2\,\varrho(\iota)}{\sqrt{2}}\right)
    \right\rangle_{\cos\iota}
  \end{split}
\end{equation}
So to get a sensitivity estimate, we need to find the $\hsens$ which
solves \eqref{e:margeff}, i.e.,
\begin{multline}
  \label{e:statsave}
  2(1-\beta)
  \\
  \approx
    \left\langle
      \erfc\left(
        \frac{\rhoth}{\sqrt{2}}
        -\frac{(\hsens)^2\,\varrho^{\text{ave}}}{\sqrt{2}}
        \frac{5}{16}(1+6\cos^2\iota+\cos^4\iota)
      \right)
    \right\rangle_{\cos\iota}
    \\
    =
    \int_0^1
    \erfc\left(
      \erfc^{-1}(2\fap)
      - \mc{S}^{\text{eff}}
      \frac{5}{16}\left[1+6\chi^2+\chi^4\right]
    \right)
    \,d\chi
  \end{multline}
so that the approximate sensitivity is
\begin{equation}
  \label{e:hsens}
  \hsens = \sqrt{\frac{\mc{S}^{\text{eff}}\sqrt{2}}{\rhoave}}
  = \left(
    (\mc{S}^{\text{eff}})^{-2}
    \langle\Xi^2\rangle^2
    \sum_{\alpha}
    \left(\widehat{\Gamma}^{\text{ave}}_{\alpha}\right)^2
  \right)^{-1/4}
\end{equation}
Equation \eqref{e:statsave} defines $\mc{S}^{\text{eff}}$ as a
specific function of $\fap$ and $\beta$, so the approximate
sensitivity correction due to marginalizing over $\cos\iota$ can be
worked out independently of the details of the search.
\begin{table*}[tbp]
  \caption{Approximate modification of search sensitivity, as a
    function of desired false alarm probability $\fap$
    (corresponding to a statistic threshold of $\rhoth$) and false
    dismissal probability $\beta$, resulting from filtering with a
    template averaged over the signal parameters $\cos\iota$ and
    $\psi$.  (The second set of $\fap$ values is chosen to correspond
    to interesting single-template false alarm probabilities with a
    trials factor of $10^8$.)
    The detectable signal amplitude $\hsens$
    \eqref{e:hsens} is proportional to $\sqrt{\mc{S}^{\text{eff}}}$.
    The table shows, for a variety of choices of $\fap$ and
    $\beta$, how the corrected factor $\sqrt{\mc{S}^{\text{eff}}}$
    calculated according to \eqref{e:statsave} compares to the
    standard expression
    $\mc{S}=\erfc^{-1}(2\fap)+\erfc^{-1}(2\beta)$ which would apply
    from filtering with known values of the parameters $\cos\iota$
    and $\psi$.  Note that using the worst-case value $\cos\iota=0$
    shows that $1<\mc{S}^{\text{eff}}/\mc{S}<3.2$.}
  \label{tab:statseff}
  \begin{center}
    \begin{tabular*}{\textwidth}{@{\extracolsep{\fill}}lcccccccccc}
\hline
\hline
\multicolumn{2}{c}{} & 
\multicolumn{3}{c}{$\mc{S}$} & \multicolumn{3}{c}{$\mc{S}^{\text{eff}}$} & \multicolumn{3}{c}{$\sqrt{\mc{S}^{\text{eff}}/\mc{S}}$}\\
\multicolumn{2}{c}{}
 & \multicolumn{3}{c}{$\beta$} & \multicolumn{3}{c}{$\beta$} & \multicolumn{3}{c}{$\beta$}\\
\cline{3-5}\cline{6-8}\cline{9-11}$\alpha$ & $\rhoth$
& $0.10$ & $0.05$ & $0.01$
& $0.10$ & $0.05$ & $0.01$
& $0.10$ & $0.05$ & $0.01$ \\
\hline
$0.10$ & $1.3$
& $1.81$ & $2.07$ & $2.55$
& $3.49$ & $4.45$ & $6.27$
& $1.39$ & $1.47$ & $1.57$\\
$0.05$ & $1.6$
& $2.07$ & $2.33$ & $2.81$
& $4.15$ & $5.16$ & $7.03$
& $1.42$ & $1.49$ & $1.58$\\
$0.01$ & $2.3$
& $2.55$ & $2.81$ & $3.29$
& $5.42$ & $6.52$ & $8.47$
& $1.46$ & $1.52$ & $1.60$\\
$10^{-9}$ & $6.0$
& $5.15$ & $5.40$ & $5.89$
& $12.73$ & $14.16$ & $16.40$
& $1.57$ & $1.62$ & $1.67$ \\
$5\times 10^{-10}$ & $6.1$
& $5.23$ & $5.48$ & $5.96$
& $12.96$ & $14.40$ & $16.64$
& $1.57$ & $1.62$ & $1.67$ \\
$10^{-10}$ & $6.4$
& $5.40$ & $5.66$ & $6.14$
& $13.48$ & $14.93$ & $17.20$
& $1.58$ & $1.62$ & $1.67$ \\
\hline
\hline
\end{tabular*}
   \end{center}
\end{table*}
We show some sample values \tref{tab:statseff} for $\fap$
and $\beta$ values between $1\%$ and $10\%$, and also for
single-template $\fap$ values corresponding to overall false alarm
probabilities in the same range, assuming a trials factor of $10^8$.  We see
that the $h_0$ sensitivity is modified by between $39\%$ and $67\%$ in
these cases.

\subsection{Scaling and comparison to directed stochastic search}

\label{e:scaling}

We consider here the behavior of \eqref{e:hsens} (or equivalently
\eqref{e:evrhoalpha}) with parameters such as the observing time
$\Tobs$ and allowed lag time $\Tmax$, which is effectively a coherence
time.  As noted in \cite{Dhurandhar:2007vb}, the detectable
\eqref{e:hsens} scales like one over the fourth root of the number of
SFT pairs included in the sum $\sum_\alpha$\footnote{Note that the
  averages here are not the weighted averages introduced in
  \sref{e:paramspace}.}:
\begin{equation}
  \begin{split}
    \hsens
    &= \left(
      (\mc{S}^{\text{eff}})^{-2}
      \langle\Xi^2\rangle^2
      \Npair
      \langle
      (\widehat{\Gamma}^{\text{ave}}_{\alpha})^2
      \rangle
    \right)^{-1/4}
    \\
    &= \left(
      \Npair\Tsft^2
      (\mc{S}^{\text{eff}})^{-2}
      \langle\Xi^2\rangle^2
      \left\langle
        \frac{4(\Gamma^{\text{ave}}_{\II\JJ})^2}{S_\II S_\JJ}
      \right\rangle
    \right)^{-1/4}
  \end{split}
\end{equation}
The approximate number of pairs for a search of data from $\Ndet$
detectors, each with observing time $\Tobs$ (so that the total
observation time is $\Ndet\Tobs$), with maximum lag time $\Tmax>\Tsft$
is
\begin{equation}
  \Npair \approx \Ndet^2\frac{\Tobs}{\Tsft}\frac{\Tmax}{\Tsft}
\end{equation}
so the sensitivity scaling is
\begin{equation}
  \hsens
  \sim
  \left(
    \Ndet^2\Tobs\Tmax
    (\mc{S}^{\text{eff}})^{-2}
    \langle\Xi^2\rangle^2
    \left\langle
      \frac{4(\Gamma^{\text{ave}}_{\II\JJ})^2}{S_\II S_\JJ}
    \right\rangle
  \right)^{-1/4}
\end{equation}

We wish to compare this sensitivity to that of the directed stochastic
search (also known as the ``radiometer'' method) defined in
\cite{Ballmer:2005uw} and used to set limits on gravitational
radiation from Sco X-1\cite{Abbott:2007tw,Abbott:2011rr}.  The
directed stochastic search is also an optimally weighted
cross-correlation search, but only includes contributions from data
taken by different detectors at the same time.  We first consider the
sensitivity of a cross-correlation search using our method with this
restriction, and then relate this to the sensitivity of the actual
directed stochastic search.  If we only allow simultaneous pairs of
SFTs, the number of pairs included in the sum \eqref{e:hsens} becomes
\begin{equation}
  \Npair^{\text{simul}} \approx \Ndet(\Ndet-1)\frac{\Tobs}{\Tsft}
\end{equation}
which makes the signal strength to which the search is sensitive
\begin{multline}
  (\hsens)^{\text{simul}}
  \\
  \sim
  \left(
    \Ndet(\Ndet-1)\Tobs\Tsft
    (\mc{S}^{\text{eff}})^{-2}
    \langle\Xi^2\rangle^2
    \left\langle
      \frac{4(\Gamma^{\text{ave}}_{\II\JJ})^2}{S_\II S_\JJ}
    \right\rangle
  \right)^{-1/4}
  \\
  \sim
    \hsens
    \left(
      \left[
        1-\frac{1}{\Ndet}
      \right]
      \frac{\Tsft}{\Tmax}
    \right)^{-1/4}
  \end{multline}
  The directed stochastic search is not quite the same as this
  hypothetical cross-correlation search with simultaneous SFTs,
  however.  Most of these differences are irrelevant or produce
  effectively identical calculations.  For instance, since the
  $\Tdiff_\alpha$ appearing in \eqref{dPhialpha} is zero for
  simultaneous SFTs, the phase difference $\Delta\Phi_\alpha=2\pi
  f_0\ddiff_\alpha$ just encodes the difference in arrival times at
  the two detectors.  Likewise, while the stochastic search assumes a
  random unpolarized signal rather than the periodic signal from a
  neutron star with unknown parameters, this has the same effect as
  our choice to use $\Gamma^{\text{ave}}_{\II\JJ}$ as the geometrical
  weighting factor.  In fact (as noted in \cite{Dhurandhar:2007vb})
  $e^{{\cmplxi}\Delta\Phi_{\II\JJ}}
  \widehat{\Gamma}^{\text{ave}}_{\II\JJ}$ is, up to a normalization,
  the overlap reduction function for the directed stochastic search.
  The one significant difference is that, since the stochastic search
  does not model the orbital Doppler modulation, it does not have access
  to the signal frequency $f_\II$ corresponding to SFT $\II$, and
  therefore cannot localize the expected signal frequency to a bin of
  width $\deltaf=\frac{1}{\Tsft}$.  Thus, instead of the optimal
  combination described by \eqref{e:zbinssum} or \eqref{e:CCbins}, it
  must sum with equal weights the contributions
  $z_{\II\kI}z^*_{\JJ\kI}$ across a coarse frequency bin of width
  $\Delta f\gtrsim \frac{2\pi a_p}{\Porb}f_0$ (see
  \eqref{s:metric-lmxb} for the definitions of the binary orbital
  parameters relevant to Doppler modulation).\footnote{This was not
    the original motivation for the coarse frequency bins in the
    stochastic cross-correlation pipeline; see for example
    \cite{Abbott:2003hr}, but it has this effect when using the method
    to search for monochromatic signals from neutron stars in binary
    systems.  Note also that it is sufficient to perform a single sum
    $\sum_\kI z_{\II\kI}z^*_{\JJ\kI}$ across the coarse bin rather
    than a double sum such as $\sum_\kI\sum_\kJ
    z_{\II\kI}z^*_{\JJ\kJ}$ because, while the frequency bin
    containing the signal is not known, it will be the same bin for
    both detectors because the unknown phase shift due to the orbit is
    the same for simultaneous SFTs.} The effect is to increase the
  variance of the cross correlation due to noise by
  $\frac{\Delta{f}}{\deltaf}=\Delta{f}\,\Tsft$ (since there are
  $\Delta{f}\,\Tsft$ bins being combined, only one of which contains a
  significant signal contribution) so that
\begin{multline}
  \label{e:stochsens}
    (\hsens)^{\text{stoch}}
    \\
    \sim
    \left(
      \Ndet(\Ndet-1)\frac{\Tobs}{\Delta f}
      (\mc{S}^{\text{eff}})^{-2}
      \left\langle
        \frac{4(\Gamma^{\text{ave}}_{\II\JJ})^2}{S_\II S_\JJ}
      \right\rangle
    \right)^{-1/4}
    \\
    \sim
    \hsens
    \left(
      \langle\Xi^2\rangle^{-2}
      \left[
        1-\frac{1}{\Ndet}
      \right]
      \frac{1}{\Delta f\Tmax}
    \right)^{-1/4}
  \end{multline}
  The appearance of the factor containing $\langle\Xi^2\rangle$ in the
comparison is because the directed stochastic search, by combining a
larger range of frequency bins, as well as techniques such as
overlapping windowed segments, avoids some of the usual leakage
issues.  On the other hand, if $\Delta f$ is chosen to maximize the
sensitivity for a given frequency, there will be similar issues with
part of the signal falling outside the coarse bin at the extremes of
Doppler modulation.

To insert concrete numbers, \eqref{e:stochsens} tells us that for a
search with data of equivalent sensitivity from three detectors, a
cross-correlation search with $\Tmax=3600\un{s}$ and
$\langle\Xi^2\rangle=0.9$ would provide an improvement in $h_0$
sensitivity over a directed stochastic search with $\Delta
f=0.25\un{Hz}$ of a factor of about $5.4$.\footnote{This does not
  include the fact that the directed stochastic method includes a
  relatively coarse search over frequency, while the model-based
  cross-correlation method must search over many more points in
  frequency and orbital parameter space, as described in
  \sref{s:metric}.  This seemingly significant increase in trials
  factor turns out to be swamped by the gain in sensitivity.  In the
  comparison above, the same signal will generate a factor of almost
  30 larger rho value in the cross-correlation search.  On the other
  hand, the $\rho$ threshold to achieve a $5\sigma$ false alarm
  probability would need to be increased only from $5$ to $7.8$ to
  overcome a trials factor of $10^8$.  Additionally, the search over
  signal parameters in the cross-correlation method allows estimates
  of those parameters.}
This is consistent with the performance of the two searches in the
Sco X-1 Mock Data Challenge\cite{ScoX1MDC},
in which the cross-correlation method was
able to detect signals with $h_0$ almost an order of magnitude lower
than those detected by the directed stochastic method.

Note that, unlike the model-based cross-correlation search, the
stochastic search is not computationally limited, with year-long
wide-band analyses being achievable on a single CPU\cite{ScoX1MDC}.
Additionally, since it does not assume a signal model (beyond sky
localization and approximate monochromaticity), it is robust against
unexpected features such as orbital parameters outside the nominally
expected range.  However, its sensitivity is fundamentally limited by
its ignorance of orbital Doppler modulation, with a maximum effective
coherence time of $\frac{1}{\Delta f} \lesssim
\frac{\Porb}{2\pi{}a_pf_0} \approx
\left(\frac{100\un{Hz}}{f_0}\right)75\un{sec}$.

\section{Parameter space behavior}

\label{e:paramspace}

So far we have implicitly assumed the parameters used to construct the
signal model \eqref{e:signalsft}, other than the amplitude parameters
$h_0$, $\cosi$, and $\psi$, were known when constructing the weighted
statistic.  In order to determine the phase evolution of the signal,
and therefore $\Phi_{\II}$ and $f_{\II}$, we need various phase-evolution
parameters $\{\lambda_i\}$.  (For example, for a neutron star at a
known sky location with a constant intrinsic signal frequency $f_0$ in
a binary orbit, these are $f_0$ and any unknown binary orbital
parameters.)  A slight error in these would lead to the $\Phi_{\II}$
appearing in $\muvec$ and that used to construct $\Wmat$ being
slightly different.  In this case we need to go back to
\eqref{e:evrho} and distinguish between the true $\Delta\Phi_{\II\JJ}$
and the one assumed in the construction of the filter.\footnote{It is
  also possible for $\widehat{\Gamma}_{\II\JJ}$ and/or $\Xi_{\II}\Xi_{\JJ}$
  to differ from their assumed values, e.g., if the search parameters
  include sky position which can change the amplitude modulation
  {\coeff}s, or a change in Doppler modulation affects the location of
  the signal frequency within the bin.  We follow the usual procedure
  of focusing on the dominant effect, which is the change in the
  expected signal phase, and thereby obtain a ``phase metric'' for the
  cross-correlation search.}  If we write these parameters as
$\{\lambda_i\}$, let the parameters assumed in constructing $\rho$ be
$\lambda_i$ and the true parameters of the signal be $\lambda_i^s$.
Let $\Delta\Phi_{\II\JJ}^s$ and $\Delta\Phi_{\II\JJ}$ be the phase
difference $\Phi_{\II}-\Phi_{\JJ}$ constructed with the true signal
parameters and the parameters assumed in $\Wmat$, respectively.  The
effect will be to reduce the expected SNR $\ev{\rho}$ from the value
given in \eqref{e:evrhoalpha} which it would attain with
$\lambda_i=\lambda_i^s$.  The modified value is
\begin{multline}
  \label{e:evrhomismatched}
  E[\rho]
    \approx h_0^2 N \langle\Xi^2\rangle
  \\
  \times\sum_{\alpha}
  \left(
    \widehat{\Gamma}_{\alpha} e^{\cmplxi(\Delta\Phi_{\alpha}^s-\Delta\Phi_{\alpha})}
    + \widehat{\Gamma}_{\alpha}^* e^{-\cmplxi(\Delta\Phi_{\alpha}^s-\Delta\Phi_{\alpha})}
  \right)
  \widehat{\Gamma}^{\text{ave}}_{\alpha}
\end{multline}
Now, for $\lambda_i$ close to $\lambda_i^s$,
\begin{multline}
  \label{e:evrhomismatched2}
  \widehat{\Gamma}_{\alpha} e^{\cmplxi(\Delta\Phi_{\alpha}^s-\Delta\Phi_{\alpha})}
  + \widehat{\Gamma}_{\alpha}^* e^{-\cmplxi(\Delta\Phi_{\alpha}^s-\Delta\Phi_{\alpha})}
  \\
  =
  2\Real \widehat{\Gamma}_{\alpha}
  \cos(\Delta\Phi_{\alpha}^s-\Delta\Phi_{\alpha})
  - 2\Imag \widehat{\Gamma}_{\alpha}
  \sin(\Delta\Phi_{\alpha}^s-\Delta\Phi_{\alpha})
  \\
  \approx
  2\Real \widehat{\Gamma}_{\alpha}
  \left(
    1 - \frac{1}{2}(\Delta\Phi_{\alpha}-\Delta\Phi_{\alpha}^s)^2
  \right)
  + 2\Imag \widehat{\Gamma}_{\alpha}
  (\Delta\Phi_{\alpha}-\Delta\Phi_{\alpha}^s)
\end{multline}
if we write the phase difference as
\begin{equation}
  \begin{split}
    \Delta\Phi_{\alpha}-\Delta\Phi_{\alpha}^s
    \approx\ &
    \sum_i\Delta\Phi_{\alpha,i} (\lambda_i-\lambda_i^s)
    \\
    &+\frac{1}{2}\sum_{i,j}\Delta\Phi_{\alpha,ij} (\lambda_i-\lambda_i^s)
    (\lambda_j-\lambda_j^s)
  \end{split}
\end{equation}
where
$\Delta\Phi_{\alpha,i}=\frac{\partial\Phi_{\alpha}}{\partial\lambda_i}$,
we obtain, to second order in the parameter difference,
\begin{multline}
  \label{e:rhoquad}
  \ev{\rho}
  \approx (h_0)^2 N \langle\Xi^2\rangle
  \left(
    2\sum_{\alpha}
    \widehat{\Gamma}^{\text{ave}}_{\alpha}\Real\widehat{\Gamma}_{\alpha}
  \right)
  \\
  \times
  \left[
    1 - \sum_i \epsilon^s_i    (\lambda_i-\lambda_i^s)
    - \sum_{i,j} g_{ij}
    (\lambda_i-\lambda_i^s)
    (\lambda_j-\lambda_j^s)
  \right]
\end{multline}
where
\begin{equation}
  \epsilon^s_i
  = -\frac
  {
    2\sum_{\alpha}
    \widehat{\Gamma}^{\text{ave}}_{\alpha}\Imag\widehat{\Gamma}_{\alpha}
    \Delta\Phi_{\alpha,i}
  }{
    2\sum_{\alpha}
    \widehat{\Gamma}^{\text{ave}}_{\alpha}\Real\widehat{\Gamma}_{\alpha}
  }
\end{equation}
and the parameter space metric is
\begin{equation}
  g_{ij}
  = \frac{1}{2}
  \frac{
    2\sum_{\alpha}
    \widehat{\Gamma}^{\text{ave}}_{\alpha}
    \left(
      \Real\widehat{\Gamma}_{\alpha}
      \Delta\Phi_{\alpha,i}
      \Delta\Phi_{\alpha,j}
      + \Imag\widehat{\Gamma}_{\alpha}\Delta\Phi_{\alpha,ij}
    \right)
  }{
    2\sum_{\alpha}
    \widehat{\Gamma}^{\text{ave}}_{\alpha}\Real\widehat{\Gamma}_{\alpha}
  }
\end{equation}
If we once again neglect the $\psi$-dependent piece of
$\Real\widehat{\Gamma}_{\alpha}$ as well as the second derivative term in
the metric, we have
\begin{equation}
  \label{e:crosscorrmetriccalc}
  \begin{split}
  g_{ij}
  &\approx \frac{1}{2}
  \frac{\sum_{\II\JJ\in\mc{P}}
    (\widehat{a}^{\II} \widehat{a}^{\JJ} + \widehat{b}^{\II} \widehat{b}^{\JJ})^2
    \Delta\Phi_{\alpha,i} \Delta\Phi_{\alpha,j}
  }
  {\sum_{\II\JJ\in\mc{P}}
    (\widehat{a}^{\II} \widehat{a}^{\JJ} + \widehat{b}^{\II} \widehat{b}^{\JJ})^2}
  \\
  &= \frac{1}{2}
  \frac{\sum_{\alpha} \widehat{\Gamma}^{\text{ave}}_{\alpha}
    \Delta\Phi_{\alpha,i} \Delta\Phi_{\alpha,j}
  }
  {\sum_{\II\JJ\in\mc{P}} \widehat{\Gamma}^{\text{ave}}_{\alpha}}
  = \frac{1}{2}\langle \Delta\Phi_{\alpha,i} \Delta\Phi_{\alpha,j}\rangle_{\alpha}
\end{split}
\end{equation}
where $\langle\cdot\rangle_{\alpha}$ indicates a weighted average with
weighting factor
$\left(\widehat{\Gamma}^{\text{ave}}_{\alpha}\right)^2$ [recall
$\widehat{\Gamma}^{\text{ave}}_{\II\JJ} \propto\left(\frac{a_\II a_\JJ
    + b_\II b_\JJ}{S_\II S_\JJ}\right)^2$] and
\begin{multline}
  \epsilon^s_i
  \approx
  \frac{2\Ap\Ac}{\Ap^2+\Ac^2}
  \\
  \times
  \frac{\sum_{\II\JJ\in\mc{P}}
    (\widehat{a}^{\II} \widehat{a}^{\JJ} + \widehat{b}^{\II} \widehat{b}^{\JJ})
    (\widehat{a}^{\II} \widehat{b}^{\JJ} - \widehat{b}^{\II} \widehat{a}^{\JJ})
    \Delta\Phi_{\II\JJ,i}
  }
  {
    \sum_{\II\JJ\in\mc{P}}
    (\widehat{a}^{\II} \widehat{a}^{\JJ} + \widehat{b}^{\II} \widehat{b}^{\JJ})^2
  }
  \\
  =
  \frac{2\Ap\Ac}{\Ap^2+\Ac^2}
  \frac{\sum_{\alpha}
    \widehat{\Gamma}^{\text{ave}}_{\alpha}\widehat{\Gamma}^{\text{circ}}_{\alpha}
    \Delta\Phi_{\alpha,i}
  }
  {\sum_{\alpha}
    \left(\widehat{\Gamma}^{\text{ave}}_{\alpha}\right)^2
  }
\end{multline}

\subsection{Systematic parameter offset}

\label{s:systematic}

The result \eqref{e:rhoquad} not only tells us how the expected SNR
falls off when the parameters $\{\lambda_i\}$ used in constructing the
statistic differ from the true signal parameters $\{\lambda_i^s\}$, it
also shows that the maximum of $\ev{\rho}$ is not actually at the signal
point $\lambda_i=\lambda_i^s$, but at the point
$\lambda_i=\lambda_i^m$ defined by
\begin{equation}
  0 = \epsilon^s_i + \sum_j 2g_{ij}(\lambda_j^m-\lambda_j^s)
\end{equation}
i.e., at
\begin{equation}
  \lambda_i^m = \lambda_i^s - \sum_j\frac{1}{2}g^{-1}_{ij} \epsilon^s_j
\end{equation}
where $\{g^{-1}_{ij}\}$ is the matrix inverse of the metric $\{g_{ij}\}$.

If the metric is approximately diagonal, so that $g^{-1}_{ii}\approx
\frac{1}{g_{ii}}$, then the offset of
the true signal parameters from the maximum value of $\ev{\rho}$ is
\begin{equation}
  \lambda_i^s - \lambda_i^m = \frac{1}{2}\frac{\epsilon^s_i}{g_{ii}}
  \approx \frac{2\Ap\Ac}{\Ap^2+\Ac^2}
  \frac{
    \sum_{\alpha}
    \widehat{\Gamma}^{\text{ave}}_{\alpha}\widehat{\Gamma}^{\text{circ}}_{\alpha}
    \Delta\Phi_{\alpha,i}
  }{
    \sum_{\alpha}
    \left(\widehat{\Gamma}^{\text{ave}}_{\alpha}\right)^2
    \Delta\Phi_{\alpha,i} \Delta\Phi_{\alpha,i}
  }
\end{equation}
This offset depends on the (generally unknown) value of the
inclination angle $\iota$ via $\Ap = \frac{1+\cos^2\iota}{2}$ and $\Ac
= \cos\iota$.  In particular it has the opposite sign for
$\iota\in(0,\pi/2)$ and $\iota\in(\pi/2,\pi)$.  For a signal detection
with unknown $\iota$, this will have the effect of a systematic error
in the measurement of the phase-evolution parameters $\{\lambda_i\}$.
(Of course, one could perform a subsequent analysis
which would produce an
estimate of $\iota$, such as a coherent followup of the signal
candidate, or a cross-correlation search using
$\cmplxi\Gamma^{\text{circ}}_{\II\JJ}$ in place of
$\Gamma^{\text{ave}}_{\II\JJ}$ in the construction of $\Wmat$.)

\subsection{Parameter space metric}
\label{s:metric}

We return now to consideration of the metric
defined by \eqref{e:crosscorrmetriccalc},
\begin{equation}
  \label{e:crosscorrmetric}
  g_{ij} =
  -
  \left.
    \frac{1}{2}\frac{\ev{\rho}_{,ij}}{\ev{\rho}}
  \right\rvert_{\lamvec=\lamvec^s}
  \approx
  \frac{1}{2}\langle \Delta\Phi_{\alpha,i} \Delta\Phi_{\alpha,j}\rangle_{\alpha}
  \ .
\end{equation}

\subsubsection{Comparison to standard expression for metric}

We can relate this to the usual notation for the phase metric.  [See,
e.g., Eq.~(5.13) of \cite{Brady:1997ji}, which was also used
in\cite{Chung:2011da}.]
\begin{equation}
  \label{e:standardmetric}
  g_{ij} =
  \langle \Phi_{,i} \Phi_{,j}\rangle
  - \langle \Phi_{,i}\rangle \langle\Phi_{,j}\rangle
\end{equation}
Note, first of all, that while the standard definition of the
parameter space metric defines the mismatch as the fractional loss in
signal-to-noise squared, our cross-correlation statistic $\rho$ is
actually the equivalent of what is usually called $\rho^2$.  This is
because it is quadratic in the signal (as is the $\mc{F}$ statistic,
and its expectation value is proportional to $h_0^2$).

The connection between \eqref{e:crosscorrmetric} and
\eqref{e:standardmetric} is made by noting that the averages in
\eqref{e:standardmetric} are over data segments, while the expression
in \eqref{e:crosscorrmetric} is a weighted average over SFT pairs,
where the weighting factor is
$(\widehat{\Gamma}^{\text{ave}}_{\alpha})^2$.  We can relate the two
in the special case where the set of pairs $\mc{P}$ contains
\emph{every} combination of SFTs (e.g., by choosing $\Tmax$ to be the
observing time), and by neglecting the influence of the weighting
factor in the cross-correlation metric.  In that case, the average can
be written as a double average over SFTs $\II$ and $\JJ$:
\begin{equation}
  \begin{split}
    g_{ij}
    &= \frac{1}{2}
    \bigl\langle
    (\Phi_{\II,i}-\Phi_{\JJ,i})(\Phi_{\II,j}-\Phi_{\JJ,j})
    \bigr\rangle_{\II\JJ\in\mc{P}}
    \\
    &= \frac{1}{2}
    \bigl\langle
    \Phi_{\II,i}\Phi_{\II,j} + \Phi_{\JJ,i}\Phi_{\JJ,j}
    - \Phi_{\II,i}\Phi_{\JJ,j} - \Phi_{\JJ,i}\Phi_{\II,j}
    \bigr\rangle_{\II\JJ\in\mc{P}}
    \\
    &= \frac{1}{2}
    \bigl(
    \langle\Phi_{\II,i}\Phi_{\II,j}\rangle_{\II}
    + \langle\Phi_{\JJ,i}\Phi_{\JJ,j}\rangle_{\JJ}
    \\
    &\phantom{
      = \frac{1}{2}
      \bigl(
    }
    - \langle\Phi_{\II,i}\rangle_{\II}\langle\Phi_{\JJ,j}\rangle_{\JJ}
    - \langle\Phi_{\JJ,i}\rangle_{\JJ}\langle\Phi_{\II,j}\rangle_{\II}
    \bigr)
    \\
    &= \langle\Phi_{\II,i}\Phi_{\II,j}\rangle_{\II}
    - \langle\Phi_{\II,i}\rangle_{\II}\langle\Phi_{\JJ,j}\rangle_{\JJ}
  \end{split}
\end{equation}
which is just \eqref{e:standardmetric}.  Note that this identification
can only be made in the case where the cross correlation includes all
pairs of SFTs (or all pairs within some time stretch).  With a
restriction such as $\abs{\tdet_\II-\tdet_\JJ}\le\Tmax$, one must
consider the weighted average over pairs, not separate averages over
SFTs.

\subsubsection{Metric for the LMXB search}
\label{s:metric-lmxb}

We now consider the explicit form of the parameter space metric for a
neutron star in a circular binary system, assuming a constant
intrinsic frequency $f_0$.  Although the actual values of phase
$\Phi_\II=\Phi(\tbin(\tmid_\II))$ and frequency
$\frac{1}{2\pi}f_\II=\left.\frac{d\Phi(\tbin(\tdet))}{d\tdet}\right\rvert_{\tdet=\tmid_\II}$
used via \eqref{e:phaselinear} to construct the expected
cross correlation $\widehat{G}_{\II\JJ}$ include relativistic
corrections, it is sufficient for the purposes of constructing the
parameter space metric to limit attention to the Roemer delay, which
gives us
\begin{equation}
  \label{e:LMXBphase}
  \begin{split}
    \Phi_\II
    &
    = \Phi_0 + 2\pi f_0
    \left(
      \tmid_\II - \frac{\rdet\cdot\khat}{c} + \frac{\rorb\cdot\khat}{c}
    \right)
    \\
    &
    = \Phi_0 + 2\pi f_0
    \left\{
      \tmid_\II - d_\II
      - a_p \sin
      \left[
        \frac{2\pi}{\Porb} (\tmid_\II-\Tasc)
      \right]
    \right\}
  \end{split}
\end{equation}
where we have defined the following:
\begin{enumerate}
\item $d_\II = \frac{\rdet\cdot\khat}{c}$, the projected distance, in
  seconds, from the solar-system barycenter to the detector, along the
  propagation direction from the source.  (Note that this depends on
  the detector, but also on the time $\tmid_\II$.)
\item $a_p=\frac{a\sin i}{c}$ is the projected semimajor axis of the
  binary orbit, in units of time.
\item $\Porb$ is the orbital period of the binary.
\item $\Tasc$ is a reference time for the orbit, defined as the
  time, measured at the solar-system barycenter, when the neutron star
  is crossing the line of nodes moving away from the solar system.
\end{enumerate}
If we use the identity
\begin{equation}
  \sin A - \sin B
  = 2 \cos\left(\frac{A+B}{2}\right) \sin\left(\frac{A-B}{2}\right)
\end{equation}
we have
\begin{multline}
  \label{dPhialpha}
  \Delta\Phi_{\alpha} = 2\pi f_0
  \biggl\{
    \Tdiff_\alpha - \ddiff_{\alpha}
    \\
    - 2\,a_p
    \,
    \sin\frac{\pi \Tdiff_\alpha}{\Porb}
    \,
    \cos
    \left[
      \frac{2\pi}{\Porb}
      (\Tbar_{\alpha}-\Tasc)
    \right]
  \biggr\}
\end{multline}
where we have defined
$\ddiff_{\II\JJ}=d_\II-d_\JJ$, $\Tdiff_{\II\JJ}=\tmid_\II-\tmid_\JJ$,
and $\Tbar_{\II\JJ}=\frac{\tmid_\II+\tmid_\JJ}{2}$.

Note that $\ddiff_{\II\JJ}$ will be much less than $\Tdiff_{\II\JJ}$
unless the SFTs $\II$ and $\JJ$ are simultaneous.  (This is because
the duration of a SFT will be long compared to the light travel time
between detectors on the Earth, and the Earth's motion is
nonrelativistic.)

We can now calculate the derivatives appearing in
\eqref{e:crosscorrmetric}:
\begin{subequations}
  \label{e:lmxbderivs}
  \begin{align}
    \begin{split}
    \frac{\partial\Delta\Phi_\alpha}{\partial f_0}
    &=
    2\pi
    \biggl\{
      \Tdiff_\alpha - \ddiff_{\alpha}
      \\
      &\phantom{
        =
        2\pi
        \biggl\{
      }
      - 2\,a_p
      \,
      \sin\frac{\pi \Tdiff_\alpha}{\Porb}
      \,
      \cos
      \left[
        \frac{2\pi}{\Porb}
        (\Tbar_{\alpha}-\Tasc)
      \right]
      \biggr\}
    \end{split}
    \\
    \frac{\partial\Delta\Phi_\alpha}{\partial a_p}
    &=
    -4\pi f_0
    \,
    \sin\frac{\pi \Tdiff_\alpha}{\Porb}
    \,
    \cos
    \left[
      \frac{2\pi}{\Porb}
      (\Tbar_{\alpha}-\Tasc)
    \right]
    \\
    \frac{\partial\Delta\Phi_\alpha}{\partial \Tasc}
    &=
    -\frac{8\pi^2 f_0 a_p}{\Porb}
    \,
    \sin\frac{\pi \Tdiff_\alpha}{\Porb}
    \,
    \sin
    \left[
      \frac{2\pi}{\Porb}
      (\Tbar_{\alpha}-\Tasc)
    \right]
    \\
    \begin{split}
    \frac{\partial\Delta\Phi_\alpha}{\partial \Porb}
    &=
    -\frac{4\pi f_0 a_p}{\Porb}
    \biggl\{
      \frac{2\pi}{\Porb}
      (\Tbar_{\alpha}-\Tasc)
      \,
      \\
      &\phantom{
        =
        -\frac{4\pi f_0 a_p}{\Porb}
        \biggl\{
      }
      \times\sin\frac{\pi \Tdiff_\alpha}{\Porb}
      \,
      \sin
      \left[
        \frac{2\pi}{\Porb}
        (\Tbar_{\alpha}-\Tasc)
      \right]
    \\
    &\phantom{
            4\pi
            \biggl\{
    }
      -\frac{\pi \Tdiff_\alpha}{\Porb}
      \,
      \cos\frac{\pi \Tdiff_\alpha}{\Porb}
      \,
      \cos
      \left[
        \frac{2\pi}{\Porb}
        (\Tbar_{\alpha}-\Tasc)
      \right]
    \biggr\}
    \end{split}
  \end{align}
\end{subequations}

\subsubsection{Approximation for long observation times}

\label{s:metric-lmxb-long}

It is relatively simple and straightforward to construct the phase
metric for a given observation; calculate the derivatives
\eqref{e:lmxbderivs} for each SFT pair and then insert them into the
weighted average \eqref{e:crosscorrmetric}.  However, we can gain insight into
the behavior of the metric if we consider an approximate form which
should be valid if the observing time (e.g., one year) is long
compared to the orbital period of the LMXB (e.g., $6.8\times
10^4\un{s}\approx 19\un{hr}$ for
Sco X-1\cite{Steeghs:2001rx,Galloway2014}).  Since the orbital period
is not commensurate with any of the relevant periods of variation such
as the sidereal or solar day [the former being relevant for
$(\Gamma^{\text{ave}}_{\alpha})^2$ and the latter for the noise
spectra], it is reasonable to assume that
$\frac{2\pi}{\Porb}(\Tbar_{\alpha}-\Tasc)$ samples all phases roughly
equally, and therefore
\begin{subequations}
  \begin{gather}
    \begin{split}
    &\left\langle
      F_\alpha
      \cos
      \left[
        \frac{2\pi}{\Porb}
        (\Tbar_{\alpha}-\Tasc)
      \right]
    \right\rangle_{\!\alpha}
    \\
    &=
    \left\langle
      F_\alpha
      \sin
      \left[
        \frac{2\pi}{\Porb}
        (\Tbar_{\alpha}-\Tasc)
      \right]
    \right\rangle_{\!\alpha}
    = 0
    \end{split}
    \\
    \left\langle
      F_\alpha
      \cos
      \left[
        \frac{2\pi}{\Porb}
        (\Tbar_{\alpha}-\Tasc)
      \right]
      \sin
      \left[
        \frac{2\pi}{\Porb}
        (\Tbar_{\alpha}-\Tasc)
      \right]
    \right\rangle_{\!\alpha}
    = 0
    \\
    \begin{split}
    &\left\langle
      F_\alpha
      \cos^2
      \left[
        \frac{2\pi}{\Porb}
        (\Tbar_{\alpha}-\Tasc)
      \right]
    \right\rangle_{\!\alpha}
    \\
    &=
    \left\langle
      F_\alpha
      \sin^2
      \left[
        \frac{2\pi}{\Porb}
        (\Tbar_{\alpha}-\Tasc)
      \right]
    \right\rangle_{\!\alpha}
    =
    \frac{1}{2}
    \left\langle
      F_\alpha
    \right\rangle_{\!\alpha}
    \end{split}
  \end{gather}
\end{subequations}
where $F_\alpha$ is any expression not involving $\Tbar_\alpha$.

We then have metric components, from \eqref{e:crosscorrmetric}, of
\begin{subequations}
  \begin{align}
    g_{f_0f_0}
    =&\, 2\pi^2
    \left\langle
      (\Tdiff_\alpha-\ddiff_\alpha)^2
    \right\rangle_{\!\alpha}
    + 4\pi^2 a_p^2
    \left\langle
      \sin^2\frac{\pi \Tdiff_\alpha}{\Porb}
    \right\rangle_{\!\alpha}
    \\
    g_{f_0a_p} =&\, 4\pi^2 f_0 a_p
    \left\langle
      \sin^2\frac{\pi \Tdiff_\alpha}{\Porb}
    \right\rangle_{\!\alpha}
    \\
    g_{f_0\Porb} =&\, -4\frac{\pi^3 f_0 a_p^2}{\Porb^2}
    \left\langle
      \Tdiff_\alpha\sin\frac{\pi \Tdiff_\alpha}{\Porb}\cos\frac{\pi \Tdiff_\alpha}{\Porb}
    \right\rangle_{\!\alpha}
    \\
    g_{a_pa_p} =&\, 4\pi^2 f_0^2
    \left\langle
      \sin^2\frac{\pi \Tdiff_\alpha}{\Porb}
    \right\rangle_{\!\alpha}
    \\
    g_{f_0\Tasc} =&\, g_{a_p\Tasc} = 0
    \\
    g_{a_p\Porb} =&\, -\frac{4\pi^3 f_0^2 a_p}{\Porb^2}
    \left\langle
      \Tdiff_\alpha\sin\frac{\pi \Tdiff_\alpha}{\Porb}
      \cos\frac{\pi \Tdiff_\alpha}{\Porb}
    \right\rangle_{\!\alpha}
    \\
    g_{\Tasc\Tasc} =&\, \frac{16\pi^4 f_0^2 a_p^2}{\Porb^2}
    \left\langle
      \sin^2\frac{\pi \Tdiff_\alpha}{\Porb}
    \right\rangle_{\!\alpha}
  \end{align}
  \begin{align}
    g_{\Tasc\Porb} =&\, -\frac{16\pi^4 f_0^2 a_p^2}{\Porb^2}
    \left(
      \frac{
        \left\langle
          \Tbar_{\alpha}
        \right\rangle_{\!\alpha}
        - \Tasc
      }
      {\Porb}
    \right)
    \left\langle
      \sin^2\frac{\pi \Tdiff_\alpha}{\Porb}
    \right\rangle_{\!\alpha}
    \\
    \begin{split}
    g_{\Porb\Porb} =&\,
    \frac{16\pi^4 f_0^2 a_p^2}{\Porb^4}
    \left\langle
      (\Tbar_{\alpha}-\Tasc)^2
    \right\rangle_{\!\alpha}
    \left\langle
      \sin^2\frac{\pi \Tdiff_\alpha}{\Porb}
    \right\rangle_{\!\alpha}
    \\
    &+ \frac{4\pi^4 f_0^2 a_p^2}{\Porb^4}
    \left\langle
      \Tdiff_\alpha^2\cos^2\frac{\pi \Tdiff_\alpha}{\Porb}
    \right\rangle_{\!\alpha}
    \end{split}
  \end{align}
\end{subequations}
The metric is not diagonal, but we can neglect the off-diagonal
elements if
\begin{equation}
  (g_{ij})^2 \ll g_{ii}\,g_{jj}
  \ .
\end{equation}
One can show that $(g_{f_0a_p})^2\ll g_{f_0f_0}\,g_{a_pa_p}$ and
$(g_{f_0\Porb})^2\ll g_{f_0f_0}\,g_{\Porb\Porb}$ as long as
\begin{equation}
  \left\langle (\Tdiff_\alpha-\ddiff_\alpha)^2\right\rangle_{\!\alpha}
  \gg a_p^2
\end{equation}
which should be the case; for Sco X-1,
$a_p=\LSCApSec\un{s}$\cite{Abbott:2006vg,Steeghs:2001rx}.
Note also that, as long as we
include cross correlations between non simultaneous SFTs,
$\left\langle (\Tdiff_\alpha-\ddiff_\alpha)^2\right\rangle_{\!\alpha}\approx
\left\langle (\Tdiff_\alpha)^2\right\rangle_{\!\alpha}$ because the
detectors are moving much slower than the speed of light.

We will also have $(g_{a_p\Porb})^2 \ll g_{a_pa_p}\,g_{\Porb\Porb}$ as
long as the square of the typical time lag $\Tdiff_\alpha$ is much less
than
$\left\langle(\Tbar_{\alpha}-\Tasc)^2\right\rangle_{\!\alpha}$,
which will be the case if the maximum allowed time lag is much less
than the length of the run.  We can see this by considering the
$\left\langle(\Tbar_{\alpha}-\Tasc)^2\right\rangle_{\!\alpha}$; if
we define
\begin{equation}
  \mu_T = \langle\Tbar_{\alpha}\rangle_{\alpha}
\end{equation}
then
\begin{equation}
  \sigma_T^2 = \langle(\Tbar_{\alpha}-\mu_T)^2\rangle_{\alpha}
\end{equation}
should be on the order of the square of the duration of the run.  In
particular, for a run of duration $\Trun$ during which the
sensitivity of the search remains roughly constant,
\begin{equation}
  \sigma_T^2 \approx \frac{1}{\Trun}\int_{-\Trun/2}^{\Trun/2} t^2 d\tdet
  = \frac{\Trun^2}{12}
  \ .
\end{equation}
But
\begin{equation}
  \left\langle(\Tbar_{\alpha}-\Tasc)^2\right\rangle_{\!\alpha}
  = \sigma_T^2 + (\mu_T-\Tasc)^2 \ge \sigma_T^2
\end{equation}

This leaves only the ratio
\begin{equation}
  \frac{(g_{\Tasc\Porb})^2}{g_{\Tasc\Tasc}g_{\Porb\Porb}}
  \approx
  \frac{
    \left(
      \left\langle
        \Tbar_{\alpha}
      \right\rangle_\alpha
      - \Tasc
    \right)^2
  }{
    \left\langle
      (\Tbar_{\alpha}-\Tasc)^2
    \right\rangle_{\!\alpha}
  }
  =
  \frac{
    (\mu_T-\Tasc)^2
  }{
    \sigma_T^2 + (\mu_T-\Tasc)^2
  }
\end{equation}
Whether or not this can be neglected seems to come down, then, to
whether the reference time $\Tasc$ falls during the run.  If it
falls outside the run, $(\mu_T-\Tasc)^2\gtrsim\sigma_T^2$ and the
off-diagonal metric element $g_{\Tasc\Porb}$ cannot be ignored.
However, it is always possible to replace one reference time
$\Tasc$ with another $\Tasc'=\Tasc + n\Porb$ separated by an
integer number $n$ of cycles, and thus it is always possible to
arrange for $(\mu_T-\Tasc')^2\le\Porb^2\ll\sigma_T^2$ and thus
obtain an approximately diagonal metric.  This comes at a cost,
though, since there will be a contribution to the uncertainty in the
new reference time due to the uncertainty in the orbital period.  If
the uncertainties in the orbital period and the original reference
time are independent, the uncertainty in the new reference time will
be given by
\begin{equation}
  \begin{split}
  \label{e:dTasceff}
  (\Delta\Tasc')^2
  &= (\Delta\Tasc)^2 + n^2(\Delta\Porb)^2
  \\
  &= (\Delta\Tasc)^2
  + \frac{(\Tasc'-\Tasc)^2}{\Porb^2}(\Delta\Porb)^2
\end{split}
\end{equation}
This will become the dominant error if
\begin{equation}
  \abs{\Tasc'-\Tasc} > \frac{\Delta\Tasc}{\Delta\Porb}\Porb
  \ .
\end{equation}
For Sco X-1, using the parameter uncertainties from
\cite{Galloway2014} (see \sref{s:ScoX1}), this is about
\begin{equation}
  \frac{\GalldTascGPS}{\GalldPorbSec} \times \GallPorbSec\un{s}
  \approx 5\un{yr}
\end{equation}
Since the $\Tasc$ quoted in \cite{Galloway2014} (chosen to minimize
their $\Delta\Tasc$) corresponds to June 2008, this will be the
case for any GW observations using Advanced LIGO and/or Advanced Virgo
data, unless additional Sco X-1 ephemeris updates are made.

\begin{figure*}[tbp]
  \begin{center}
    \includegraphics[width=0.45\textwidth]{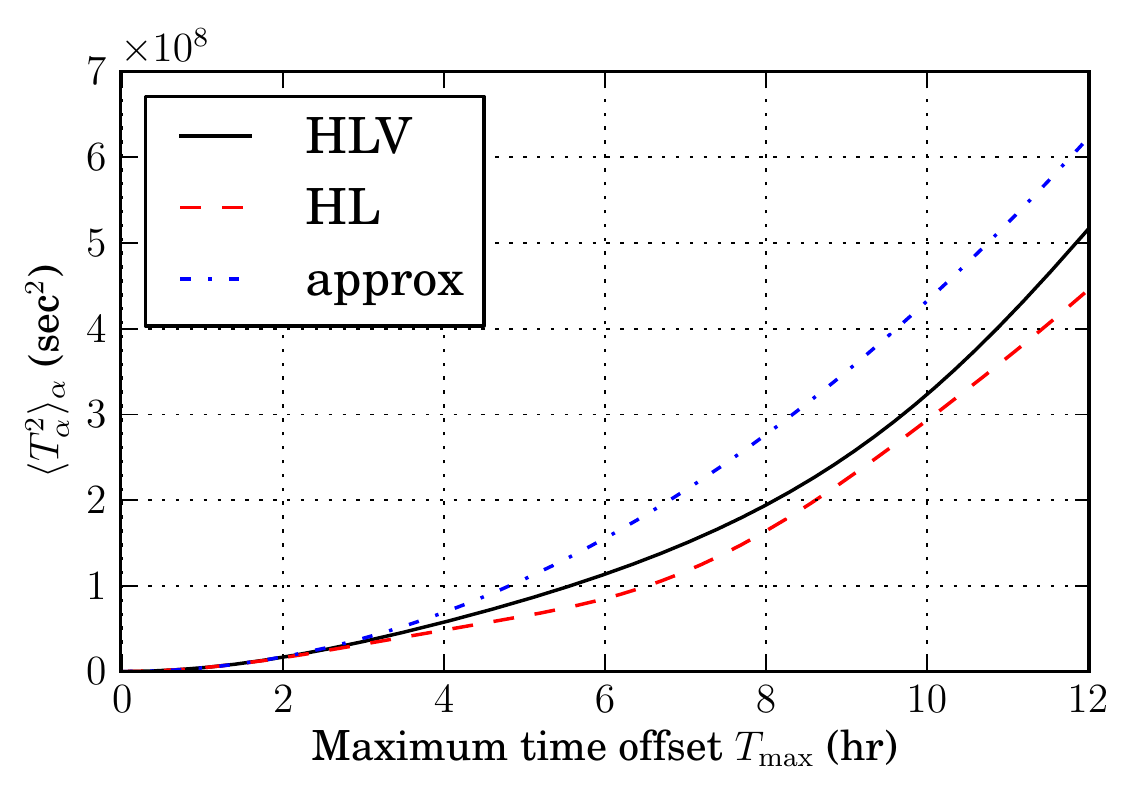}
    \includegraphics[width=0.45\textwidth]{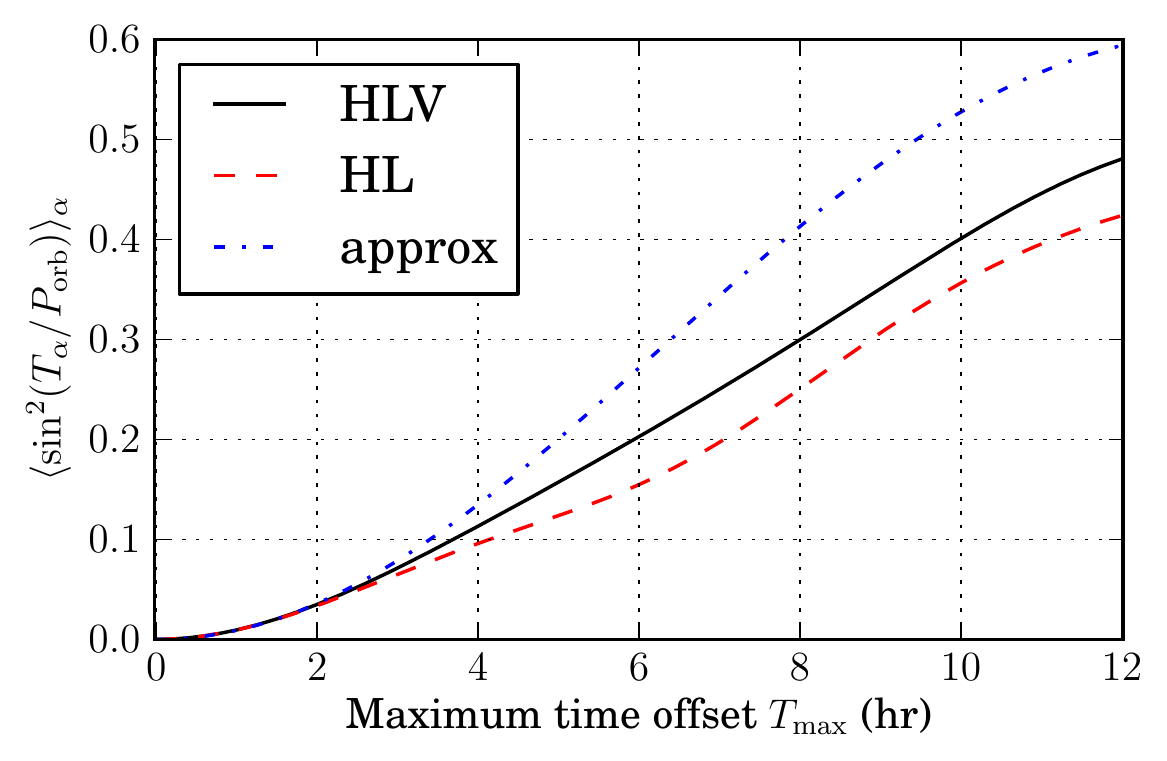}
  \end{center}
  \caption{Plot of weighted averages $\left\langle \Tdiff_\alpha^2
    \right\rangle_{\!\alpha}$ and $\left\langle \sin^2\frac{\pi
        \Tdiff_\alpha}{\Porb} \right\rangle_{\!\alpha}$ appearing in the
    metric components \eqref{e:gdiag} as a function of maximum allowed
    lag time $\Tmax$.  The dotted lines show the approximate values
    \eqref{e:Tsqavg} and \eqref{e:sinsqavg} neglecting the variation
    of the weighting factor.  The solid line (labeled HLV) shows the
    value for a search using detectors at the LIGO Hanford, LIGO
    Livingston, and Virgo sites, assuming a source at the sky position
    of Sco X-1, and that all detectors have the same sensitivity at
    the relevant frequency, and all sidereal times are evenly sampled.
    The dashed line (HL) shows the same thing for a search using only
    the LIGO detectors at Hanford and Livingston.  The actual weighted
    averages (and therefore the number of templates needed to cover
    the parameter space) are less than the approximate ones, because
    the geometrical factor $(\Gamma^{\text{ave}}_{\alpha})^2$ weights
    smaller lag times more.}
  \label{f:Tsq_sinsq}
\end{figure*}

Subject to the aforementioned approximations, the metric can be
treated as diagonal with non-negligible elements
\begin{subequations}
  \label{e:gdiag}
  \begin{align}
    \label{e:gdiagff}
    g_{f_0f_0}
    &\approx 2\pi^2
    \left\langle
      \Tdiff_\alpha^2
    \right\rangle_{\!\alpha}
    \\
    g_{a_pa_p} &= 4\pi^2 f_0^2
    \left\langle
      \sin^2\frac{\pi \Tdiff_\alpha}{\Porb}
    \right\rangle_{\!\alpha}
    \\
    g_{\Tasc\Tasc} &= \frac{16\pi^4 f_0^2 a_p^2}{\Porb^2}
    \left\langle
      \sin^2\frac{\pi \Tdiff_\alpha}{\Porb}
    \right\rangle_{\!\alpha}
    \\
    \label{e:gPP}
    g_{\Porb\Porb} &\approx
    \frac{16\pi^4 f_0^2 a_p^2}{\Porb^4}
    \ \sigma_T^2
    \left\langle
      \sin^2\frac{\pi \Tdiff_\alpha}{\Porb}
    \right\rangle_{\!\alpha}
  \end{align}
\end{subequations}
The quantities $\left\langle \Tdiff_\alpha^2 \right\rangle_{\!\alpha}$ and
$\left\langle\sin^2\frac{\pi
    \Tdiff_\alpha}{\Porb}\right\rangle_{\!\alpha}$ which appear in the
parameter space metric are constructed by a weighted average over SFT
pairs.  If we consider a search which includes all pairs up to a
maximum time lag of $\Tmax$, the parameter space resolution, and
therefore the required number of templates, will depend on $\Tmax$.
We can get a rough estimate on this dependence by assuming that
we can write
\begin{equation}
  \langle f(\Tdiff_\alpha)\rangle_\alpha
  \sim \frac{1}{2\Tmax} \int_{-\Tmax}^{\Tmax} f(\tdet)\,d\tdet
\end{equation}
which assumes $\Tobs\gg\Tmax\gg\Tsft$ so that we can replace the sum over
specific lags with an integral, and neglects the variation of
$(\widehat{\Gamma}^{\text{ave}}_{\alpha})^2$ from pair to pair.
Subject to this approximation, we have
\begin{equation}
  \label{e:Tsqavg}
  \left\langle \Tdiff_\alpha^2\right\rangle_\alpha
  \sim \frac{1}{2\Tmax} \int_{-\Tmax}^{\Tmax} t^2\,d\tdet
  = \frac{\Tmax^2}{3}
\end{equation}
and\footnote{Note that for $\Tmax\ll \Porb$ (coherent integration
  times small compared to the binary orbital period), the factor
  $\left\langle\sin^2\frac{\pi \Tdiff_\alpha}{\Porb}\right\rangle_\alpha$
  tends to $\frac{\pi^2\Tmax^2}{3\Porb^2}$ (so the number of templates in
  each direction grows like the coherent integration time), while for
  $\Tmax\gg \Porb$, coherent integration times long compared to the
  binary orbital period, it tends to a constant $\frac{1}{2}$, so the
  growth in number of templates in the $a_p$ and $\Tasc$ directions
  saturates.  This is analogous to an effect described in
  \cite{Prix:2006wm}.}
\begin{equation}
  \begin{split}
  \label{e:sinsqavg}
  \left\langle\sin^2\frac{\pi \Tdiff_\alpha}{\Porb}\right\rangle_\alpha
  &\sim
  \frac{1}{2\Tmax} \int_{-\Tmax}^{\Tmax} \sin^2\frac{\pi t}{\Porb}\,d\tdet
      \\
  &= \frac{1}{2}
  \left(
    1 - \sinc \frac{2\Tmax}{\Porb}
  \right)
  \end{split}
\end{equation}
where once again $\sinc x=\frac{\sin\pi x}{\pi x}$.  Note that this is
only a rough approximation, since increasing the time offset
$\Tdiff_\alpha$ between a pair of SFTs from the same instrument (or from
well-aligned instruments like the LIGO detectors in Hanford and
Livingston) will tend to decrease the expected cross correlation as
the detectors are rotated out of alignment with each other.  We
confirm this by comparing the approximate expressions to more accurate
values calculated using the geometry of the LIGO and Virgo detectors
and the sky position of Scorpius X-1, in \fref{f:Tsq_sinsq}.

Note that some care needs to be taken when comparing our metric
expressions to those in \cite{Leaci:2015bka}.  For example, combining
\eqref{e:gdiagff} with \eqref{e:Tsqavg} gives us
$g_{f_0f_0}\approx2\pi^2\frac{\Tmax^2}{3}$, which seems at odds with
the analogous expression in e.g., Eq.~(61) of \cite{Leaci:2015bka},
where the corresponding metric element is
$\pi^2\frac{(\Delta{T})^2}{3}$.  The difference is that the
semicoherent search in \cite{Leaci:2015bka} is defined by combining
distinct coherent segments of length $\Delta{T}$, which makes the
mean squared difference
\begin{multline}
  \frac{1}{(\Delta{T})^2}\int_0^{\Delta{T}} \int_0^{\Delta{T}} (t-t')^2\,dt\,dt'
  \\
  = \frac{1}{(\Delta{T})^2}\int_{-\Delta{T}}^{\Delta{T}}
  \int_{\abs{\Delta{t}}/2}^{\Delta{T}-\abs{\Delta{t}}/2}
  (\Delta{t})^2\,d\tbar\,d\Delta{t}
  \\
  = \frac{1}{(\Delta{T})^2}\int_{-\Delta{T}}^{\Delta{T}}
  (\Delta{t})^2(\Delta{T}-\abs{\Delta{t}})\,d\Delta{t}
  \\
  = \left(
    \frac{2}{3}-\frac{2}{4}
  \right)
  (\Delta{T})^2
  = \frac{1}{6}(\Delta{T})^2
\end{multline}
whereas our maximum lag rule $\abs{t-t'}<\Tmax$ gives a mean square
time difference
\begin{multline}
  \frac{
    \int_{0}^{\Trun}\int_{\max(t'-\Tmax,0)}^{\min(t'+\Tmax,\Trun)}(t-t')^2 \,dt\,dt'
  }{
    \int_{0}^{\Trun}\int_{\max(t'-\Tmax,0)}^{\min(t'+\Tmax,\Trun)} dt\,dt'
  }
  \\
  =
  \frac{
    \int_{-\Delta{T}}^{\Delta{T}}
    \int_{\abs{\Delta{t}}/2}^{\Trun-\abs{\Delta{t}}/2}
    (\Delta{t})^2\,d\tbar\,d\Delta{t}
  }{
    \int_{-\Delta{T}}^{\Delta{T}}
    \int_{\abs{\Delta{t}}/2}^{\Trun-\abs{\Delta{t}}/2}
    d\tbar\,d\Delta{t}
  }
  \\
  =
  \frac{(2/3)\Tobs\Tmax^3 - (2/4)\Tmax^4}
  {2\Tobs\Tmax - \Tmax^2}
      \approx \frac{1}{3}\Tmax^2
\end{multline}
where the assumption $\Tmax\ll\Tobs$ gives us the result \eqref{e:Tsqavg}.

\section{Implications of deviation from signal model}

\label{s:deviations}

So far, we have assumed that the underlying signal model contained in
\eqref{e:signalsft2}, along with the phase evolution
\eqref{e:LMXBphase} is correct, although some of the parameters may
be unknown.  We consider two effects which violate this
assumption, and their potential impacts on the expected SNR
\eqref{e:evrhoalpha}.  These are (1) spin wandering, in which the
frequency is not a constant $f_0$ but varies slowly and unpredictably
with time and (2) the impact of higher terms in the Taylor expansion of
$\Phi(\tbin(\tdet))$ about $\tdet=\tmid_\II$, which are neglected in the linear
phase model \eqref{e:phaselinear}.  The former effect will place a
potential limit on the coherence time $\Tmax$ by providing an
intrinsic limit to the frequency resolution, whereas the latter will
constrain our choice of SFT length $\Tsft$ in order that neglected
phase acceleration effects not cause too much loss of SNR.

\subsection{Spin wandering}
\label{s:spinwander}

We have assumed so far that the LMXB is in approximate equilibrium,
where the spin-up torque due to accretion is balanced by the spin-down
due to gravitational waves.  Even if this is true on average, the
balance will not be perfect, and the spin frequency will ``wander''.
This means that rather than a constant frequency $f_0$ appearing in
\eqref{e:LMXBphase}, there will be a time-varying frequency $f(\tbin)$,
where $\tbin = t - \frac{\rdet\cdot\khat}{c} +
\frac{\rorb\cdot\khat}{c}$ is the time measured in the neutron star's
rest frame.  Thus the phase difference between SFTs $\II$ and $\JJ$ will be,
rather than just $\Delta\Phi_{\II\JJ}=2\pi f_0[\tbin_\II-\tbin_\JJ]$,
\begin{equation}
  \Delta\Phi_{\II\JJ}^{\text{true}}
  = \Phi_\II-\Phi_\JJ = 2\pi\int_{\tbin_\II}^{\tbin_\JJ} f(\tbin)\,d\tbin
\end{equation}
We can consider the loss of SNR due to the existence of spin
wandering, compared to what we would expect if the frequency truly were
constant.  Qualitatively, there are two reasons for loss of SNR:
first, on short time scales, the change in frequency could disrupt the
coherence between the two SFTs in a pair being cross correlated;
second, on longer time scales, the spin could wander enough that the
SNR is distributed over different frequency templates.

To quantify the loss of SNR we follow a calculation analogous to that
in \sref{e:paramspace}, e.g., in \eqref{e:evrhomismatched} and
\eqref{e:evrhomismatched2}, to obtain
\begin{equation}
  \label{e:spinwanderloss}
  \frac{\ev{\rho}^{\text{ideal}}-\ev{\rho}}{\ev{\rho}^{\text{ideal}}}
  \approx \frac{1}{2}
  \left\langle
    \left(\Delta\Phi_{\alpha}^{\text{true}}-\Delta\Phi_{\alpha})\right)^2
  \right\rangle_{\alpha}
\end{equation}
where $\langle\cdot\rangle_{\alpha}$ is a weighted average over SFT
pairs with weighting factor
$\left(\widehat{\Gamma}^{\text{ave}}_{\alpha}\right)^2$ as before.
To estimate $\left\langle
  \left(\Delta\Phi_{\alpha}^{\text{true}}-\Delta\Phi_{\alpha})\right)^2
\right\rangle_{\alpha}$ we assume that the wandering is slow enough
that we can expand $f(\tbin)$ in a Taylor series about
$\tbinbar_{\II\JJ}=(\tbin_\II+\tbin_\JJ)/2$:
\begin{multline}
  f(\tbin) \approx f(\tbinbar_{\II\JJ})
  + \dot{f}(\tbinbar_{\II\JJ})(\tbin-\tbinbar_{\II\JJ})
  \\
  \min(\tbin_\II,\tbin_\JJ) \le \tbin \le \max(\tbin_\II,\tbin_\JJ)
\end{multline}
Then
\begin{multline}
  \Delta\Phi_{\II\JJ}^{\text{true}} - \Delta\Phi_{\II\JJ}
  = 2\pi\int_{\tbin_\II}^{\tbin_\JJ} [f(\tbin)-f_0]\,d\tbin
  \\
  \approx 2\pi \left(
    [f(\tbinbar_{\II\JJ}) - f_0] \tbindiff_{\II\JJ} + \dot{f}(\tbinbar_{\II\JJ})
    \frac{(\tbindiff_{\II\JJ})^2}{2}
  \right)
  \ ,
\end{multline}
where $\tbindiff_{\II\JJ}=\tbin_\II-\tbin_\JJ$
Subject to reasonable assumptions about the randomness of the spin
wandering, \eqref{e:spinwanderloss} can be written in the form
\begin{equation}
\label{e:spinwanderlossapprox}
  \begin{split}
    \frac{\ev{\rho}^{\text{ideal}}-\ev{\rho}}{\ev{\rho}^{\text{ideal}}}
    \approx&
    \ 2\pi^2
    \left\langle
      [f(\tbinbar_{\alpha}) - f_0]^2
    \right\rangle_{\alpha}
    \left\langle
      (\tbindiff_{\alpha})^2
    \right\rangle_{\alpha}
    \\
    &+
    \frac{\pi^2}{2}
    \left\langle
      [\dot{f}(\tbinbar_{\alpha})]^2
    \right\rangle_{\alpha}
    \left\langle
      (\tbindiff_{\alpha})^4
    \right\rangle_{\alpha}
    \\
    \approx&
    \ 2\pi^2
    \left\langle
      [f(\Tbar_{\alpha}) - f_0]^2
    \right\rangle_{\alpha}
    \left\langle
      (\Tdiff_{\alpha})^2
    \right\rangle_{\alpha}
    \\
    &+
    \frac{\pi^2}{2}
    \left\langle
      [\dot{f}(\Tbar_{\alpha})]^2
    \right\rangle_{\alpha}
    \left\langle
      (\Tdiff_{\alpha})^4
    \right\rangle_{\alpha}
  \end{split}
\end{equation}
where in the last expression we have used the fact that since $a_p$ and
$\ddiff_{\alpha}$ are small, $\abs{\tbin_\II-\tmid_\II}\ll\Tmax$.  The two
terms in \eqref{e:spinwanderlossapprox} quantify the effects we
predicted at the beginning of the section.  The second term describes
a loss of SNR due to the neutron star spin not being constant during
the time spanned by a SFT pair, while the first term indicates a loss
due to the mismatch between contributing frequencies and the frequency
of a single template.  [In fact, the first term is just
$g_{f_0f_0}\left\langle [f(\Tbar_{\alpha}) - f_0]^2
\right\rangle_{\alpha}$.[  Note that we are free to choose the $f_0$
which maximizes the SNR for a given instantiation of spin wandering,
which will be $f_0=\left\langle
  f(\Tbar_{\alpha})\right\rangle_{\alpha}$, so
\begin{equation}
  \left\langle
    [f(\Tbar_{\alpha}) - f_0]^2
  \right\rangle_{\alpha}
  =
  \left\langle
    [f(\Tbar_{\alpha}) - \langle f(\Tbar_{\alpha})\rangle_{\alpha}]^2
  \right\rangle_{\alpha}
\end{equation}
is the weighted variance of $f(\tdet)$ over the observing time.

To get a quantitative estimate of the effects of spin wandering,
consider a model where the neutron star spins up or down linearly with
typical amplitude $\fdotdrift$, changing on a time scale $\Tdrift$
where $\Tmax\ll\Tdrift\ll\Tobs$.  For simplicity, also neglect the
impact of the weighting factor
$\left(\widehat{\Gamma}^{\text{ave}}_{\alpha}\right)^2$,
so that $\langle\Tdiff_\alpha^2\rangle\approx \frac{\Tmax^2}{3}$ and
$\langle\Tdiff_\alpha^4\rangle\approx \frac{\Tmax^4}{5}$.  Then
\begin{equation}
  \left\langle
    [\dot{f}(\Tbar_{\alpha})]^2
  \right\rangle_{\alpha}
  \lesssim \fdotdrift^2
\end{equation}
and
\begin{equation}
  \begin{split}
  \left\langle
    [f(\Tbar_{\alpha}) - \langle f(\Tbar_{\alpha})\rangle_{\alpha}]^2
  \right\rangle_{\alpha}
  &\lesssim
  \left\langle
    \abs{\frac{\Tbar_{\alpha}-\Tmid}{\Tdrift}}
    \left(\Tdrift\fdotdrift\right)^2
  \right\rangle_{\alpha}
  \\
  &\approx
  \frac{\Tobs\Tdrift}{4}\fdotdrift^2
  \end{split}
\end{equation}
Combining these results, we have
\begin{equation}
  \frac{\ev{\rho}^{\text{ideal}}-\ev{\rho}}{\ev{\rho}^{\text{ideal}}}
  \lesssim
  \frac{\pi^2}{6}
  \Tobs\Tdrift\fdotdrift^2
  \Tmax^2
  +
  \frac{\pi^2}{10}
  \fdotdrift^2
  \Tmax^4
\end{equation}
So, in order to avoid a fractional loss in SNR of more than $\mism$,
one would need to limit the lag time to
\begin{equation}
  \Tmax \le
  \min
  \left(
    \frac{\sqrt{6\mism}}{\pi} \left(\fdotdrift\sqrt{\Tobs\Tdrift}\right)^{-1}
    ,
    \sqrt{\frac{\sqrt{10\mism}}{\pi}}\fdotdrift^{-1/2}
  \right)
\end{equation}
For example, if $\fdotdrift=10^{-12}\un{Hz/s}$, $\Tdrift=10^6\un{s}$,
$\Tobs=1\un{Yr}$, and $\mism=0.1$, the first limit is about
$44,\!000\un{s}$ and the second is $320,\!000\un{s}$.  So in that case
spin wandering would become an issue if $\Tmax\gtrsim 12\un{hr}$.

Note that this is somewhat less than the estimate $\Delta{T}\lesssim
3\un{day}$ given in \cite{Leaci:2015bka}.  The source of this apparent
discrepancy is a combination of the distinction between the coherent
segment length $\Delta{T}$ and the maximum lag time $\Tmax$, described
in \sref{s:metric-lmxb-long}, and the rough nature of some estimates
used in \cite{Leaci:2015bka}.  That work compares the change in
frequency $\fdotdrift\sqrt{\Tobs\Tdrift/2}$ to the frequency
resolution, which they give as $\sim 1/\Delta{T}$.  This is
effectively an order of magnitude estimate, since it effectively
assumes $\mism=1$, and also leaves out the numerical factor in
$1/\sqrt{g_{f_0f_0}}=\sqrt{3}/(\pi\Delta{T})$.  On the other hand,
their frequency drift is the expected drift from the middle of the run
to the end; averaging the drift over the run gives an effective change
of $(\fdotdrift\sqrt{\Tobs\Tdrift})/2$.  Including these three effects
to do a calculation analogous to the one here would give a factor of
$\pi\sqrt{5/3}\approx 4$ reduction on the estimated tolerable segment
length to
$\Delta{T}\lesssim 2\sqrt{3\mu}/\pi
\left(\fdotdrift\sqrt{\Tobs\Tdrift}\right)^{-1}\approx
62,\!000\un{s}\approx 17\un{hr}$.  Of course, the assumptions of
$\fdotdrift$ and $\Tdrift$ given above are uncertain and somewhat
arbitrary, so our 12-hour number should also not be viewed as an exact
constraint on the method.

\subsection{SFT length}
\label{s:sftlength}

Most searches for continuous gravitational waves have used short
Fourier transforms with a duration $\Tsft$ of
$30\un{min}=1800\un{s}$.  The limiting factor which sets a maximum on
the reasonable $\Tsft$ is the accuracy of the linear phase
approximation \eqref{e:phaselinear}.

If we consider higher order terms in the phase expansion, we have
\begin{multline}
  \Phi(\tbin(\tdet))
  \approx \Phi_\II + 2\pi f_\II (t-\tmid_\II)
  + \frac{1}{2}\ddot{\Phi}(\tmid_\II)(t-\tmid_\II)^2
  \\
  + \frac{1}{3!}\dddot{\Phi}(\tmid_\II)(t-\tmid_\II)^3
  + \frac{1}{4!}\ddddot{\Phi}(\tmid_\II)(t-\tmid_\II)^4
  + \ldots \ .
\end{multline}
The effect of these corrections is to modify \eqref{e:signalsft2}
to
\begin{widetext}
\begin{multline}
  \cft{h}_{\II\kI} \approx h_0 (-1)^\kI
  e^{\cmplxi\Phi_\II} \frac{F^{\II}_+ \Ap - \cmplxi F^{\II}_\times \Ac}{2}
  \\
  \times
  \int_{\tmid_\II-\Tsft/2}^{\tmid_\II+\Tsft/2}
  e^{-\cmplxi 2\pi(f_\kI-f_\II)(t-\tmid_\II)}
  \exp\left(
    \cmplxi
    \left[
      \frac{\ddot{\Phi}(\tmid_\II)}{2}(t-\tmid_\II)^2
      + \frac{\dddot{\Phi}(\tmid_\II)}{3!}(t-\tmid_\II)^3
      + \frac{\ddddot{\Phi}(\tmid_\II)}{4!}(t-\tmid_\II)^4
    \right]
  \right)
  d\tdet
  \\
  \approx
  h_0 (-1)^\kI
  e^{\cmplxi\Phi_\II} \frac{F^{\II}_+ \Ap - \cmplxi F^{\II}_\times \Ac}{2}
  \Tsft
  \left[
    I_0(\kappa_{\II\kI})
    + i \frac{\ddot{\Phi}(\tmid_\II)}{2} I_2(\kappa_{\II\kI})\Tsft^2
    + i \frac{\dddot{\Phi}(\tmid_\II)}{3!} I_3(\kappa_{\II\kI})\Tsft^3
  \right.
  \\
  \left.
    +
    \left(
      i \frac{\ddddot{\Phi}(\tmid_\II)}{4!}
      - \frac{[\ddot{\Phi}(\tmid_\II)]^2}{8}
    \right)
    I_4(\kappa_{\II\kI})\Tsft^4
  \right]
\end{multline}
\end{widetext}
where
\begin{equation}
  \label{e:Indef}
  I_n(\kappa) \equiv \int_{-1/2}^{1/2} x^n e^{-\cmplxi 2\pi\kappa x}\,dx
  = \left(\frac{\cmplxi}{2\pi}\right)^n \frac{d^n}{d\kappa^n}\sinc(\kappa)
\end{equation}
Note that for even $n$, $I_n(\kappa)$ is real and even, while for odd
$n$, it is imaginary and odd.

We can then construct, as a replacement for \eqref{e:meancompts},
\begin{equation}
  \begin{split}
    \mu_\II &= \frac{1}{\Xi_{\II}}
    \sum_{\kI\in\mc{K}_{\II}}(-1)^\kI I_0(\kappa_{\II\kI})\cft{h}_{\II\kI}
    \\
    &\approx h_0
    e^{\cmplxi\Phi_\II} \frac{F^{\II}_+ \Ap - \cmplxi F^{\II}_\times \Ac}{2}
    \frac{Q_{\II}}{\Xi_{\II}}\sqrt{\frac{2\Tsft}{S_\II}}
  \end{split}
\end{equation}
where
\begin{equation}
  \begin{split}
    Q_{\II} =&\,
    \ \Xi_{\II}^2
    + i \frac{\ddot{\Phi}(\tmid_\II)}{2} \Sigma_{\II 02}\,\Tsft^2
    + i \frac{\dddot{\Phi}(\tmid_\II)}{3!} \Sigma_{\II 03}\,\Tsft^3
    \\
    &+
    \left(
      i \frac{\ddddot{\Phi}(\tmid_\II)}{4!}
      - \frac{[\ddot{\Phi}(\tmid_\II)]^2}{8}
    \right)
    \Sigma_{\II 04}\,\Tsft^4
  \end{split}
\end{equation}
and
\begin{equation}
  \label{e:S0ndef}
  \Sigma_{\II 0n} = \sum_{\kI\in\mc{K}_\II} I_0(\kappa_{\II\kI})I_n(\kappa_{\II\kI})
\end{equation}
The expectation value \eqref{e:evrho} of the statistic thus becomes,
including the correction for higher phase derivatives and finite SFT
length,
\begin{equation}
  \ev{\rho}
  \approx Nh_0^2\,2\sum_{\II\JJ\in\mc{P}}
  \widehat{\Gamma}^{\text{ave}}_{\II\JJ}
  \Real\left(Q_\II Q_\JJ^*\widehat{\Gamma}_{\II\JJ}\right)
\end{equation}
As in \sref{s:kappa} we assume that the sum over pairs evenly and
independently samples the fractional frequency offset $\kappabest_\II$
from each SFT, which means we can replace $Q_\II$ and $Q_\JJ$ with
\begin{equation}
  \begin{split}
  \langle Q_\II\rangle_{\kappa}
  =&\ \langle\Xi^2\rangle
  + i \frac{\ddot{\Phi}(\tmid_\II)}{2} \langle \Sigma_{02}\rangle\,\Tsft^2
  \\
  &+
  \left(
    i \frac{\ddddot{\Phi}(\tmid_\II)}{4!}
    - \frac{[\ddot{\Phi}(\tmid_\II)]^2}{8}
  \right)
  \langle \Sigma_{04}\rangle\,\Tsft^4
\end{split}
\end{equation}
where the fact that $I_3(\kappa)$ is odd in $\kappa$ means that the
average $\langle \Sigma_{03}\rangle$ vanishes.

Now,
\begin{multline}
    \Real\left(Q_\II Q_\JJ^*\widehat{\Gamma}_{\II\JJ}\right)
    =
    \Real\left(Q_\II Q_\JJ^*\right)\Real\widehat{\Gamma}_{\II\JJ}
    - \Imag\left(Q_\II Q_\JJ^*\right)\Imag\widehat{\Gamma}_{\II\JJ}
    \\
    \approx
    \Real\left(Q_\II Q_\JJ^*\right)
    \frac{5}{2}\frac{\Ap^2+\Ac^2}{2}
    \widehat{\Gamma}^{\text{ave}}_{\II\JJ}
    - \Imag\left(Q_\II Q_\JJ^*\right)
    \frac{5\Ap\Ac}{2}\Gamma^{\text{circ}}_{\II\JJ}
\end{multline}
We assume that the impact of the second piece is small\footnote{In
  particular, it is suppressed by averaging non-positive-definite
  antenna patterns, although the same combination is the source of
  systematic errors in parameter estimation.} and focus only on
$\Real\left(Q_\II Q_\JJ^*\right)$, which leads to a fractional loss of
SNR of
\begin{multline}
  1 - \frac{\ev{\rho}}{\ev{\rho}_{\text{ideal}}}
  = \frac{\langle\Xi^2\rangle^2-\langle\Real\left(Q_\II Q_\JJ^*\right)\rangle}
  {\langle\Xi^2\rangle^2}
  \\
  =
  \left(
    \frac{\langle \ddot{\Phi}_\II^2\rangle+\langle \ddot{\Phi}_\JJ^2\rangle}{8}
    \frac{\langle \Sigma_{04}\rangle}{\langle\Xi^2\rangle}
    - \frac{\langle \ddot{\Phi}_\II\ddot{\Phi}_\JJ\rangle}{4}
    \frac{\langle \Sigma_{02}\rangle^2}{\langle\Xi^2\rangle^2}
  \right) \Tsft^4
\end{multline}
Differentiating \eqref{e:LMXBphase} gives
\begin{equation}
  \ddot{\Phi}_\II = 2\pi f_0 \ddot{d}_\II - \frac{(2\pi)^3}{\Porb^2}
  f_0 a_p \sin
  \left[
    \frac{2\pi}{\Porb} (\tmid_\II-\Tasc)
  \right]
\end{equation}
We can neglect the first term,
since the acceleration due to the Earth's orbit is
$\mc{O}(10^{-11}\un{s}^{-1})$ and that due to the Earth's rotation is
$\mc{O}(10^{-10}\un{s}^{-1})$.  In comparison, for Sco X-1,
\begin{equation}
  a_p \left(\frac{2\pi}{\Porb}\right)^2 = \orbitAcc\un{s}^{-1}
\end{equation}
If we assume, as in the metric calculation, that the average over
pairs evenly samples the orbital phase, then
\begin{equation}
  \langle\ddot{\Phi}_\II^2\rangle+\langle \ddot{\Phi}_\JJ^2\rangle
  = \frac{(2\pi)^6f_0^2a_p^2}{\Porb^4}
\end{equation}
Using the identity
\begin{equation}
  \sin A\sin B = \frac{1}{2}\left[\cos(A-B)-\cos(A+B)\right]
\end{equation}
we can calculate
\begin{multline}
  \left\langle
    \sin
    \left[
      \frac{2\pi}{\Porb} (\tmid_\II-\Tasc)
    \right]
    \sin
    \left[
      \frac{2\pi}{\Porb} (\tmid_\JJ-\Tasc)
    \right]
  \right\rangle
  \\
  =
  \frac{1}{2}
  \left(
    \left\langle
      \cos\frac{2\pi\Tdiff_\alpha}{\Porb}
    \right\rangle_{\!\alpha}
    -
    \left\langle
      \cos\frac{4\pi(\Tbar_\alpha-\Tasc)}{\Porb}
    \right\rangle_{\!\alpha}
  \right)
\end{multline}
so the fractional loss in SNR is
\begin{multline}
  \label{e:shortsftmismatch}
  1 - \frac{\ev{\rho}}{\ev{\rho}_{\text{ideal}}}
  \\
  \approx
  \frac{8\pi^6f_0^2a_p^2}{\Porb^4}
  \left(
    \frac{\langle \Sigma_{04}\rangle}{\langle\Xi^2\rangle}
    - \frac{\langle \Sigma_{02}\rangle^2}{\langle\Xi^2\rangle^2}
    \left\langle
      \cos\frac{2\pi\Tdiff_\alpha}{\Porb}
    \right\rangle_{\!\alpha}
  \right) \Tsft^4
\end{multline}
The factors $\langle \Sigma_{04}\rangle$ and $\langle \Sigma_{02}\rangle$ can be
calculated by using \eqref{e:S0ndef} along with
\begin{equation}
  \label{e:I2}
  I_2(\kappa) =
    \frac{\sin\pi\kappa}{4\pi\kappa}
    + \frac{\cos\pi\kappa}{2(\pi\kappa)^2}
    - \frac{\sin\pi\kappa}{2(\pi\kappa)^3}
\end{equation}
and
\begin{equation}
  \label{e:I4}
  I_4(\kappa) =
    \frac{\sin\pi\kappa}{16\pi\kappa}
    + \frac{\cos\pi\kappa}{4(\pi\kappa)^2}
    - \frac{3\sin\pi\kappa}{4(\pi\kappa)^3}
    - \frac{3\cos\pi\kappa}{2(\pi\kappa)^4}
    + \frac{3\sin\pi\kappa}{2(\pi\kappa)^5}
\end{equation}
and averaging numerically over $\kappa$ given the number of frequency
bins included.  In \tref{tab:sftcontribs}, we show the two {\coeff}s
appearing in \eqref{e:shortsftmismatch}, for various choices of the
number $m$ of included frequency bins (see also
\tref{tab:bincontribs}).
\begin{table}[tbp]
  \caption{The {\coeff}s $\langle \Sigma_{04}\rangle/\langle\Xi^2\rangle$
    and $\langle \Sigma_{02}\rangle^2/\langle\Xi^2\rangle^2$ appearing
    in \eqref{e:shortsftmismatch}, for various choices of the number $m$
    of included frequency bins, where $\langle\Sigma_{0n}\rangle$ is the
    mean value of
    $\Sigma_{0n}(\kappa)=\sum_{s=-\hint{(m-1)/2}}^{\lint{(m-1)/2}} I_0(\kappa+s)I_n(\kappa+s)$,
    averaged over $-\frac{1}{2}\le\kappa\le\frac{1}{2}$, and
    $I_n(\kappa)$ is defined in \eqref{e:Indef} with
    $I_0(\kappa)=\sinc\kappa=\frac{\sin\pi\kappa}{\pi\kappa}$, and
    $I_2(\kappa)$ and $I_4(\kappa)$ are given by \eqref{e:I2} and
    \eqref{e:I4}.  Note that the value of
    $\langle\Xi^2\rangle\equiv\langle I_{00}\rangle$ is tabulated
    in \tref{tab:bincontribs}.
  }
  \label{tab:sftcontribs}
  \begin{center}
    \begin{tabular*}{\columnwidth}{@{\extracolsep{\fill}}lcccccc}
\hline
\hline
$m$ & 1 & 2 & 3 & 4 & 5 & 6 \\
\hline
$\langle \Sigma_{04}\rangle/\langle\Xi^2\rangle$ & 0.0107 & 0.0086 & 0.0099 & 0.0100 & 0.0106 & 0.0108 \\
$\langle \Sigma_{02}\rangle^2/\langle\Xi^2\rangle^2$ & 0.0056 & 0.0042 & 0.0052 & 0.0055 & 0.0059 & 0.0060 \\
\hline
\hline
\end{tabular*}
   \end{center}
\end{table}

Note that for the cross-correlation search, choosing shorter SFTs does
not directly impact the sensitivity.  For the same allowed lag time,
searches with different SFT lengths should have approximately the same
sensitivity.  We can see this by considering the SNR for a given
signal amplitude $h_0$, for example, from \eqref{e:evrhoXi}.
Since
\begin{equation}
  \widehat{\Gamma}^{\text{ave}}_{\II\JJ}
  = \Gamma_{\II\JJ}
  \frac{2\Tsft}{\sqrt{S_\II S_\JJ}}
\end{equation}
the quantity $(\widehat{\Gamma}^{\text{ave}}_{\II\JJ})^2$ inside the
sum is proportional to $(\Tsft)^2$.  However, for a fixed maximum time
lag $\Tmax$, the number of terms in the sum will be proportional to
$(\Tsft)^{-2}$ and the resulting expected SNR will be approximately
independent of $\Tsft$.  (For example, halving the SFT length will mean each
SFT pair contributed one-fourth as much to the sensitivity, but will
double the number of SFTs and thus quadruple the number of SFT pairs.)

\begin{figure}[tbp]
  \begin{center}
    \includegraphics[width=\columnwidth]{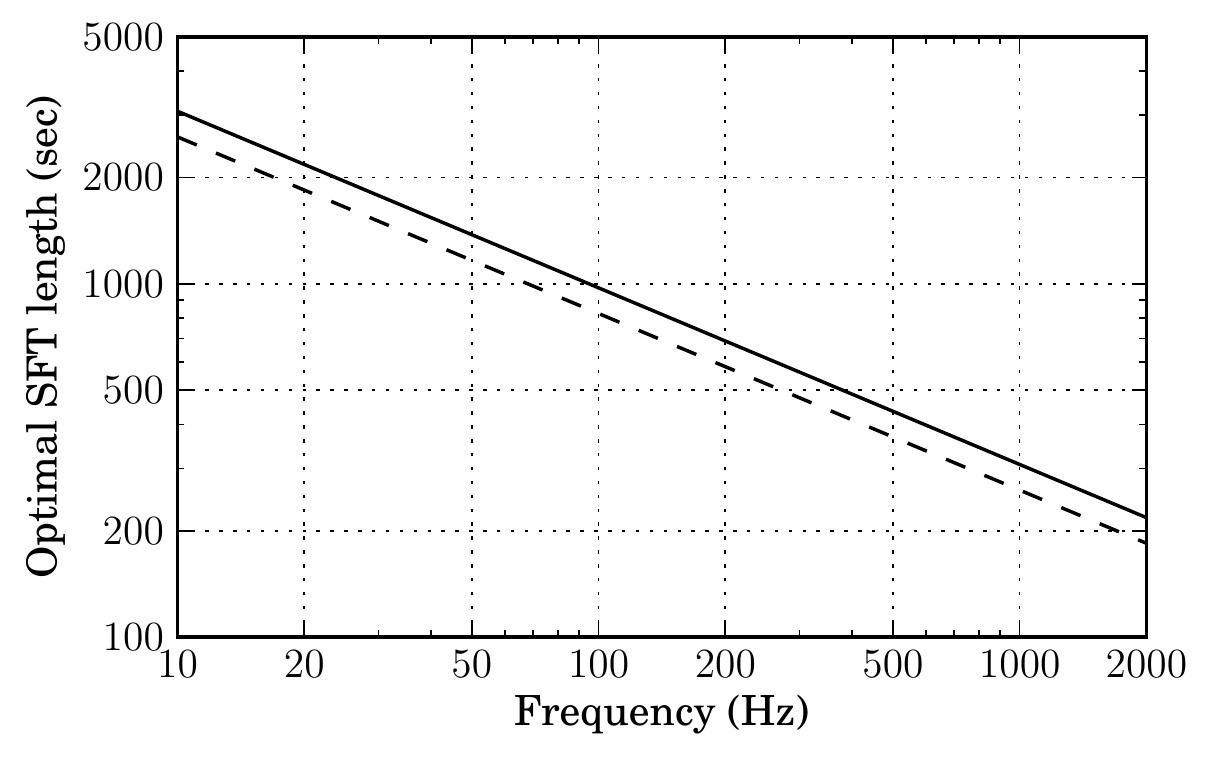}
  \end{center}
  \caption{The optimal SFT length $\Tsft$, defined in \eqref{e:optsft}
    and \eqref{e:Adef}, as a function of frequency, for a signal with
    the most likely orbital parameters for Sco X-1, as given in
    \tref{tab:ScoX1}, assuming that $d=3$, i.e., the density of points
    in parameter space grows as the third power of the coherence time
    $\Tmax$.  This is appropriate for a search over, e.g., frequency
    $f_0$, projected semimajor axis $a_p$, and time of ascension
    $\Tasc$ (when the uncertainty in the period $\Porb$ is small
    enough that a single value may be assumed), in the case where
    $\Tmax$ is small compared with $\Porb$.  The solid line represents
    a more optimistic scenario where the average cosine appearing in
    the second term of \eqref{e:Adef} is approximately unity, which
    should also be the case if $\Tmax\ll\Porb$.  The dashed line
    represents a worst-case scenario where the average is
    approximately zero.  The optimal SFT length maximizes the expected
    SNR in \eqref{e:evrhoopt} and represents a balance between two
    competing effects: if $\Tsft$ is too large, phase acceleration
    will lead to a loss in SNR compared to the ideal
    formula \eqref{e:evrhoalpha}; if $\Tsft$ is too small, the large
    number of SFT pairs in the computation will lead to a restriction
    on the possible $\Tmax$ achievable at fixed computing cost, and
    reduce the ideal SNR itself.}
  \label{f:sftopt}
\end{figure}

On the other hand, by increasing the number of SFT pairs, using a
shorter SFT length will mean increasing computing cost at the same
$\Tmax$.  If the computing budget is fixed, the sensitivity gained by
reducing the mismatch \eqref{e:shortsftmismatch} will be offset by the
loss of sensitivity, in the form of a lower
$\ev{\rho}_{\text{ideal}}$, resulting from a smaller $\Tmax$.
Following the reasoning in \sref{s:kappa}, if the computing cost
scales like the number of templates (which scales like $\Tmax^d$)
times the number of SFT pairs (which scales like
$\Tmax\Tobs\Tsft^{-2}$), then the overall sensitivity for a fixed
observing time $\Tobs$ scales like $\Tmax^{d+1}\Tsft^{-2}$, and
therefore the restriction at constant computing cost will be
$\Tmax\propto\Tsft^{\frac{2}{d+1}}$.  Since the sensitivity scales
with the square root of the number of SFT pairs, we have
$\ev{\rho}^{\text{ideal}}\propto\Tsft^{\frac{1}{d+1}}$ and
\begin{equation}
  \label{e:evrhoopt}
  \ev{\rho}\propto \Tsft^{\frac{1}{d+1}}
  \left(
    1 - A f_0^2 \Tsft^4
  \right)
\end{equation}
where
\begin{equation}
  \label{e:Adef}
  A \approx
  \frac{8\pi^6a_p^2}{\Porb^4}
  \left(
    \frac{\langle \Sigma_{04}\rangle}{\langle\Xi^2\rangle}
    - \frac{\langle \Sigma_{02}\rangle^2}{\langle\Xi^2\rangle^2}
    \left\langle
      \cos\frac{2\pi\Tdiff_\alpha}{\Porb}
    \right\rangle_{\!\alpha}
  \right)
\end{equation}
is the mismatch scaling appearing in
\eqref{e:shortsftmismatch}.\footnote{Of course $A$ still depends on
  $\Tmax$ through
  $\left\langle\cos\frac{2\pi\Tdiff_\alpha}{\Porb}\right\rangle$, but
  if $\Tmax$ is small compared to $\Porb$, which we are assuming in
  the scaling of number of templates with $\Tmax$, this average is
  approximately unity.}  The sensitivity at fixed computing cost is
thus maximized when
\begin{equation}
  1 - (4d+5) A f_0^2 \Tsft^4 = 0
\end{equation}
i.e., when the mismatch due to SFT length is
\begin{equation}
  \mism =  A f_0^2 \Tsft^4 = \mism^{\text{opt}} = \frac{1}{4d+5}
\end{equation}
The corresponding optimal SFT length is
\begin{equation}
  \label{e:optsft}
  \Tsft = ([4d+5]A)^{-1/4} f_0^{-1/2}
\end{equation}
For example, if $d=3$, $\mism^{\text{opt}}=\frac{1}{17}\approx 0.059$.
In \fref{f:sftopt}, we show this optimal SFT length for $d=3$, using
$a_p=\LSCApSec\un{s}$ and $\Porb=\GallPorbSec\un{s}$ (the most likely
values for Sco X-1).  The solid line shows the most optimistic
scenario, in which
$\left\langle\cos\frac{2\pi\Tdiff_{\!\alpha}}{\Porb}\right\rangle_\alpha
\approx 1$ (which will be the case for $\Tmax\ll\Porb$) and the dashed
line shows the most pessimistic scenario, in which the average goes to
zero.

\section{Conclusions and outlook}

\label{s:ScoX1}

In this paper we have explored details of the model-based
cross-correlation search for periodic gravitational waves, focusing
on its application to signals from neutron stars in binary systems
(LMXBs) and Scorpius X-1 in particular.  We have carefully considered
the impact of spectral leakage (in \sref{s:kappa}) and the
implications of unknown amplitude parameters (in \sref{s:hsens}) on
the sensitivity of the method.  We have also produced expressions for
the parameter space metric of the search (in \sref{s:metric}), at
varying levels of approximation, and a systematic offset in the
parameters of a detected signal related to the unmeasured inclination
angle of the neutron star to the line of sight (in
\sref{s:systematic}).  In \sref{s:spinwander} we estimated the effects
of ``spin wandering'' caused by deviations from equilibrium in the
torque balance configuration, and in \eqref{s:sftlength} we consider
the appropriate SFT duration needed to avoid significant loss of SNR
due to unmodeled phase acceleration.

We have shown (in \sref{e:scaling}) that the method produces an
improvement in strain sensitivity over the directed stochastic search
method which inspired it; this is roughly proportional to the fourth
root of the product of the coherence time of the model-based search
and the frequency bin size for the stochastic search.  A mock data
challenge \cite{ScoX1MDC} has been carried out by comparing the
performance of the available search methods, including the model-based
cross-correlation search, on simulated signals injected into Gaussian
noise.  As reported elsewhere \cite{ScoX1MDC,CrossCorrMDC}, the
cross-correlation search is the most sensitive one currently implemented.

To give an estimate of expected sensitivity for data from detectors
such as Advanced LIGO and Advanced Virgo, it is necessary to make some
suppositions about the parameters of the search, especially the time
$\Tmax$ over which SFTs are coherently cross correlated.  Since this
drives both the sensitivity and computing cost, the choice of $\Tmax$
will depend on available computing resources, and will likely vary
with frequency in order to optimize the distribution of computing
resources where they can be most effective.  In \cite{CrossCorrMDC},
we performed searches with $9\un{min}\le\Tmax\le 90\un{min}$ for a
range of frequency bands covering a total of $500\un{Hz}$ distributed
in $f_0\in[50,1455]\un{Hz}$, using moderate computational resources.
On the other hand, in \sref{s:spinwander}, we considered spin wandering
effects which might lead to a significant loss of SNR for a search
with $\Tmax\gtrsim 12\un{hr}$ for a one-year observation.

\begin{figure}[tbp]
  \begin{center}
    \includegraphics[width=\columnwidth]{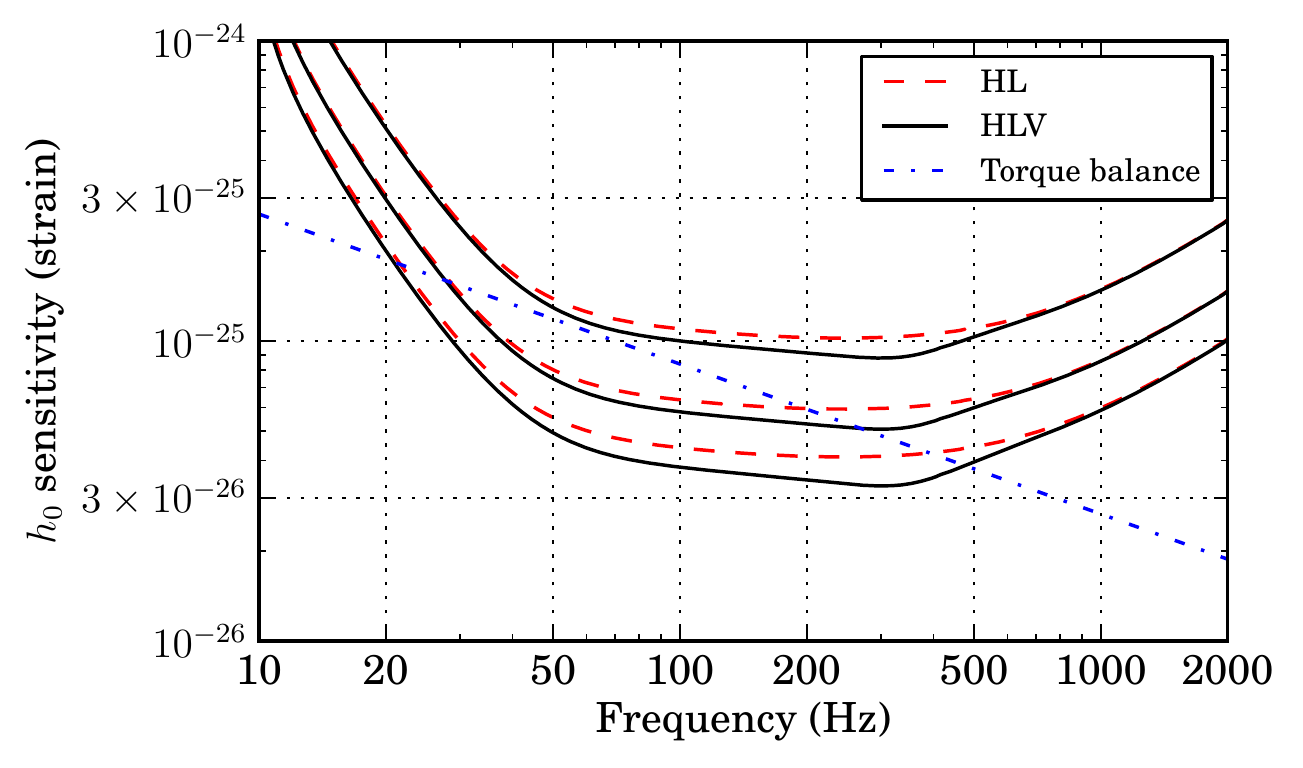}
  \end{center}
  \caption{Expected sensitivity \eqref{e:hsens} for a search of one
    year of coincident data from either the two LIGO detectors
    (labeled HL) or the three LIGO+Virgo detectors (labeled HLV), at
    design sensitivity.  The value plotted is the observable $h_0$ at
    5\% false dismissal probability, assuming an overall false alarm
    probability of 5\% and a trials factor of $10^8$ for a
    single-template false alarm probability of $5\times 10^{-10}$
    (i.e., see \sref{s:hsens} and \tref{tab:statseff}).  The three
    curves in each set, are, from top to bottom, for
    $\Tmax=6\un{min}$, $60\un{min}$ and $10\un{hr}$.  They are
    compared to the signal strength \eqref{e:torquebal} predicted by
    the torque balance argument\cite{Bildsten:1998ey}.}
  \label{f:sensitivity}
\end{figure}
In \fref{f:sensitivity}, we show the projected sensitivity
\eqref{e:hsens} of a search using one year of data, either from the
two advanced LIGO detectors in Hanford, WA and Livingston, LA, or from
the two advanced LIGO detectors plus the Virgo detector in Cascina,
Italy, all operating at their projected design sensitivity.  We show
the sensitivity of three hypothetical searches, with
$\Tmax=6\un{min}$, $60\un{min}$ or $600\un{min}=10\un{hr}$, and
compare the observable $h_0$ (at a 5\% false dismissal probability,
assuming a single-template false alarm probability of $5\times
10^{-10}$, corresponding to an overall 5\% false alarm probability and
a trails factor of $10^8$, as described in \sref{s:hsens} and
\tref{tab:statseff}).
For comparison, we show a representative signal strength predicted by
the torque balance argument\cite{Bildsten:1998ey,Watts:2008qw}.  By
assuming that the spin-down torque due to gravitational waves is
balanced by the spin-up torque due to accretion, estimated using the
observed x-ray flux, it is possible to estimate the strength of the
gravitational-wave signal as a function of the neutron star spin
frequency $\nu_s$\cite{Watts:2008qw}:
\begin{equation}
\begin{split}
  h_0
  \approx
  &\ 
  3\times 10^{-27}
  \left(
    \frac{F_X}{10^{-8}\un{erg}\un{cm}^{-2}\un{s}^{-1}}
  \right)^{1/2}
  \left(
    \frac{\nu_s}{300\un{Hz}}
  \right)^{-1/2}
  \\
  &\times
  \left(
    \frac{R}{10\un{km}}
  \right)^{3/4}
  \left(
    \frac{M}{1.4M_{\odot}}
  \right)^{-1/4}
\end{split}
\end{equation}
The spin frequency of Sco X-1 is unknown, but $\nu_s$ values inferred
for other LMXBs from pulsations or burst oscillations range from
$50\un{Hz}$ to $600\un{Hz}$, so we consider the sensitivity over a wide
range of GW frequencies.
For Sco X-1, using the observed x-ray flux
$F_X=3.9\times10^{-7}\un{erg}\un{cm}^{-2}\un{s}^{-1}$ from
\cite{Watts:2008qw}, and assuming that the GW frequency $f_0$ is twice
the spin frequency $\nu_s$ (as would be the case for GWs generated by
anisotropies in the neutron star), the torque balance value is
\begin{equation}
  \label{e:torquebal}
  h_0 \approx \htorque
  \left(
    \frac{\nu_s}{300\un{Hz}}
  \right)^{-1/2}
  \ ,
\end{equation}
which is the reference curve plotted in \fref{f:sensitivity}.
\begin{figure}[tbp]
  \begin{center}
    \includegraphics[width=\columnwidth]{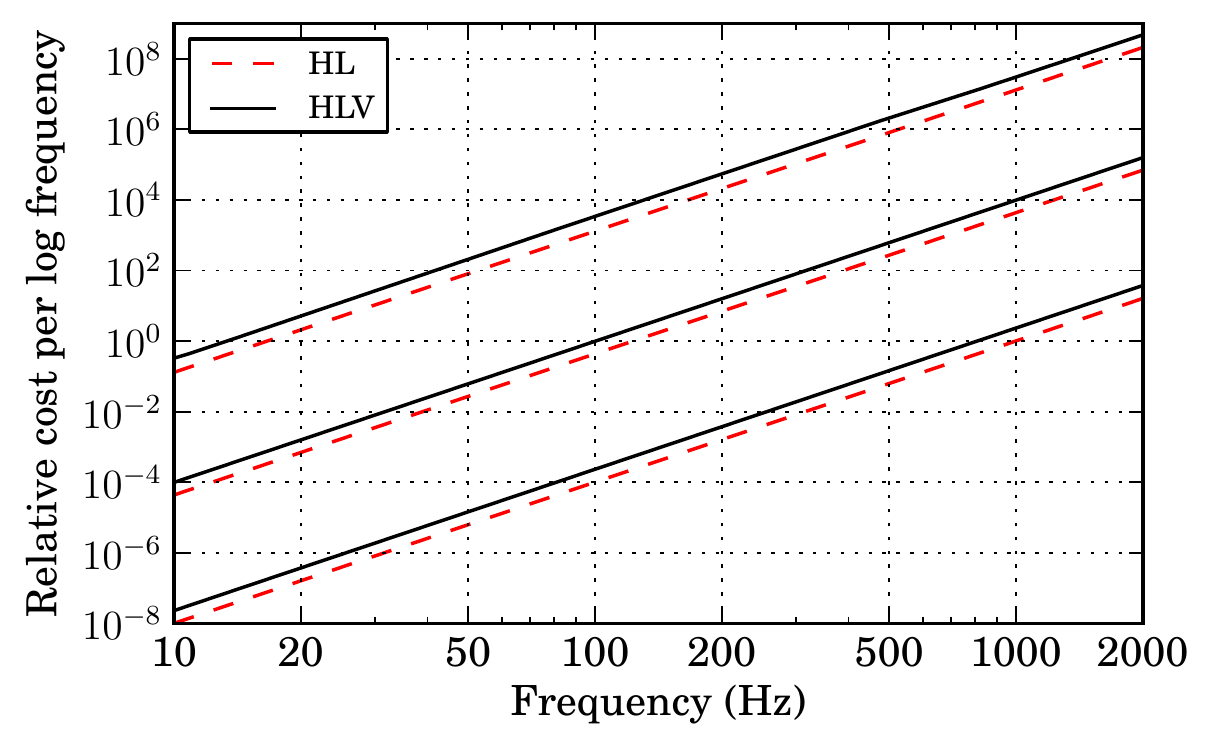}
  \end{center}
  \caption{Relative scaling of expected computing cost per logarithmic
    frequency interval for a search of one year of coincident data
    from either the two LIGO detectors (labeled HL) or the three
    LIGO+Virgo detectors (labeled HLV).  The three curves in each
    set, are, from top to bottom, for $\Tmax=10\un{hr}$, $60\un{min}$
    and $6\un{min}$.  The calculation assumes that the computing cost
    scales with the number of SFT pairs times the number of points in
    parameter space.  It also assumes that the optimal SFT length
    $\Tsft$ given by \eqref{e:optsft} and \eqref{e:Adef} has been
    chosen at each frequency, and that we are searching over frequency
    and two orbital parameters.  The approximate scaling is as
    $f_0^4\Tmax^4$, so for instance a $\Tmax=60\un{min}$ search from
    $100$ to $200\un{Hz}$ would consume the same resources as a
    $\Tmax=6\un{min}$ search from $1000$ to $2000\un{Hz}$.  For
    reference, the mock data analysis in \cite{CrossCorrMDC}, which
    was accomplished in approximately $20,\!000$ CPU-days, covered a
    set of roughly logarithmically-spaced frequency bands totaling
    $250\un{Hz}$ spread from $50\un{Hz}$ to $1375\un{Hz}$ at a range
    of $\Tmax$ values from $9$ to $90\un{min}$.}
  \label{f:compcost}
\end{figure}
We see that for a three-detector, one-year analysis, a signal at the
torque balance limit should be detectable for $30\un{Hz}\lesssim
f_0\lesssim 300\un{Hz}$ with $\Tmax=60\un{min}$ (which is already
computationally manageable at most frequencies), and if one could
increase to $\Tmax=600\un{min}$ through algorithmic improvements,
programming optimization, and/or application of additional resources,
that range could be broadened to $20\un{Hz}\lesssim f_0\lesssim
500\un{Hz}$.  The best-case $h_0$ sensitivity of $\sensHLVmin$ for the
$60$\,min search is consistent with the results of the Sco X-1 MDC
\cite{CrossCorrMDC,ScoX1MDC}, where a cross-correlation search with
$9\un{min}\le\Tmax\le 90\un{min}$ was able to detect signals with
$h_0\gtrsim 5\times 10^{-26}$.

The choice of $\Tmax$ will in part be constrained by computing cost;
in \fref{f:compcost} we show the approximate relative computing cost
scaling for the six searches considered (one year of data from either
the two LIGO detectors or the two LIGO detectors and Virgo, with
a maximum allowed lag time of $\Tmax=6\un{min}$, $60\un{min}$ or
$600\un{min}=10\un{hr}$.  The computing cost is assumed to be
proportional to the number of SFT pairs times the number of parameter
space points to be searched, and we plot the relative cost per
logarithmic frequency interval.  We also assume that at each frequency
the SFT length is chosen to be the optimal SFT length given by
\eqref{e:optsft} and \eqref{e:Adef}.  Roughly speaking, the number of
SFT pairs will scale as $f_0\Tmax$ (since the optimal SFT length
scales as $\Tmax^{-1/2}$), and the density of templates in parameter
space will scale as $f_0^2\Tmax^3$.  The density of points per
logarithmic frequency interval introduces another factor of $f_0$, so
the quantity plotted, cost per unit frequency interval, scales
approximately as $f_0^4\Tmax^4$.  This means that, for example, a
$\Tmax=60\un{min}$ search from $100$ to $200\un{Hz}$ would consume the
same resources as a $\Tmax=6\un{min}$ search from $1000$ to
$2000\un{Hz}$ or a $\Tmax=600\un{min}$ search from $10$ to $20\un{Hz}$.

Finally, we consider one possible avenue for enhancement of the
cross-correlation method.  As explained in \sref{s:meanvar}, the fact
that we filter with $G^{\text{ave}}_{\II\JJ}$ means that the method
provides an estimate of $h_0^{\text{eff}}$, a function of $h_0$ and
$\cos\iota$ defined in \eqref{e:h0eff}, rather than $h_0$.  If we had
a method of independently estimating $\cos\iota$, or in fact any other
combination of $h_0$ and $\cos\iota$ besides $h_0^{\text{eff}}$, we
could obtain a better measurement of $h_0$.  In
\cite{Dhurandhar:2007vb}, a method was proposed to obtain estimates of
$h_0\Ap$ and $h_0\Ac$, but a more effective procedure would seem to be
adding a second statistic which uses
$\cmplxi\Gamma^{\text{circ}}_{\II\JJ}$ [see \eqref{e:abba}] in place
of $\Gamma^{\text{ave}}_{\II\JJ}$ and therefore observes the quantity
$h_0^2\Ap\Ac$; between this and the original $h_0^{\text{eff}}$
estimate, we would be able to disentangle $h_0$ and $\cos\iota$.
This prospect bears further investigation.

\begin{acknowledgments}
  We wish to thank Duncan Galloway, Evan Goetz, Badri Krishnan, Grant
  David Meadors, Chris Messenger, Reinhard Prix, and Keith Riles for
  helpful discussions and comments.
    S.S.\ and Y.Z.\ acknowledge the hospitality of the Center for
  Computational Relativity and Gravitation at Rochester Institute of
  Technology, and the Max Planck Institute for Gravitational Physics
  (Albert Einstein Institute) in Hannover, respectively.  This work
  was supported by NSF Grants No.~PHY-0855494 and No.~PHY-1207010.
    This paper has been assigned LIGO Document No.~\dcc.
\end{acknowledgments}

\appendix

\section{Effects of Nontrivial Windowing}
\label{app:windowing}

\subsection{General formulation}

As noted in \sref{s:sft}, the construction of Fourier transformed data
is often done with a window function $w(\tnorm)$, as in
\eqref{e:sftwin}, as opposed to the unwindowed (or
nearly-rectangularly-windowed) data considered in the main body of the
text.  This appendix considers the impact on the search method and its
sensitivity of using a nontrivial window function, which is
investigated in greater detail in \cite{T1200431-v1}.

The use of windowing for Fourier transforms affects the expected
signal and noise contributions to the data.  For the signal
contribution, Eq.~\eqref{e:signalsft} becomes
\begin{equation}
  \label{e:signalsftwin}
  \cft{h}_{\II\kI} \approx h_0 (-1)^\kI
  e^{\cmplxi\Phi_\II} \frac{F^{\II}_+ \Ap - \cmplxi F^{\II}_\times \Ac}{2}
  \delta^w_{\Tsft}(f_\kI-f_\II)
\end{equation}
where $\delta^w_{\Tsft}(f_\kI-f_\II)$ is the generalization of the
finite time delta function defined in \eqref{e:deltarect}:
\begin{multline}
  \label{e:deltawingen}
  \delta^w_{\Tsft}(f_\kI-f_\II)
  \\
  = \int_{\tmid_\II-\Tsft/2}^{\tmid_\II+\Tsft/2}
  w\left(\frac{\tdet-\tmid_\II}{\Tsft}\right)
  \,e^{-\cmplxi2\pi(f_\kI-f_\II)(\tdet-\tmid_\II)}\,d\tdet
  \\
  = \Tsft \int_{1/2}^{1/2}
  w(\tnorm)\,e^{-\cmplxi2\pi\kappa_{\II\kI}\tnorm}\,d\tnorm
  \equiv\Tsft\,\xi^w(\kappa_{\II\kI})
\end{multline}
with $\kappa_{\II\kI}=(f_\kI-f_\II)\Tsft$ as before.  The noise
contribution is modified by replacing \eqref{e:expectation_rectangular}
with
\begin{equation}
  \ev{\cft{n}_{\II\kI}^w \cft{n}^{w*}_{\JJ\kJ}}
  \approx
  \delta_{\II\JJ}\,\gamma^w_{\kI\kJ}\,\Tsft\,\frac{S_\II}{2}
\end{equation}
where
\begin{equation}
  \begin{split}
    \label{e:gammawindow}
    \gamma^w_{\kI\kJ} &= \frac{(-1)^{\kI-\kJ}}{\Tsft}
    \int_{-\infty}^{\infty}
    \delta^w_{\Tsft}(f_{\kI}-f)\,\delta^{w*}_{\Tsft}(f_{\kJ}-f)
    \,df
    \\
    &= (-1)^{\kI-\kJ}\int_{-1/2}^{1/2} e^{\cmplxi2\pi(\kI-\kJ)\tnorm}
    [w(\tnorm)]^2\,d\tnorm
  \end{split}
\end{equation}
Note that the diagonal elements of this matrix are equal to the
mean square of the window function:
\begin{equation}
  \label{e:gammadiag}
  \gamma^w_{\kI\kI} = \int_{-1/2}^{1/2} [w(\tnorm)]^2\, d\tnorm
  \equiv \wsqbar
\end{equation}

\begin{figure}[tbp]
  \begin{center}
    \includegraphics[width=\columnwidth]{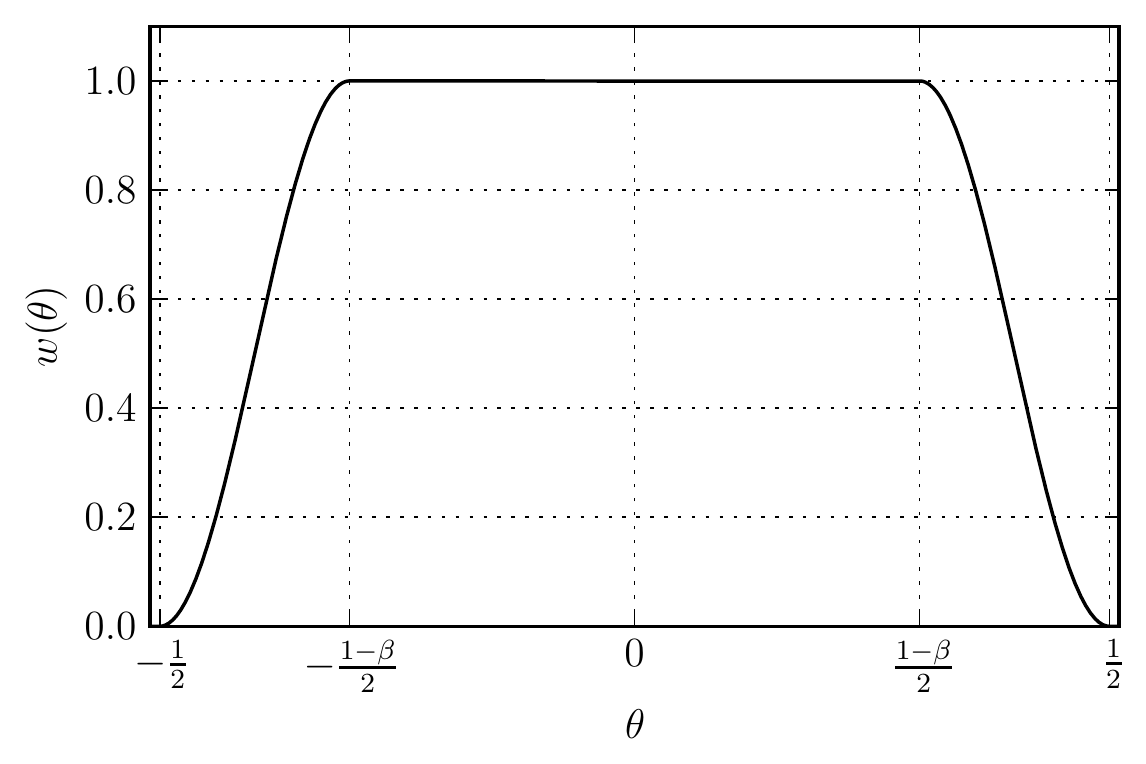}
  \end{center}
  \caption{The general Tukey window $w_\beta(\theta)$ as defined in
    \eqref{e:windowform} for a generic value of the parameter
    $\beta\in[0,1]$, where $\beta$ is the fraction of the window
    length taken up by the transitions from $0$ to $1$ and back.}
  \label{f:beta_tukeywin}
\end{figure}

If we define
\begin{equation}
  z^w_{\II{\kI}} = \cft{x}^w_{\II{\kI}} \sqrt{\frac{2}{\Tsft S_\II}}
\end{equation}
as in \eqref{e:zKkdef}, we have
\begin{multline}
  \label{e:meanzwin}
  \ev{z^w_{\II{\kI}}} = \mu^w_{\II{\kI}}
  \\
  \approx h_0 (-1)^\kI\xi^w(\kappa_{\II\kI})
  e^{\cmplxi\Phi_\II} \frac{F^{\II}_+ \Ap - \cmplxi F^{\II}_\times \Ac}{2}
  \sqrt{\frac{2\Tsft}{S_\II}}
\end{multline}
and
\begin{equation}
  \ev{(z^w_{\II{\kI}}-\mu^w_{\II{\kI}})
    (z^w_{\JJ{\kJ}}-\mu^w_{\JJ{\kJ}})^*}
  = \delta_{\II\JJ}\,\gamma^w_{{\kI}{\kJ}}
  \ .
\end{equation}
We then modify \eqref{e:zbinssum} to
\begin{equation}
  z^w_\II
  \equiv \frac{1}{\Xi^w_{\II}}
  \sum_{\kI\in\mc{K}_{\II}}\sum_{\kI'\in\mc{K}_{\II}}
  (-1)^\kI\xi^{w*}\!(\kappa_{\II\kI})(\gamma^w)^{-1}_{\kI\kI'}\,z^w_{\II\kI'}
\end{equation}
where $\{(\gamma^w)^{-1}_{\kI\kJ}\}$ are the elements of the matrix
inverse of $\{\gamma^w_{\kI\kJ}\}$, and
\begin{equation}
  \label{e:Xiwdef}
  \Xi^w_\II
  = \sqrt{
    \sum_{\kI\in\mc{K}_{\II}}\sum_{\kI'\in\mc{K}_{\II}}
    (-1)^{\kI-\kI'}\xi^{w*}\!(\kappa_{\II\kI})
    \,(\gamma^w)^{-1}_{\kI\kI'}\,
    \xi^w(\kappa_{\II\kI'})
  }
\end{equation}
ensures that the normalization \eqref{e:covcompts} holds as before.
Then the derivation proceeds as before, with $\Xi^w_\II$ replacing
$\Xi_\II$, and, in particular, the expected SNR \eqref{e:evrhoXi} becomes
\begin{equation}
  \label{e:evrhoXiwin}
  \ev{\rho}
  \approx
  (h^{\text{eff}}_0)^2
  \langle(\Xi^w)^2\rangle
  \sqrt{
    2\sum_{\II\JJ\in\mc{P}}
    \left(\widehat{\Gamma}^{\text{ave}}_{\II\JJ}\right)^2
  }
\end{equation}

\subsection{Results for specific windows}

\begin{figure}[tbp]
  \begin{center}
    \includegraphics[width=\columnwidth]{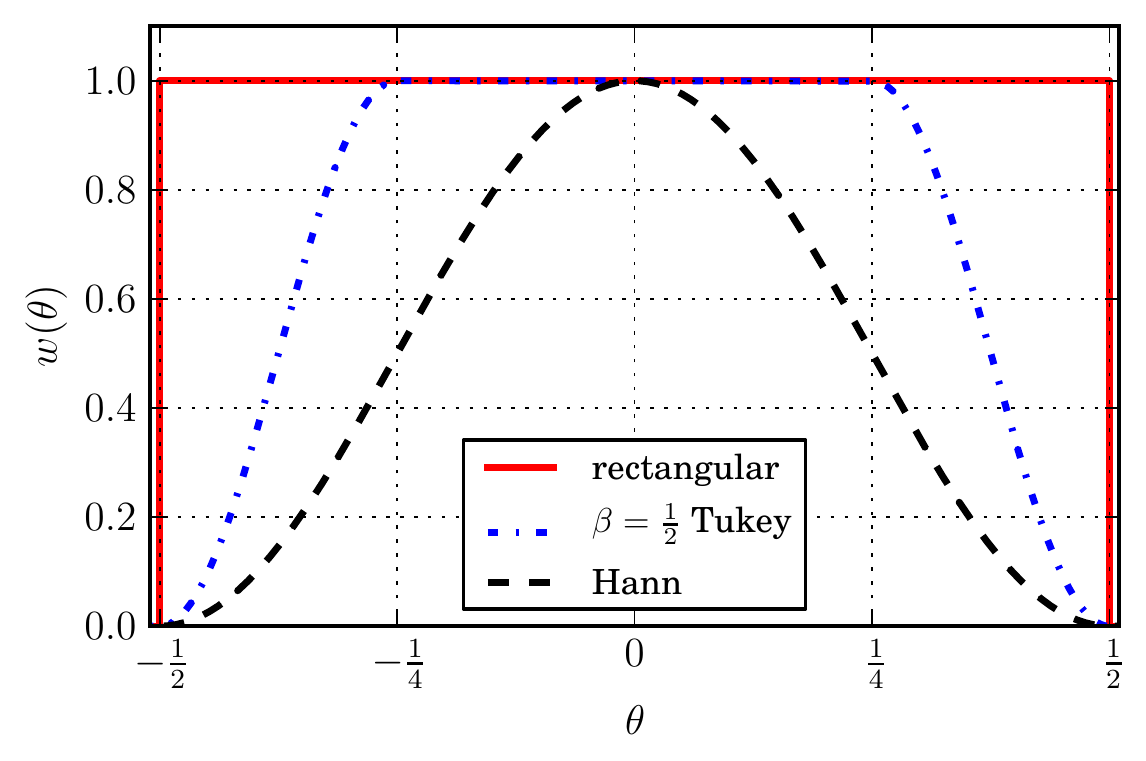}
  \end{center}
  \caption{Specific versions of the general Tukey window
    $w_\beta(\theta)$ as defined in \eqref{e:windowform}: the
    rectangular window $w^{\text{rect}}(\theta)=w_0(\theta)$, a
    canonical ($\beta=\frac{1}{2}$) Tukey window $w_{1/2}(\theta)$,
    and the Hann window $w^{\text{Hann}}(\theta)=w_1(\theta)$.}
  \label{f:windows_all}
\end{figure}

We now consider the consequences of the modification
\eqref{e:evrhoXiwin} by investigating the form of
$\xi^w(\kappa)=\Tsft^{-1}\delta^w_{\Tsft}(\kappa/\Tsft)$ defined in
\eqref{e:deltawingen} and $\gamma^w_{\kI\kJ}$ defined in
\eqref{e:gammawindow} for specific nonrectangular window choices.  We
consider the general family of Tukey windows, defined using an
adjustable parameter $0\le\beta\le 1$ by
\begin{equation}
  \label{e:windowform}
  w_\beta(\tnorm) =
  \begin{cases}
    \frac{1}{2}
    \left(
      1-\cos \frac{\pi}{\beta}(2\tnorm+1)
    \right)
    &
    -\frac{1}{2} \le \tnorm \le -\left(\frac{1-\beta}{2}\right)
    \\
    1
    &
    -\left(\frac{1-\beta}{2}\right) \le \tnorm \le \left(\frac{1-\beta}{2}\right)
    \\
    \frac{1}{2}
    \left(
      1-\cos \frac{\pi}{\beta}(2\tnorm-1)
    \right)
    &
    \left(\frac{1-\beta}{2}\right) \le \tnorm \le \frac{1}{2}
  \end{cases}
  \ .
\end{equation}
The general form of the Tukey window is illustrated in
\fref{f:beta_tukeywin}.
This family includes at its extremes the rectangular window
($\beta=0$) and the Hann window ($\beta=1$).  In practical
applications it is also common to use a Tukey window with a small
finite parameter $\beta\ll1$ rather than a pure rectangular window.
These two specific cases are shown in \fref{f:windows_all}, along with
a Tukey window with $\beta=\frac{1}{2}$.

\begin{figure}[tbp]
  \begin{center}
    \includegraphics[width=\columnwidth]{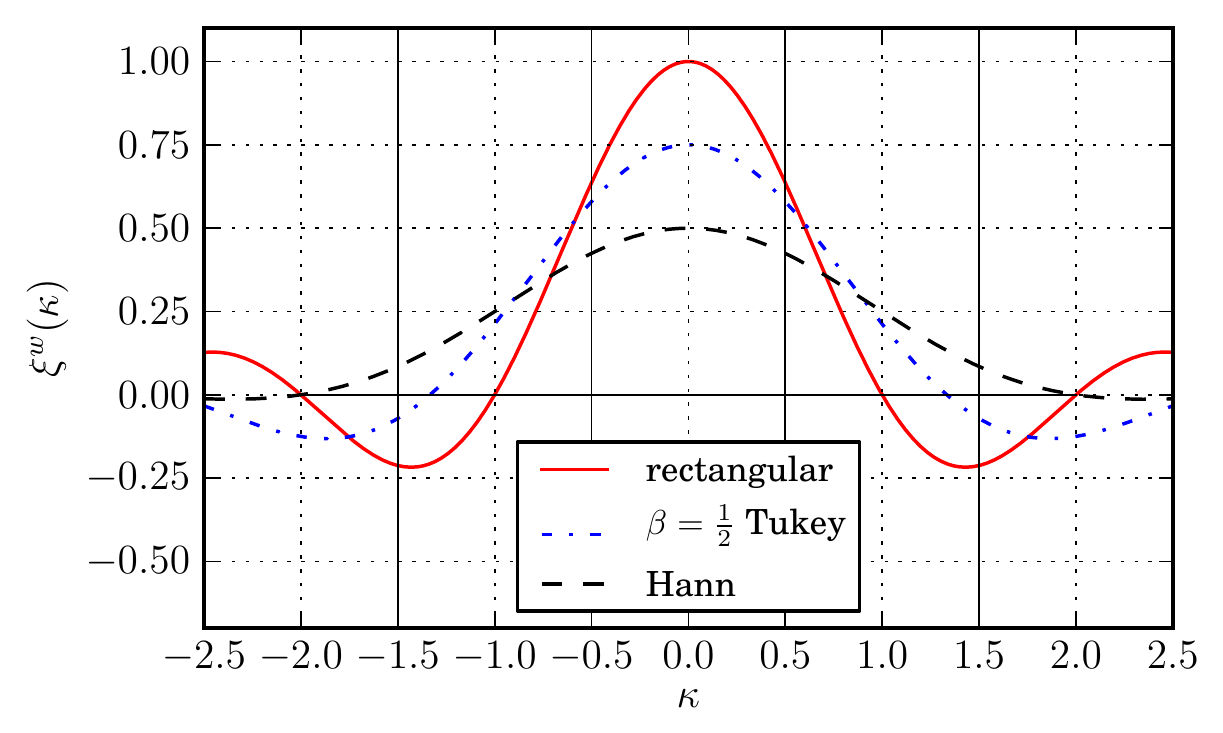}
  \end{center}
  \caption{The window leakage function
    $\xi^w(\kappa)=\Tsft^{-1}\delta^w_{\Tsft}(\kappa/\Tsft)$ defined
    in \eqref{e:deltawingen} for the windows shown in
    \fref{f:windows_all}.  The explicit formulas are given in
    \eqref{e:rectdelta} for the rectangular window,
    \eqref{e:tukeydelta} for the canonical ($\beta=\frac{1}{2}$) Tukey
    window, and \eqref{e:Hanndelta} for the Hann window.  Note that
    the version for rectangular-windowed data is just
    $\xi^{\text{rect}}(\kappa)=\Tsft^{-1}\delta_{\Tsft}(\kappa/\Tsft)
    =\sinc(\kappa)=\frac{\sin\pi\kappa}{\pi\kappa}$, which is the
    finite-time delta function plotted in \fref{f:deltaT}.}
  \label{f:deltaT_all}
\end{figure}

We can insert the general form of $w_\beta(\theta)$ from
\eqref{e:windowform} into \eqref{e:deltawingen} to obtain
\begin{multline}
  \label{e:windowdelta}
  \xi^w_\beta(\kappa)
  =  \frac{1}{2}\sinc \kappa + \frac{1}{2}(1-\beta)\sinc (\kappa[1-\beta])
  \\
  + \frac{\beta}{4}
  \sin \left(\pi \kappa \left[1-\frac{\beta}{2}\right]\right)
  \left[
    \sinc \left(\frac{1 - \beta\kappa}{2}\right)
    - \sinc \left(\frac{1 + \beta\kappa}{2}\right)
  \right]
  \ ;
\end{multline}
the ``interesting'' values of $\beta$ also have somewhat simpler
explicit forms.  For the rectangular window ($\beta=0$), which was
considered in the main body of the paper, we have
\begin{equation}
  \label{e:rectdelta}
  \xi^w_0(\kappa)
  = \xi^{\text{rect}}(\kappa) = \sinc\kappa
  \ ;
\end{equation}
for the Hann window ($\beta=1$), we have
\begin{equation}
  \label{e:Hanndelta}
  \xi^w_1(\kappa)
  = \xi^{\text{Hann}}(\kappa)
  = \frac{1}{2}\sinc \kappa
  + \frac{1}{4}\sinc (1-\kappa)
  + \frac{1}{4}\sinc (1+\kappa)
  \ ;
\end{equation}
and for the canonical ($\beta=\frac{1}{2}$) Tukey window, we have
\begin{multline}
  \label{e:tukeydelta}
  \xi^w_{1/2}(\kappa)
  = \xi^{\text{Tukey}}(\kappa)
  \\
  = \frac{1}{2}\sinc \kappa
  - \frac{1}{4}\sinc (2+\kappa)
  - \frac{1}{4}\sinc (2-\kappa)
  \\
  + \frac{1}{4}\sinc \frac{\kappa}{2}
  + \frac{1}{8}\sinc \left(1+\frac{\kappa}{2}\right)
  + \frac{1}{8}\sinc \left(1-\frac{\kappa}{2}\right)
\end{multline}
We plot these three functions in \fref{f:deltaT_all}.

\begin{figure}[tbp]
  \begin{center}
    \includegraphics[width=\columnwidth]{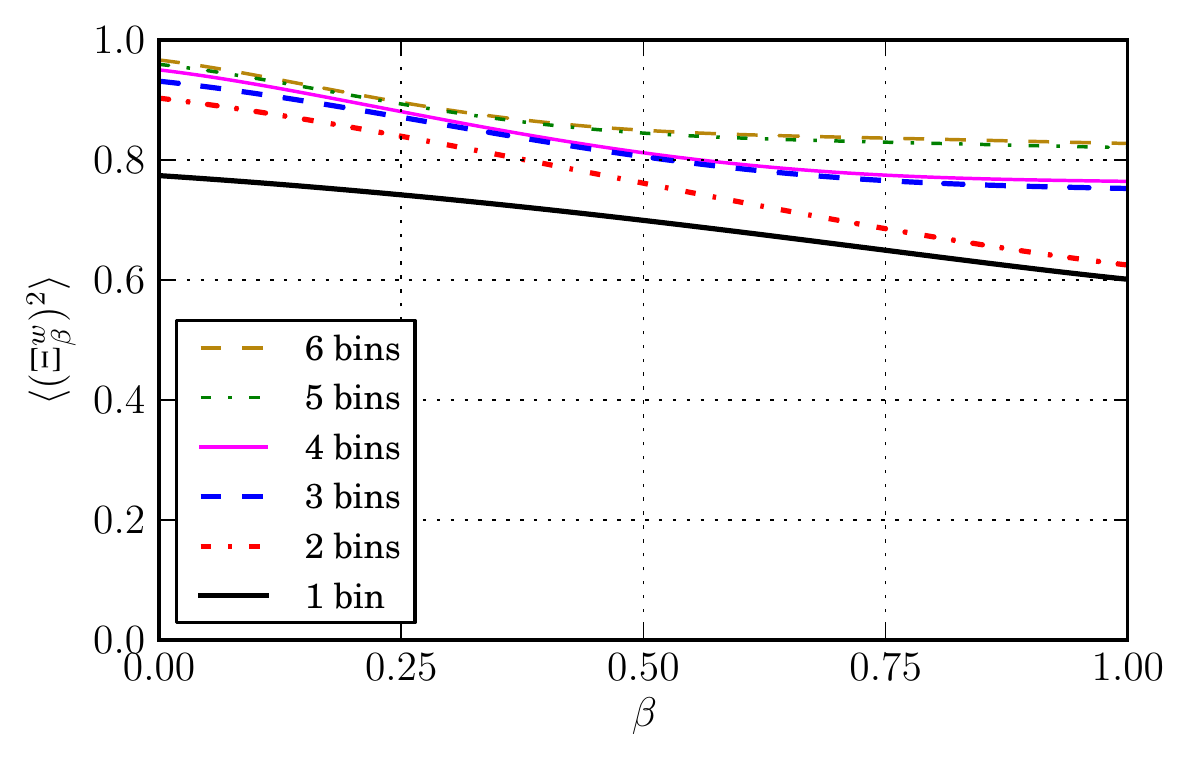}
  \end{center}
  \caption{The leakage factor $\langle(\Xi^w)^2\rangle$ appearing in
    \eqref{e:evrhoXiwin} for a search using between one and six bins
    from each SFT, assuming a general Tukey window from the family
    \eqref{e:windowform}.  We see that, for any number of bins, the
    most sensitive search is when $\beta=0$, i.e., for rectangular
    windows.  In particular, when a single bin is used from each SFT,
    we have $\langle\Xi^2\rangle=\Xisqaverect$ for rectangular
    windowing ($\beta=0$),
    $\langle(\Xi^{w}_\beta)^2\rangle=\XisqaveTukey$ for a canonical
    ($\beta=\frac{1}{2}$) Tukey window, and
    $\langle(\Xi^{\text{Hann}})^2\rangle=\XisqaveHann$ for Hann
    windowing ($\beta=1$).  Note that the $\beta=0$ value on each
    curve is just the corresponding ``cumulative'' number from
    \tref{tab:bincontribs}.}
  \label{f:beta_Xiave}
\end{figure}

To evaluate the factor of $\langle(\Xi^w)^2\rangle$ appearing in
\eqref{e:evrhoXiwin}, we need to construct the matrix
$\{\gamma^w_{\kI\kJ}\}$ via \eqref{e:gammawindow}.  Substituting
\eqref{e:windowform} into \eqref{e:gammawindow}, we can find
\begin{multline}
  \label{e:gammabeta}
  (\gamma^{w}_{\beta})_{\kI\kJ}
  =
  (-1)^{\kI-\kJ}(1-\beta)\sinc\left[(\kI-\kJ)(1-\beta)\right]
  \\
  + \frac{3}{8}\beta\sinc\left[(\kI-\kJ)\beta\right]
  \\
  - \frac{1}{4}\beta\sinc\left[(\kI-\kJ)\beta-1\right]
  - \frac{1}{4}\beta\sinc\left[(\kI-\kJ)\beta+1\right]
  \\
  + \frac{1}{16}\beta\sinc\left[(\kI-\kJ)\beta-2\right]
  + \frac{1}{16}\beta\sinc\left[(\kI-\kJ)\beta+2\right]
  \ .
\end{multline}
We can see that, for the rectangular case $\beta=0$, we get
$(\gamma^w_0)_{\kI\kJ}=\delta_{\kI\kJ}$ as before, while for the Hann
case $\beta=1$ we have
\begin{equation}
  \gamma^{\text{Hann}}_{\kI\kJ}
  =
  \frac{3}{8} \delta_{\kI,\kJ}
  - \frac{1}{4}\delta_{\kI,\kJ-1} - \frac{1}{4} \delta_{\kI,\kJ+1}
  + \frac{1}{16}\delta_{\kI,\kJ-2} + \frac{1}{16}\delta_{\kI,\kJ+2}
  \ .
\end{equation}
The diagonal elements for general $\beta$ are
\begin{equation}
  \label{e:gammadiagbeta}
  (\gamma^{w}_{\beta})_{\kI\kI} = 1 - \frac{5}{8}\beta = \wsqbarbeta
\end{equation}
as in \eqref{e:gammadiag}.  This means that, in the special case where
the set of bins $\mc{K}_{\II}$ from each SFT is just the ``best bin''
$\kbest_\II$ defined in \eqref{e:bestbin}, the matrix
$\{\gamma^w_{\kI\kJ}\}$ just has a single element
$\gamma^w_{\kbest_\II\kbest_\II} = 1 - \frac{5}{8}\beta$, and
\begin{equation}
  (\Xi^w_\II)^2
  = \frac{\abs{\xi^w_\beta(\kappabest_{\II})}^2}{1 - \frac{5}{8}\beta}
\end{equation}
where $\xi^w_\beta(\kappa)$ is defined in \eqref{e:windowdelta}.  In
general, though, we need to invert the matrix \eqref{e:gammabeta} and
then average $(\Xi^w_\II)^2$ defined in \eqref{e:Xiwdef} over possible
values of $\kappabest_{\II}$.  We plot the results in
\fref{f:beta_Xiave} as a function of $\beta$, for cases where we take
the ``best'' $m$ bins from each SFT.  We see that, for any number of
bins, $\langle(\Xi^w)^2\rangle$ is a maximum for $\beta=0$, i.e.,
rectangular windowing.
The $\beta=0$ values are just the ``cumulative'' entries from
\tref{tab:bincontribs} for the corresponding number of bins.
Specifically, for the single-bin case, when $\beta=0$, we have
$\langle\Xi^2\rangle=\Xisqaverect$ (as seen in the $m=1$ entry of
\tref{tab:bincontribs}), when $\beta=\frac{1}{2}$, we have
$\langle(\Xi^{\text{Tukey}})^2\rangle=\XisqaveTukey$, and when
$\beta=1$, we have $\langle(\Xi^{\text{Hann}})^2\rangle=\XisqaveHann$.
These values also appear in \cite{T1200431-v1}, which explains in more
detail the relevant phenomenon.  While the dropoff from the maximum
value of $(\Xi^w_\II)^2$ to its average value is greatest for
rectangular windowing, the maximum value and the average value are
also greatest for the rectangular window.

A common approach to handle the loss of signal associated with
Hann-windowed data is to divide the data into overlapping
Hann-windowed data segments, as in \cite{Goetz:2011bd}.  For the
present search, however, it is easier just to include more bins from
the rectangularly windowed Fourier transform, if desired, to increase
the sensitivity of the search.  The only drawback to that is a slight
increase in computational time, but this increase is much smaller than
what would arise from almost doubling the number of SFTs by the use of
overlapping windows.

\begin{figure*}[tbp]
  \begin{center}
    \includegraphics[width=0.45\textwidth]{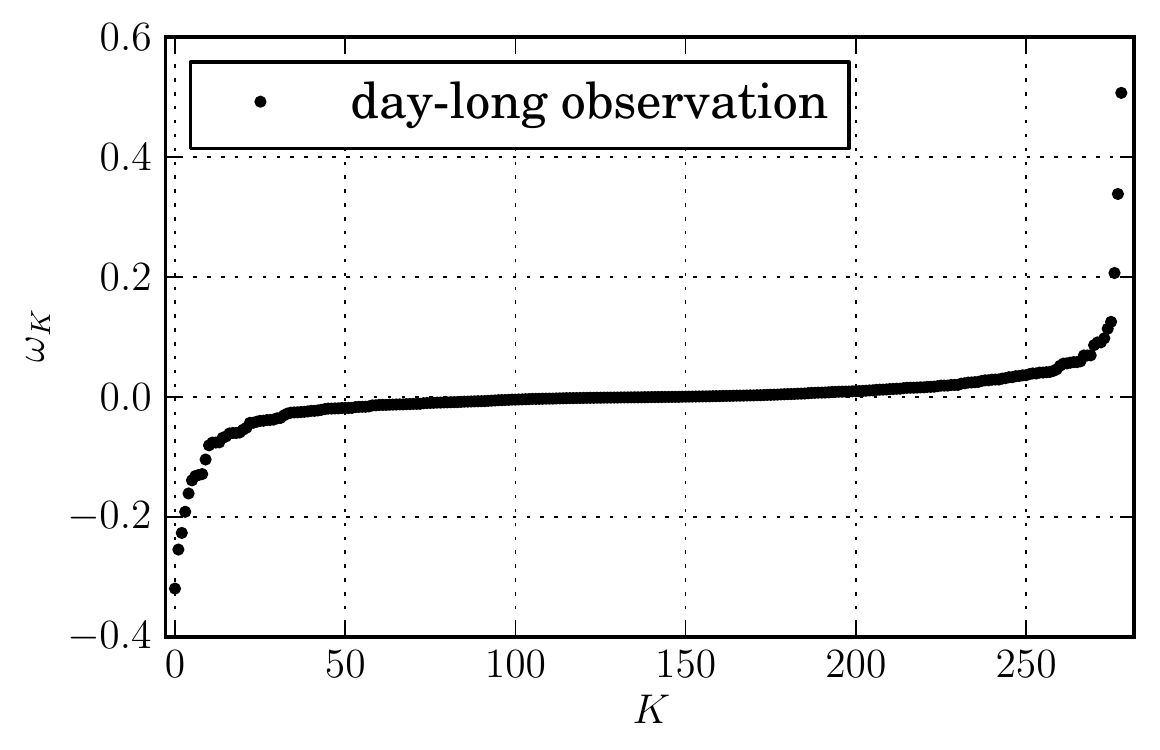}
    \includegraphics[width=0.45\textwidth]{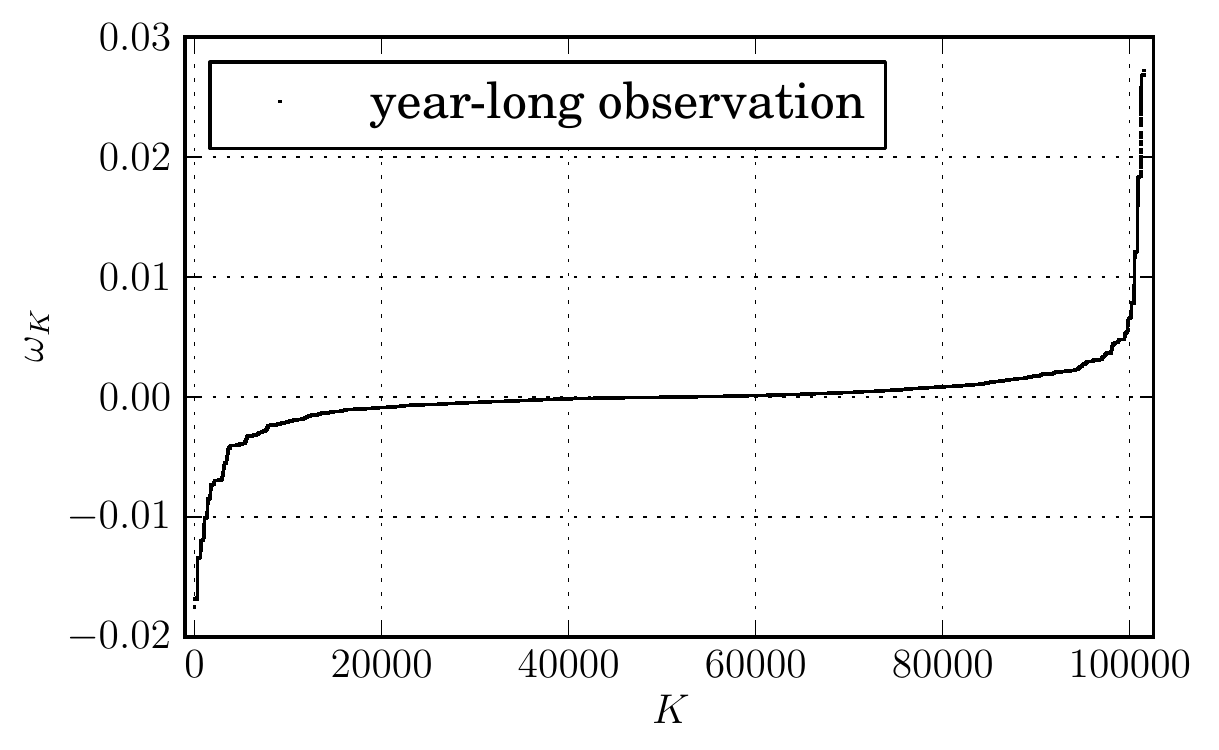}
  \end{center}
  \caption{Eigenvalues $\{\omega_\II\}$ of the weights matrix $\Wmat$
    defined in\eqref{e:Wdef} for two scenarios.
    On the left, we show one day of
    observation with the LLO, LHO and Virgo detectors, assuming equal
    sensitivity, with $\Tsft=900\un{s}$ and $\Tmax=3600\un{s}$.
    On the right, we show one year of observation under the same conditions,
    constructed as the union of 365 such days, spread throughout the
    year.  In both cases, the start and end of each day include data
    gaps of 900--1800\un{s}, randomly and independently generated
    for each detector.}
  \label{f:omega}
\end{figure*}

\section{Probability distribution for cross-correlation statistic in Gaussian noise}

\label{app:stats}

In this appendix, we consider the detailed statistical properties of
the cross-correlation statistic \eqref{e:CCbins} in the presence of
Gaussian noise.  If the noise contribution to $\cft{x}_{\II\kI}$ is
Gaussian, the definitions \eqref{e:zKkdef} and \eqref{e:zbinssum}
imply that $\zvec-\muvec$ is a circularly symmetric Gaussian random
vector \cite{Gallager} with zero mean, unit covariance and zero
pseudocovariance, as described in \eqref{e:zvecstats}.  If
$\{\omega_{\II}\}$ and $\{\vvec_{\II}\}$ are the eigenvalues and
eigenvectors, respectively, of the Hermitian weighting matrix $\Wmat$
defined in \eqref{e:Wdef} , so that
\begin{equation}
  \Wmat = \sum_{\II} \vvec_{\II} \omega_{\II} \adj{\vvec}_{\II}
\end{equation}
then the statistic is
\begin{equation}
  \rho = \sum_{\II} \adj{\zvec}\vvec_{\II} \omega_{\II} \adj{\vvec}_{\II}\zvec
  = \sum_{\II} \omega_{\II} \abs{\adj{\vvec}_{\II}\zvec}^2
  \ .
\end{equation}
The conditions $\Tr(\Wmat)=0$ and $\Tr(\Wmat^2)=1$ imply that
$\sum_\II\omega_\II=0$ and $\sum_\II\omega_\II^2=1$.  To give an
example of the typical form of the eigenvalues, we present in
\fref{f:omega} two typical sets of eigenvalues, one assuming a day-long
observation with three detectors, assuming $\Tsft=900\un{s}$ and
$\Tmax=3600\un{s}$, the other combining 365 such observations with
randomly staggered starting times to simulate a year-long observation,
assuming LIGO Livingston, Hanford and Virgo detectors with identical
and stationary noise spectra.\footnote{Note that since
  $\widehat{G}^{\text{ave}}_{\II\JJ}=e^{\cmplxi\Phi_{\II}}\widehat{\Gamma}^{\text{ave}}_{\II\JJ}e^{-\cmplxi\Phi_{\JJ}}$,
  a matrix made of the $\{\widehat{\Gamma}^{\text{ave}}_{\II\JJ}\}$
  has the same eigenvalues as one made of the
  $\{\widehat{G}^{\text{ave}}_{\II\JJ}\}$.  If the noise PSDs are
  (approximately) the same for all SFTs, it is also equivalent to using
  the eigenvalues of a metric made of the
  $\{\Gamma^{\text{ave}}_{\II\JJ}\}$.}

Each $\adj{\vvec}_{\II}\zvec$ is an independent circularly symmetric
Gaussian random variable with zero mean and unit variance, which means its
real and imaginary parts are independent Gaussian random variables
with mean zero and variance $\frac{1}{2}$.  Thus
$\abs{\adj{\vvec}_{\II}\zvec}^2$ is $\frac{1}{2}$ times a $\chi^2(2)$
random variable, i.e., it is an exponential random variable with unit
rate parameter.  The characteristic function is thus
\begin{equation}
  \varphi_{\II}(t) = \ev{e^{{\cmplxi}t\abs{\adj{\vvec}_{\II}\zvec}^2}}
  = \frac{1}{1-{\cmplxi}t}
\end{equation}
which means that the characteristic function of the cross-correlation
statistic is
\begin{equation}
  \label{e:charfcn}
  \begin{split}
  \varphi(t) &= \ev{
    \exp\left(
      {\cmplxi}t\sum_{\II} \omega_{\II} \abs{\adj{\vvec}_{\II}\zvec}^2
    \right)
  }
  \\
  &= \prod_{\II} \varphi_{\II}(\omega_{\II} t)
  = \frac{1}{\prod_{\II}(1-{\cmplxi}\omega_{\II} t)}
  \end{split}
\end{equation}
This allows a straightforward computation of the exact probability density function for the
statistic $\rho$ as
\begin{equation}
  f(\rho|h_0=0) =
  \begin{cases}
    \sum_{\II,\,\omega_{\II}>0}
    \frac{\omega_{\II}^{-1}e^{-\rho/\omega_{\II}}}{\prod_{\JJ\ne\II}(1-\omega_{\JJ}/\omega_{\II})}
    & \rho>0 \\
    \sum_{\II,\,\omega_{\II}<0}
    \frac{-\omega_{\II}^{-1}e^{\rho/\omega_{\II}}}{\prod_{\JJ\ne\II}(1-\omega_{\JJ}/\omega_{\II})}
    & \rho<0
  \end{cases}
\end{equation}
which is a a mixture of exponential distributions.  To get the false
alarm probability $\fap$ at a threshold $\rhoth>0$, we calculate
\begin{equation}
  \begin{split}
  \fap &\equiv P(\rho>\rhoth|h_0=0)
  = \int_{\rhoth}^{\infty} f(\rho|h_0=0)\,d\rho
  \\
  &= \sum_{\II,\,\omega_{\II}>0}
  \frac{e^{-\rhoth/\omega_{\II}}}{\prod_{\JJ\ne\II}(1-\omega_{\JJ}/\omega_{\II})}
  \end{split}
\end{equation}
The problem with this expression is that the denominator can get very
small, and the signs of the terms alternate.  To see this, assume that
we have ordered the eigenvalues so that
\begin{equation}
  \omega_N > \omega_{N-1} > \cdots > \omega_{\II_0} > 0 > \omega_{\II_0-1}
  > \cdots > \omega_1
\end{equation}

\begin{figure}[t!]
  \begin{center}
    \includegraphics[width=\columnwidth]{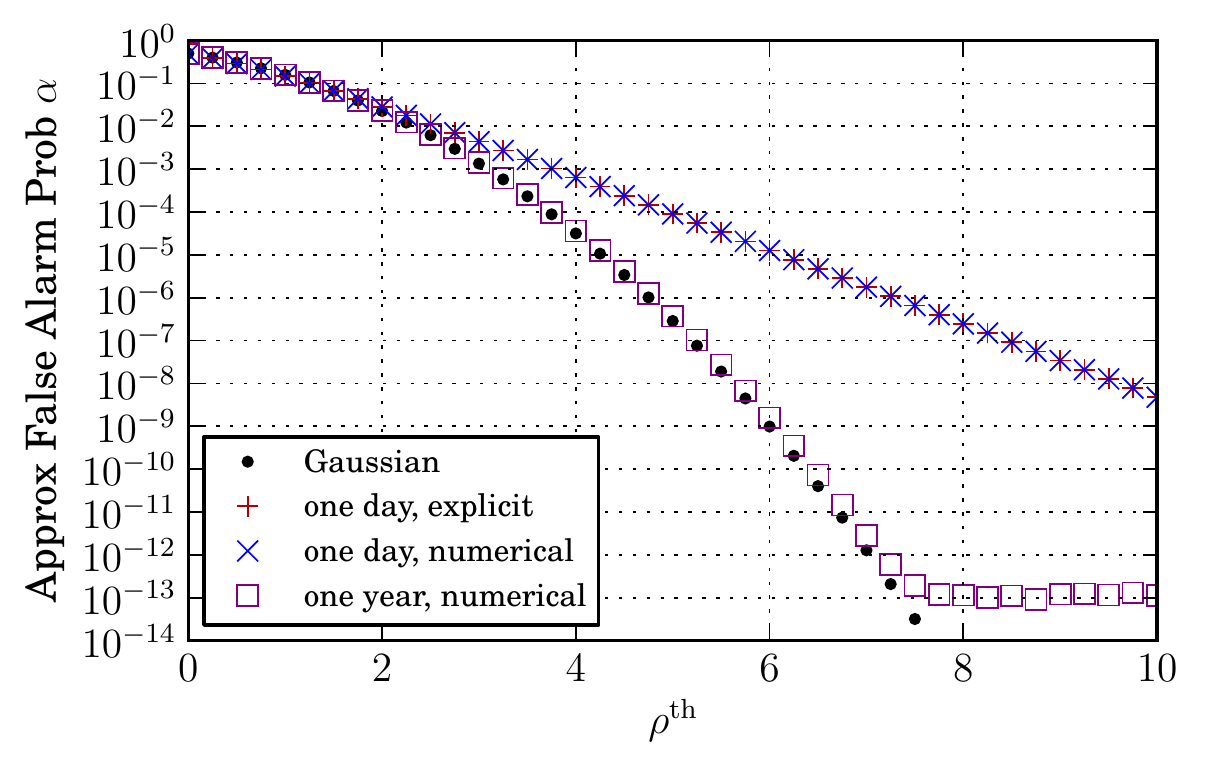}
  \end{center}
  \caption{False alarm probabilities for the cross-correlation
    statistic in the day-long and year-long scenarios considered in
    \fref{f:omega}, using the explicit formula \eqref{e:pval-exact} as
    well as numerical integration of \eqref{e:fapnumerical}, along
    with the probabilities we would get if we assumed the statistic to be
    Gaussian.  For a day-long observation (with three detectors,
    $\Tsft=900\un{s}$ and $\Tmax=3600\un{s})$, both methods give
    comparable results, but the Gaussian approximation is invalid for
    single-template false alarm probabilities below about $10^{-2}$.  Note
    that for large signal values, a single exponential term dominates.
    For a year-long observation, practical calculation with
    \eqref{e:pval-exact} is impossible due to underflow issues.  The
    numerical integration of \eqref{e:fapnumerical} becomes unstable
    for false alarm probabilities below $10^{-12}$, but not before quantifying
    deviations from the Gaussian approximation even for a year-long
    observation.}
  \label{f:FAcomp}
\end{figure}
Then
\begin{multline}
    \prod_{\JJ\ne\II}\left(1-\frac{\omega_{\JJ}}{\omega_{\II}}\right)
    = \left[
      \prod_{\JJ=1}^{\II-1} \left(1-\frac{\omega_{\JJ}}{\omega_{\II}}\right)
    \right]
    \left[
      \prod_{\JJ=\II+1}^{N} \left(1-\frac{\omega_{\JJ}}{\omega_{\II}}\right)
    \right]
    \\
    =
    (-1)^{N-\II}
    \left[
      \prod_{\JJ=1}^{\II-1} \left(1-\frac{\omega_{\JJ}}{\omega_{\II}}\right)
    \right]
    \left[
      \prod_{\JJ=\II+1}^{N} \left(\frac{\omega_{\JJ}}{\omega_{\II}}-1\right)
    \right]
\end{multline}
and the false alarm probability is
\begin{multline}
  \label{e:pval-exact}
  \fap
  = \sum_{\II=\II_0}^N
  (-1)^{N-\II}
  e^{-\rhoth/\omega_{\II}}
  \\
  \times
  \left[
    \prod_{\JJ=1}^{\II-1} \left(1-\frac{\omega_{\JJ}}{\omega_{\II}}\right)
  \right]^{-1}
  \left[
    \prod_{\JJ=\II+1}^{N} \left(\frac{\omega_{\JJ}}{\omega_{\II}}-1\right)
  \right]^{-1}
\end{multline}
The last two factors can be very large, and are larger when the
eigenvalues are closer together.  (Recall that $N$ is the number of
SFTs, which is approximately $\Tobs/\Tsft$, so there are many factors
appearing in the product.)

Given the numerical problems with the exact false alarm probability
\eqref{e:pval-exact} when the number of SFTs is large, it is sometimes
necessary to use an alternate approach.  We can perform a calculation
analogous to that in \cite{Goetz:2011bd}, based on the method of
\cite{Davies:1973,Davies:1980}.  This uses the Gil-Pelaez
expression\cite{Gil-Pelaez:1951} to construct a cumulative
distribution directly from the characteristic function
\eqref{e:charfcn} according to
\begin{equation}
  \label{e:fapnumerical}
  \fap
  = \frac{1}{2} + \frac{1}{\pi} \int_0^\infty
  \Imag\left(\varphi(t)\,e^{-\cmplxi t\rhoth}\right)\frac{dt}{t}
\end{equation}
We can then find the false alarm probability by numerical integration
of \eqref{e:fapnumerical}.  Results of both of these methods are shown
in \fref{f:FAcomp}, for the two scenarios considered in
\fref{f:omega}.  Both methods produce consistent results for a
day-long observation, and illustrate deviation of the false alarm probability
from the Gaussian value for $\rhoth\gtrsim 2$.  For the year-long
observation, explicit evaluation of \eqref{e:pval-exact} is impossible
because of underflow in the cancellations, but numerical integration
of \eqref{e:fapnumerical} works until the false alarm probability goes below
$10^{-12}$ or so.  False alarm probabilities are considered in detail for a
wider range of observing scenarios in \cite{CrossCorrMDC}.

\end{document}